\documentclass[iop,appendixfloats,numberedappendix]{emulateapj}

\usepackage[caption=false]{subfig}
\usepackage{graphics}
\usepackage{graphicx}
\usepackage{natbib}
\usepackage{doi}
\usepackage{rotfloat}
\usepackage{setspace}
\usepackage{amssymb}
\usepackage{gensymb}
\usepackage{textcomp}
\usepackage{morefloats}
\restylefloat{figure}

\newcommand{\extravspace}{\rule{0pt}{2.7ex}}

\newcommand{\hh}{H$_{2}$}
\newcommand{\um}{$\mu$m~}
\newcommand{\kms}{\,\mbox{km s$^{-1}$}}
\newcommand{\OI}{[{O}{I}]}

\newcommand{\OIII}{[{O}{III}]}
\newcommand{\SII}{[{S}{II}]}
\newcommand{\NII}{[{N}{II}]}

\newcommand{\OII}{[{O}{II}]}

\newcommand{\Ha}{H$\alpha\,$}
\newcommand{\Hb}{H$\beta\,$}

\newcommand{\HII}{{H}{II}}

\begin{document}

%to get margins right:
\voffset=-0.6in

\title{Galaxy Mergers Drive Shocks: an Integral Field Study of GOALS galaxies} 

\author{J. A. Rich \altaffilmark{1,2},  L. J. Kewley\altaffilmark{3} \&  M. A. Dopita\altaffilmark{3,4}}
\email{jrich@ipac.caltech.edu}
\altaffiltext{1}{IPAC, California Institute of Technology, 1200 E. California Blvd., Pasadena, CA 91125}
\altaffiltext{2}{Observatories of the Carnegie Institution of Washington, 813 Santa Barbara St., Pasadena, CA 91101}
\altaffiltext{3}{Research School of Astronomy and Astrophysics, Australian National University, Cotter Rd., Weston ACT 2611, Australia}
\altaffiltext{4}{Astronomy Department, Faculty of Science, King Abdulaziz University, PO Box 80203, Jeddah, Saudi Arabia}

\date{\today}

\begin{abstract}
We present an integral field spectroscopic study of radiative shocks in 27 nearby ultraluminous and luminous infrared galaxies (U/LIRGs) from the Great Observatory All-sky LIRG Survey, a subset of the Revised Bright Galaxy Sample. Our analysis of the resolved spectroscopic data from the Wide Field Spectrograph (WiFeS) focuses on determining the detailed properties of the emission line gas, including a careful treatment of multi-component emission line profiles. The resulting information obtained from the spectral fits are used to map the kinematics of the gas, sources of ionizing radiation and feedback present in each system. The resulting properties are tracked as a function of merger stage. Using emission line flux ratios and velocity dispersions, we find evidence for widespread, extended shock excitation in many local U/LIRGs. These low-velocity shocks become an increasingly important component of the optical emission lines as a merger progresses. We find that shocks may account for as much as half of the \Ha luminosity in the latest-stage mergers in our sample. We discuss some possible implications of our result and consider the presence and effects of AGN on the spectra in our sample.
\end{abstract}


\section{Introduction}
As gas rich starburst galaxies and massive mergers, ultraluminous and luminous infrared galaxies (U/LIRGs) in the local universe provide valuable insight and context for phenomena that impact galaxy formation and evolution at high redshift. ULIRGs, with total L$_{IR}$/L$_{\odot}>10^{12}$ are overwhelmingly represented by systems in the latest stages of a merger between two gas-rich galaxies \citep{Sanders88,Veilleux95,Murphy96,Genzel98,Rigopoulou99,Scoville00,Kewley01b,Spoon06,Desai07}. The strong infrared luminosity (L$_{IR}$) in local LIRGs is powered by star formation and in some cases AGN, both driven by galaxy mergers and interactions. LIRGs (L$_{IR}$/L$_{\odot}>10^{11}$) in the high redshift Universe may commonly be mergers, as supported by observational evidence of disturbed morphologies in Submillimetre Galaxies (SMGs) and an increase in merger activity with redshift ~\citep{Blain02,Dasyra08,deRavel09, Bundy09}. In the local Universe, however, a significant fraction ($\sim20-40\%$) of LIRGs show little or no evidence of ongoing or past strong interactions (\citealt{Ishida04},~Howell et al. in prep).  In order to fully understand the processes that govern star formation and black hole assembly a clear and complete picture of the evolution U/LIRGs in the local Universe is essential.  

Integral field spectroscopy (IFS) furnishes one of the most powerful tools for the study of nearby U/LIRGs. Through its ability to both spatially and spectrally map a galaxy, it provides a wealth of spatially resolved information tracing the kinematics of the gas and stars, variations in the radiation field and sources of ionizing photons, metallicity and chemistry in the ISM, maps of extinction and gas flows.

IFS studies in recent years have enabled the discovery of extended shock excitation in several nearby galaxies, notably in U/LIRGs and other massive galaxies. \citet{Monreal06} identified extended LINER-like emission resulting from shocks in a small sample of ULIRGs. \citet{Sharp10} found widespread shock-excitation in the extended emission associated with outflows in a study of several galactic wind-bearing galaxies. A larger IFS study found a significant fraction of tidally induced shock excitation in nearby LIRGs \citep{Monreal10}. \citet{Farage10} discovered extended shocks caused by gas accreting on to a giant brightest cluster galaxy (BCG). Finally, \citet{Rich10} found extended shock excitation caused by a galactic wind in the M82-like galaxy NGC 839 and widespread shocks in the late-stage LIRG mergers NGC 3256 and IC 1623 \citep{Rich11}. In \citet{Farage10}, \citet{Rich10} and \citet{Rich11}, new slow shock models were employed to analyze the shocked gas. In all of the above cases, shock excitation exhibits characteristics of extended LINER-like emission with broadened line profiles.

The shock excitation seen in the LIRGs NGC 839, NGC 3256 and IC 1623 is associated with moderately broadened emission line profiles and enhanced emission line ratios of lower ionization species such as [SII]$\lambda\lambda6717,6731\AA$~ and [OI]$\lambda6300\AA~$ \citep{Monreal10,Rich11}. The shocks themselves are due to major gas flows caused by the merger process.  During a major merger, gas is driven inwards by tidal forces, which can result in shock excitation (e.g. \citealt{Farage10}). This infalling gas fuels massive bursts of star formation and AGN activity which in turn drive massive galactic outflows and further shocks into the ISM and beyond. In an effort to determine the contribution from shocks in U/LIRGs and the relationship between shocks and merger progress, we have searched for similar signatures in a larger IFS sample of 10 nearby U/LIRGs \citep{Rich12}. We find many of the galaxies in our sample, especially those in later-stage mergers, exhibit evidence of significant ongoing shock excitation in their optical spectra. To characterize the relative contribution from shocks throughout the merger process, we extend and expand upon the analyzes from our previous work in \citet{Rich10,Rich11} with a sample of U/LIRGs covering the entire merger sequence. 

This paper presents a comprehensive study of the emission line properties of the Wide Field Spectrograph (WiFeS) Great Observatory All-Sky LIRG Survey (GOALS) sample (WiGS), with an emphasis on understanding the shock emission observed. The sample, observations and analysis are described in Sections 2 and 3. The emission line ratio maps and diagnostic diagrams for the sample are  discussed in Section 4. A summary of the velocity dispersion distributions for the WiFeS GOALS sample is given in Section 5. Section 6 summarizes the various diagnostics as a function of merger stage. The total contribution from shocks as a function of merger stage, the potential impact of shock excited gas on observational interpretation, and the potential overlap with AGN/composite galaxies is discussed in Section 7. Our conclusions are presented in Section 8. 

We use the cosmological parameters assumed in the summary of the GOALS sample~\citet{Armus09},  based on the five-year WMAP results \citet{Hinshaw09}: H$_{0}$=70~km~s$^{-1}$Mpc$^{-1}$, $\Omega_{\mathrm{V}}$=0.72, and $\Omega_{\mathrm{M}}$=0.28.

\section{Sample, Observations and Data Reduction}
Our targets are U/LIRGs drawn from the Great Observatory All-Sky LIRG Survey (GOALS) sample \citep{Armus09}.  GOALS is a multi-wavelength survey of the brightest 60$\mu$m extragalactic sources in the local universe ($log(L_{IR}/L_{\sun}) > 11.0$) with redshifts z $<$ 0.088. GOALS is a complete subset of the IRAS Revised Bright Galaxy Sample (RBGS) \citep{Sanders03}. Objects in GOALS cover the full range of nuclear spectral types and interaction stages and may serve as useful analogs for comparison with high-redshift galaxies.

The GOALS U/LIRGs represent an entire array of merger stages from isolated starburst systems to massive post-mergers as well as galaxies in compact groups. The GOALS systems include starburst galaxies, AGN, composite systems, LINERs and E+a galaxies. While the ULIRGs in GOALS are composed entirely of major mergers, the lower luminosity LIRGs represent a variety of systems-some past or future ULIRGs, others simply undergoing intense bursts of star formation driving up the infrared luminosity temporarily \citep{Ishida04,Armus09}. Our sample consists of targets observable from the south.

\subsection{Observations}
Our observations were conducted with the Wide Field Field Spectrograph at the Mount Stromlo and Siding Spring Observatory 2.3m telescope. WiFeS is a, dual beam, image-slicing integral field unit (IFU) described in detail by \citet{Dopita07} and \citet{Dopita10}. We have previously presented analysis of smaller portions of the WiFeS GOALS dataset in \citet{Rich10,Rich11,Rich12}. In short, a single pointing from our dataset provides a data cube $25\arcsec \times 38\arcsec$, sampled with $1\arcsec$ spaxels in double binning mode. The typical seeing was $\sim1.5\arcsec$. The setup used for our observations produces a spectral coverage of 3700-5700 \AA~for the blue spectra, at a spectral resolution of R$\sim3000$ ($100$ km s$^{-1}$) and 5700-7000 \AA\ for the red spectra at a resolution of R$\sim7000$ ($40$ km s$^{-1}$). Thus, the data have sufficient spectral resolution in the red to allow detailed dynamical studies, while the total wavelength coverage (3700-7000 \AA) is sufficient to support excitation and chemical abundance analyzes.

The data were taken over 5 separate observing runs in July, August and September 2009 and March and May 2010. Observations were begun shortly after the instrument was commissioned, prior to the availability of nod-and-shuffle mode. Hence data taken in 2009 are taken primarily in classical observing mode with separate observations of the sky throughout each evening, while observations carried out in 2010 primarily use nod-and-shuffle mode for sky subtraction. Table A.1 lists the systems and dates observed, with the number of individual WiFeS pointings and total exposure times for each pointing, divided among two to five exposures. Figure A1 shows our WiFeS pointings overlaid on Digitized Sky Survey (DSS) images for each system listed in Table A.1. We observe a total of 27 GOALS systems with a wide variety of properties described in section 2.3. Observations were generally aligned and/or mosaiced to cover the entire galaxy or galaxies in each system.

\subsection{Data Reduction}
The data were reduced and flux calibrated using the WiFeS pipeline, briefly described in \citet{Dopita10}, which uses IRAF routines adapted primarily from the Gemini NIFS data reduction package. Cosmic ray removal was performed on the raw data frames prior to reduction with the ``dcr'' routine \citep{Pych04}. 

The bias subtraction is somewhat complicated by the use of quad-readout with four amplifiers on each ccd to decrease chip read-out time. Additionally, there is a slight slope and instability in the bias across each region of the chip. Bias frames are taken immediately before and after each set of observations and a 2-D fit of the surface is subtracted from the temporally nearest object data in order to avoid adding additional noise to the data. Any resulting residual is accounted for with a fit to unexposed regions of the detector.

Quartz lamp flats are used to account for the response curve of the chip and twilight sky flats are used to correct for illumination variation along each of the slitlets.  Spatial calibration is carried out by placing a thin wire in the filter wheel and illuminating the slitlet array with a continuum lamp. This procedure defines the center of each slitlet. The individual spectra have no spatial distortion because the camera corrects the small amount of distortion introduced by the spectrograph. Thus only low-order spatial mapping of the slitlets is required. 

Wavelength calibration is performed using CuAr and NeAr arc lamp observations to provide sufficient lines in both the blue and red arms of the camera. Arc lamp data were taken in between sets of object exposures.  Each of the 25 slitlets is then rectified by the pipeline into a full data cube (one for each camera) sampled on a common wavelength scale.  

Telluric absorption features were removed from the resulting red data cubes using observations of B-stars or featureless white dwarfs (typically also used as flux standards) taken at similar air mass. The effects of differential atmospheric refraction are calculated and corrected by the pipeline for the blue data cubes.

Flux calibration of each individual data cube was carried out using flux standards observed throughout each night. Standard stars are taken from the WiFeS observing manual \citep{Bessell99}. Individual data cubes were flux calibrated using the standards observed nearest in time and air mass.

The individual data cubes are binned by 2 pixels in the spatial direction to increase signal to noise and produce square spatial pixels (spaxels) 1\arcsec x 1\arcsec. Observations taken in march 2009 were binned during the data reduction process, observations taken in May 2009 and later were binned on-chip.  Typical seeing achieved at SSO during our observations is $\sim1.5$\arcsec, with some variation, on par with the spaxel size for our data cubes. Individual data cubes thus consist of 25x38 1\arcsec square spaxels, producing over 900 spectra. In practice, a few rows of spaxels are trimmed from the top and bottom of the individual cubes, corresponding to the edges of the slitlets.

Finally, the individual reduced, flux-calibrated data cubes are median combined and sampled to a common spatial grid using overlapping features found in each pointing. This is done for both single pointing and for mosaiced observations. Combined data cubes were aligned astrometrically by comparing a pseudo `r-band' image generated using the red spectrum from each spaxel with either DSS R-band or HST data when available. 

\subsection{The WiFeS GOALS Sample}
In all, 27 systems comprising nearly 40 individual galaxies were observed. THe sample was selected to cover a range of infrared luminosities, merger stages and nuclear activity. Basic sample properties are shown in Table 1.

The final sample comprises 4 ULIRGs and 23 LIRGs. This represents a slightly larger fraction of ULIRGs than the GOALS sample itself, which has 23 ULIRGs and 198 LIRGs. We apply the merger stage classification used in \citet{Yuan10}, adopted from \citet{Veilleux02}. Isolated systems (``iso'') have no neighbors within a projected distance of 100 kpc, wide pairs (``a'') are separated by a distance of greater than 10 kpc, close pairs (``b'') are separated by less than 10 kpc and latest-stage, coalesced mergers are a combination of the original 'diffuse', 'compact' and 'old' merger class (``cde'')-defined by an increase in compactness in K-band and decreasing to no evidence of any tidal features.  This somewhat coarse sampling of the merger process is necessary given the small number in each class: two isolated systems, five wide pairs, ten close pairs and nine late-stage mergers. None the less, we are still able to trace bulk changes in the properties of our sample as a function of merger stage. Individual notes about each system are provided in the appendix.

To avoid contamination by AGN ionization in our analysis of shocks, four galaxies with strong optical, mid-IR and X-ray AGN signatures are removed from the various merger stage bins and treated separately. The isolated galaxy IRAS F21453-3511, the closely interacting pair IRAS F23128-5919 and the coalesced mergers IRAS 13120-5453 \&  F16164-0746 are all dominated by an AGN, as detected via [Ne V] in mid-IR Spitzer spectroscopy \citep{Petric11} and X-ray color/spectroscopy from Chandra observations \citep{Farrah07,Iwasawa09,Iwasawa11}. These galaxies are not included in their respective merger classes for the analysis of shocks for the remainder of the paper. The possibility of lower-level AGN activity in other systems is further discussed in Section 7.3.

Once the AGN are removed, there are three isolated galaxies (``iso''), four close pairs (``a''), nine close pairs (``b'') and seven coalesced systems (``cde''). The average log(L$_{IR})$ for the four classes are $11.37\pm0.04, 11.37\pm0.26, 11.62\pm0.23 and 11.85\pm0.33 L_{\odot}$ respectively, and $11.86\pm0.4 L_{\odot}$ for the four AGN.

\section{Data Analysis}
The key data products required for our work are accurate emission line fluxes and emission line kinematics at the highest possible spatial resolution, with as few confounding spectroscopic fitting parameters as possible. In order to tailor the final data product derived from the cubes to the needs of the analysis, in-house IDL routines combined with existing IDL software were used to analyze the spectra from each final data cube.

%
% Table of WiFeS Observations
%

\begin{table}
\caption{WiFeS IFU GOALS Sample (WIGS) \label{table_observations}}
\centering
\begin{tabular}{lllcc}
IRAS Name & \multicolumn{1}{c}{L$_{IR}$/L$_{\odot}$} & \multicolumn{1}{c}{z} & \multicolumn{1}{c}{pc/\arcsec} & \multicolumn{1}{c}{Merger Stage} \\

\multicolumn{1}{c}{(a)} & \multicolumn{1}{c}{(b)} & \multicolumn{1}{c}{(c)} & \multicolumn{1}{c}{(d)} & \multicolumn{1}{c}{(e)} \\
\hline\extravspace
%  Name     & LIR    #    & z       &pc/arc& Stage & Mr & Mk  \\      
F01053-1746 & 11.71  & 0.0201  & 399  & b\\
F02072-1025 & 11.01  & 0.0129  & 256  & a\\
F06076-2139 & 11.65  & 0.0374  & 742  & b\\
 08355-4944 & 11.62  & 0.0259  & 543  & cde\\
F10038-3338 & 11.78  & 0.0341  & 707  & cde\\
F10257-4339 & 11.64  & 0.0094  & 185  & b\\
F12043-3140 & 11.43  & 0.0232  & 517  & b\\
F12592+0436 & 11.68  & 0.0375  & 784  & cde\\
 13120-5453 & 12.32  & 0.0308  & 653  & cde\\
F13373+0105 & 11.62  & 0.0226  & 502  & a\\
F15107+0724 & 11.35  & 0.0130  & 308  & iso\\
F16164-0746 & 11.62  & 0.0272  & 588  & cde\\
F16399-0937 & 11.63  & 0.0270  & 584  & b\\
F16443-2915 & 11.37  & 0.0209  & 465  & a\\
F17138-1017 & 11.49  & 0.0173  & 392  & cde\\
F17207-0014 & 12.46  & 0.0428  & 878  & cde\\
F17222-5953 & 11.41  & 0.0208  & 456  & iso\\
 17578-0400 & 11.48  & 0.0140  & 322  & a\\
F18093-5744 & 11.62  & 0.0173  & 383  & b\\
F18293-3413 & 11.88  & 0.0182  & 401  & b\\
F18341-5732 & 11.35  & 0.0156  & 344  & iso\\
F19115-2124 & 11.93  & 0.0487  & 976  & b\\
F20551-4250 & 12.06  & 0.043   & 860  & cde\\
F21330-3846 & 11.14  & 0.0191  & 400  & b\\
F21453-3511 & 11.42  & 0.0162  & 341  & iso\\
F22467-4906 & 11.84  & 0.0430  & 853  & cde\\
F23128-5919 & 12.06  & 0.0446  & 884  & b\\
\hline
\hline
\end{tabular}
\begin{quote}
(a) IRAS identifier from \citet{Sanders03,Armus09} (b) log(L$_{IR}$/L$_{\odot}$) \citep{Sanders03,Armus09} (c) systemic redshift \citep{Sanders03} (d) Spatial Scale (e) \citet{Veilleux02} Merger Stage, taken from \citet{Yuan10} where possible.
\end{quote}
\end{table}
\begin{figure*}[htbp]
\centering
\includegraphics[width=\textwidth, trim=0cm 0cm 0cm 0.51cm, clip=true,]{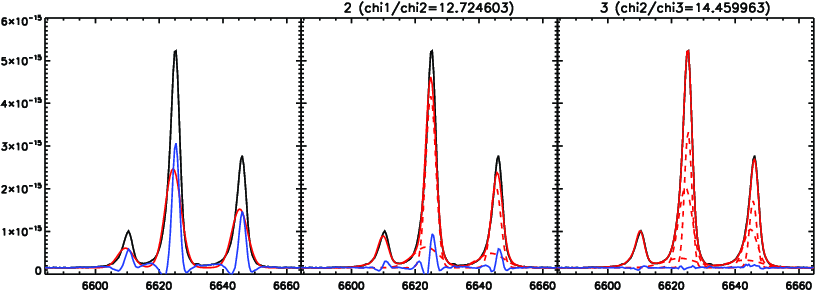}
\caption{Example of one, two and three component fits applied to the \NII+\Ha complex prior to fitting the entire spectrum. This is an example of a spaxel from IRAS F10257-4339 where a 3-component fit was chosen. The black line represents the data, the red line the best fit and the blue line the residuals. Individual components are plotted with a dashed red line.}
\end{figure*}

\subsection{Spectral Fitting}
Each data cube was analyzed using the UHSPECFIT software package \citep{Rupke10b, Rich10, Zahid11}. The WiFeS UHSPECFIT package cycles through the red and blue cubes for a single object, fitting the blue and red spectra from a given spaxel simultaneously. For each spectrum, the package performs an initial fit to the continuum and emission lines, checks whether any emission is found in the strongest lines (\Ha and \NII) and if so performs a second, final fit using the first fit as input estimators. After fitting every spectrum in the data cube in this fashion, the final data product is a data structure with spectral fits as well as all of the parameters necessary to reconstruct the emission line component fits for each line of each spectrum, as well as the errors on those parameters. The details of the fitting routine are further described in the following subsections.

\begin{figure*}%[htb]
\centering
{\includegraphics[width=0.75\textwidth]{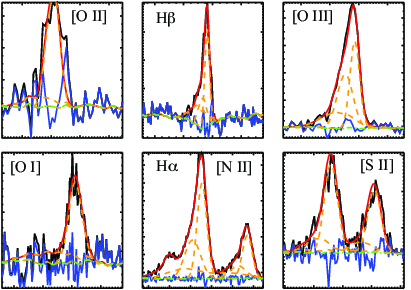}}
\caption{Example of a 3-component fit to a spectrum from a single spaxel in IRAS F23128-5919. Several strong emission lines are shown. The total fit is plotted in red over the data in black, individual emission components are orange dashed lines, the continuum fit is a green dashed line and the residual fit between the total fit and the emission fit is shown in blue. The individual component (orange lines) for the \NII~6548 line have been removed to decrease confusion.}
\end{figure*}

\subsection{Continuum Fitting}
The first step in fitting a spectrum in all cases was to fit and subtract a stellar continuum using stellar population synthesis models from \citet{Gonzalez05}. This is done in order to obtain a pure emission line spectrum while accounting more accurately for stellar features that underlie the emission lines in several systems. An IDL routine, IBACKFIT \citep{Moustakas06}, fits a linear combination of stellar templates to the stellar continuua. IBACKFIT employs a least-squares analysis using the Levenberg-Marquardt algorithm via MPFIT \citep{Markwardt09}. IBACKFIT was applied in each case only to the blue spectrum, which holds the majority of the stellar spectral features in the WiFeS data. The best-fitting combination of templates was then combined with a functional fit and extrapolated to the entire data cube. No information from stellar template fitting (i.e. stellar population age, metallicity, etc.) was derived for the work in this paper.

\subsection{Emission Line Fitting}
The continuum-subtracted spectrum is then fit with several gaussian emission lines using routines built using the MPFIT package. The list of emission lines is chosen by the user prior to running UHSPECFIT. The emission lines fit in the WiFeS GOALS data include \OII~3727,3729\AA, H$\epsilon$, H$\delta$, H$\gamma$, H$\beta$, \OIII~4959,5007\AA, \OI~6300,6364\AA, \NII~6548,6583\AA, H$\alpha$ and \SII~6717,6731\AA. The \OII~ and \SII~ doublet flux ratios were allowed to vary within the upper and lower theoretical limits and the flux ratios of the \OIII, \OI\ and \NII\ doublets were fixed to their prescribed theoretical values (i.e. \citealt{Osterbrock89}). Emission line gaussian peaks are constrained to be greater than or equal to zero.

One, two or three Gaussian components are fit to every emission line, to account for the complex line profiles seen in several of the WiFeS GOALS systems. Prior to running UHSPECFIT on a galaxy, mpfit was used to fit one, two and three gaussian components to the \NII+\Ha lines. An example of this process is given in Figure 1. Each resulting one, two and three gaussian component fit was checked by eye. Upon inspection, the best fit was chosen based on both quantitative comparison of the respective $\chi^{2}$ as well as a qualitative assessment of the goodness of fit and consistency with features in surrounding fits within a single galaxy, similar to the method employed by \citet{Westmoquette11,Westmoquette12}. 

If the $\chi^{2}$ does not markedly improve when an extra gaussian component is added, fewer components are used to avoid over-fitting the emission lines. Consideration is also given to emission line fits in neighboring spaxels within a single cube when choosing whether to add additional gaussian components. The best fit emission line widths and redshifts are then passed to UHSPECFIT, which holds them fixed while simultaneously varying the remaining parameters. The resulting emission line fits to the rest of the spectrum were then again inspected by eye to ensure poorly and erroneously fit lines are rejected. This method resulted in more reliable, consistent fits with UHSPECFIT.

All emission lines are fit simultaneously, but the redshift and velocity width of each gaussian component are fixed to be the same value for every emission line. The relative velocities are fixed separately for the red and blue data cubes to allow for variation in wavelength calibration and resolution between the red and the blue spectra. This assumption implies that the line-emitting gas in a single spaxel is produced in the same region (e.g. \citealt{Kewley01b}). Figure 2 shows an example of a successful multi-component emission line fit to key diagnostic lines in a single spectrum.

The errors in the parameters used in fitting the emission lines are calculated by the fitting code, which includes propagation of the variance spectra. The parameters include the gaussian widths (velocity dispersions) ratio of dispersions between the blue and red spectra, individual gaussian peaks, redshift and any slight deviation between the blue and red wavelength calibrations. 

\subsection{Data Products}
Once a galaxy has been run through UHSPECFIT, we generate maps of the recessional velocity, velocity dispersion and emission line fluxes of each emission line component as well as maps of the total emission line flux of several strong lines. In addition to manual inspection of the emission line fits, a final S/N of 5 is used as a cutoff to create the total flux maps and a S/N of 3 cutoff as measured between component peaks and continuum noise is applied to the individual component maps. In all, $\sim10,000$ spectra are analyzed in this paper, with $\sim17,000$ individual gaussian components above the S/N cutoff. This corresponds to about $2100$ individual gaussian components in the ``iso'' galaxies, $3300$ in ``a'', $8300$ in ``b'', $1300$ in ``cde'' and $1750$ in the AGN. The lower number of gaussian components in the cde bin despite the higher number of galaxies reflects both the small spatial extent of the emission line gas and the higher average distance to the coalesced merger galaxies.

%\end{sidewaysfigure*}
\begin{figure*}[htbp]
\centering
\includegraphics[width=0.8\textwidth]{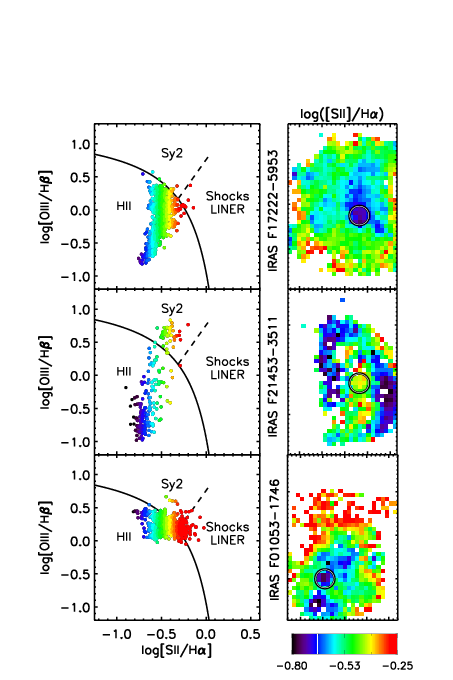}
\caption{Example line ratio diagnostic diagrams and maps from IRAS F17222-5953, F21453-3511 and F01053-1746. Points in the diagnostic diagrams are color coded with respect to the \SII/\Ha values shown in the line ratio maps. The center of each galaxy, corresponding to the peak in R-band emission, is marked with a circle in the line ratio map for reference. IRAS F17222 is dominated by star formation, F21453 is an AGN blended with star formation (\citealt{Davies14}) and F01053 is a mixture of radiative shocks and star formation (\citealt{Rich11}). These three galaxies are chosen to show key features of galaxies in our sample with each source of ionizing radiation. The remaining diagnostic diagrams and maps are shown in the appendix. }
\end{figure*}

\section{Emission Line Gas Properties}
Our emission line maps and diagrams provide an overview of the distribution of spaxels spaxels dominated by enhanced low ionization species emission as well as some indication of the underlying power source and metallicity. For our purposes, the maps and diagnostic diagrams are most useful in spatially separating regions dominated by \HII-region emission, shock emission and AGN emission. The observed emission line ratios are governed by the overall flux of ionizing photons, the hardness of the radiation field and the metallicity of the gas. By taking emission-line ratios that are sensitive to abundance and with a variety of ionization potentials, the source of ionizing radiation can be determined (e.g. \citealt{BPT81,VO87,Kewley01a,Kauffmann03,Kewley06}). With a harder radiation field than young stars or radiative shocks, AGN-ionized gas has generally higher log(\OIII/\Hb) emission line ratios than HII-region and shock emission, with log(\OIII/\Hb) of order unity to a factor of several times unity. \HII-regions exhibit a range of emission line ratios, with low ionization species-to-\Ha ratios (log(\NII/\Ha)\textless-0.4, log(\SII/\Ha)\textless-0.3 and log(\OI/\Ha)\textless-1.0). \HII-regions show a large range of log(\OIII/\Hb) emission line ratios (log(\OIII/\Hb from -1.0 to 0.8 from high to low abundance respectively) as a result of less efficient cooling leading to higher ionization parameters and thus higher \OIII/\Hb. Shocks with velocities of $\sim100-200$ km~s$^{-1}$, on the other hand correspond to enhancements in lower ionization lines (\NII, \SII, \OI), and some enhancement in \OIII/\Hb in higher metallicity systems \citep{Rich11}.

\subsection{Distinguishing Power Sources}
Both the distribution of line ratios and the values of the line ratios themselves provide useful information about the source(s) of ionizing radiation in a galaxy. Emission line ratio maps are generated using the total line fluxes spaxel by spaxel, which effectively traces the dominant power source. Line ratio diagnostic diagrams provide a more detailed perspective of the sources of ionizing radiation at work in a given system \citep{BPT81,VO87,Kewley01a,Kauffmann03,Kewley06}. When combined with the spatial information from the line ratio maps, a clear picture of one or more power sources present in a galaxy is revealed.

Figure 3 shows how the diagrams and line ratios maps appear for three systems with very different sources of ionizing radiation. Some of the galaxies in our sample appear to be dominated entirely by HII region emission, with no indication of any other source of ionizing radiation. IRAS F17222-5953 is a good example of such a system, as seen in Figure 3. The line ratios fall entirely within the HII-region portion of the diagnostic diagram. The \OIII/\Hb and \NII/\Ha ratios fall as a function of radius due to the presence of a metallicity gradient \citep{Rich12}. The gradient and line ratios are entirely consistent with other star forming galaxies \citep{Kewley10, Rupke10b}.

Some of the galaxies in our parent sample exhibit a blend of strong AGN activity blended with star formation. IRAS F21453-3511, shown in Fig. 3, is a good example of this behavior. The diagnostic diagrams in these galaxies have very strong \OIII/\Hb, moving into the Seyfert/AGN portion of the diagnostic diagram. Line ratio maps tend to show the highets emission line ratios in the nucleus, corresponding to the AGN, decreasing with distance from the nucleus \citep{Davies14, Leslie14}.

Finally, some galaxies show evidence of star formation blended with composite/LINER-like emission. This is shown in figure 3 with the close pair IRAS F01053-3511, which shows extended LINER-like emission well outside the nuclear regions, which are dominated primarily by star formation. Extended composite/LINER emission alone, however, does not confirm the presence of shocks \citep{Rich10, Rich11}. Extended low-ionization emission regions (LIERs) can be caused by a number of mechanisms, including heating by post-AGB stars \citet{Sarzi10,Annibali10,Eracleous10} and diffuse ionized emission (e.g. \citealt{Blanc09}).

\subsection{Line Ratio Maps}
Example emission line ratio maps are shown in Figure 3, with the remaining galaxy emission-line maps given in the Appendix. Emission line ratio maps are generated using the total line fluxes spaxel by spaxel, which effectively traces the dominant power source. For instance, in Figure 3 large portions of the merging system IRAS F01053-1746 are dominated by emission line ratios consistent with star formation while there is evidence of enhanced emission line ratios in areas of lower \Ha surface brightness (as traced by the contours). As is noted in~\citet{Rich11}, even in some portions of IRAS F01053-1746 dominated by \HII~region emission there may be a contribution to the emission from shocks that can be separated kinematically, but the total flux will show \HII~region-like line ratios overall.

Systems dominated entirely by \HII~region emission include IRAS F13373+0105 W \& E, IRAS F17222-5953, IRAS F16443-2915 N \& S. These systems exhibit line ratios consistent with \HII~region emission (see also section 4.2) across the entire line ratio map, with no indication of any other significant source of ionizing flux.

In some systems the line ratio map provides an immediate indication that there may be an AGN present. In systems with a strong AGN, the nuclear regions tend to be dominated by strong line ratios in every map, including \OIII/\Hb. We are able to resolve and separate regions of star formation and AGN ionization in the maps of IRAS F21453-3511 and IRAS F23128-5919, both of which show a combination of Seyfert II nuclei and ongoing star formation \citep{Davies14, Leslie14}. 

In many of the emission line maps, galaxies show enhanced low-ionization line ratios away from the nucleus and in inter-galaxy regions with \HII-region like line ratios in the nuclear regions: this radial rise in line ratios is opposite to the drop in line ratios seen in AGN-dominated systems. As shown in \citet{Rich10,Rich11}, underlying shock excitation combined with a lower contribution from star formation with increasing radius reveals itself as stronger low-ionization line ratios away from the nucleus. Some examples include IRAS F12043-3140, IRAS F15107+0724, IRAS F17138-1017, IRAS F20551-4250 and IRAS F21330-384 In some cases this may be enhanced by galactic winds \citep{Sharp10}, as in IRAS F02072-1025 and IRASF F10257-4339 \citep{Rich10,Rich11}. Dust lanes also serve to bury ongoing star formation, enhancing shock-like line ratios, shock-like ratios. especially where shocks are caused by outflowing gas, the passage of dense clouds of gas through a substrate of more diffuse ISM, or through cloud-cloud collisions. This is evident in IRAS F01053-1746, IRAS F10257-4339 and IRAS F18293-3413. Some of the latest stage mergers are entirely dominated by enhanced line ratios consistent with shock excitation, such as IRAS F12592+0436 and IRAS F22467-4906. Shocks are not restricted to portions of the galaxy dominated by shock emission, however, and can be spread across merging galaxies even where the optical emission is dominated by star formation \citep{Rich11}.

\subsection{Line Ratio Diagnostic Diagrams}
Line ratio diagnostic diagrams provide a more detailed perspective of the sources of ionizing radiation at work in a given system \citep{BPT81,VO87,Kewley01a,Kauffmann03,Kewley06}. When combined with the spatial information from the line ratio maps, a clearer picture of the combined power sources is revealed. Figure 3 shows example line diagnostic diagrams using the total emission line flux for three galaxies with disparate sources of ionizing radiation including star forming regions, AGN and radiative shocks. The diagrams for the remaining 30 data cubes, including diagrams with individual line components rather than total flux, are shown in the appendix.

Traditionally, emission line ratio diagnostic diagrams have been applied to nuclear spectra of U/LIRGs to investigate the presence of AGN (e.g. \citealt{Yuan10}). Systems with resolved AGN activity provide a good example of the power of IFS. IRAS F21453-3511 has a cluster of spaxels in the AGN portion of all three diagnostic diagrams corresponding to the central regions seen in the emission line ratio maps. The rest of the spaxels form a sequence stretching towards the \HII~region portion of the diagnostic diagrams as the line ratios decrease away from the nucleus.

The combination of line ratio maps and diagnostic diagrams provides a way to disambiguate the difference between AGN enhanced emission and shock enhanced line ratios. \citet{Rich10, Rich11} discuss in detail this method for the systems IRAS F02072-1025, F01053-1746, F10257-4339 but the same effect can be seen in, for example, IRAS F12043-3140 and IRAS F21330-3846. The latter two systems show a sequence of spaxels in the line diagnostic diagrams indicating enhanced line ratios consistent with some contribution from non-HII region emission. The enhanced line ratios in the emission line maps of these systems are not consistent with AGN but with shock excitation due to the merger process.

As described in \citet{Rich11}, the presence of slow shocks leads to LINER-like line ratios in the diagnostic diagrams. The \SII/\Ha and \OI/\Ha ratios in particular are enhanced in the presence of shocks due to their low ionization potential, as is seen in great detail in IRAS F01053-1746 and F10257-4339 \citep{Rich11}. LINER-like line ratios are seen in extranuclear regions in some other systems as well (e.g. IRAS F20551-4250). Two of our systems, IRAS F13373+0105 and F18341-5732 exhibit evidence of nuclear LINER activity, which are more likely associated with the traditional low-luminosity AGN explanation for LINER emission, whether it be photo-excited \citep{Ferland83,Halpern83,Ho99,Kewley06,Ho09,Eracleous10} or shock excited \citep{Fosbury78,BPT81,Dopita15}.

The underlying metallicity of the gas also affects the line ratios seen in the diagnostic diagrams. IRAS F17222-5953 and 13373+0105 W are good examples of this effect: both galaxies have an intrinsic metallicity gradient which create the sequence of values seen in the \NII/\Ha diagrams \citep{Rich12}. Most of the galaxies that show \HII~region emission have spaxels that correspond to intrinsically high metallicity gas, with the exceptions being IRAS F01053-1746 and F18093-5744 C\&N.

\begin{figure*}[htpb!]
\centering
{\Large \textbf{IRAS F01053-1746}}
{\includegraphics[width=\textwidth]{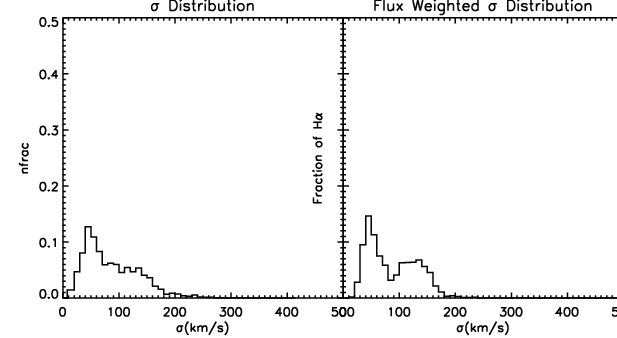}}
{\includegraphics[width=\textwidth]{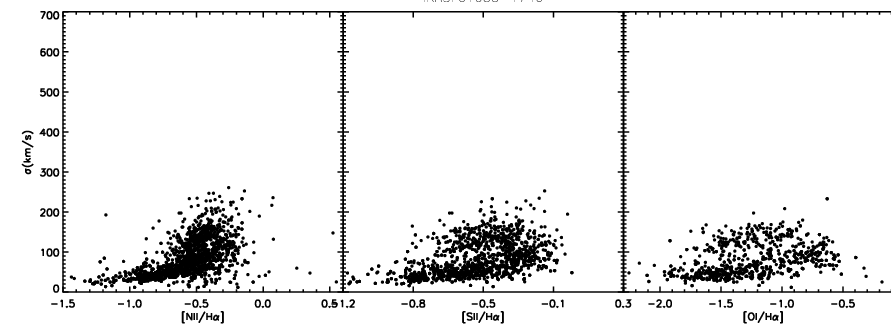}}
\caption{Velocity dispersion distribution and comparison with line ratios. The top left panel shows the histogram for the fraction of the total number of components for a given velocity dispersion The top right panel shows a similar histogram, but with the fraction of the total \Ha flux for a given velocity dispersion. Systems with a stronger contribution from shocks show a larger fraction of $\sigma \gtrsim$90\kms.
The bottom three panels show various diagnostic emission line ratios vs. velocity dispersion. Lower emission line ratios are, on average, dominated by lower-$\sigma$ star forming regions. The same diagrams for the remaining portion of the sample are in the Appendix.}
\end{figure*}

\section{Velocity Dispersions}
Velocity dispersion provides a separate, independent test for the presence of shock emission \citep{Rich11}. Here we examine the velocity dispersions of the emission line gaussian components. Velocity dispersions associated with \HII~regions are typically a few tens of \kms, while low-velocity shocks and AGN line profiles can have velocity dispersions in excess of 100 \kms~\citep{Epinat10,Rich11}. In their discussions of the high velocity dispersions in the SINS sample, \citet{Genzel08} note that the turbulent star formation, shocks and stellar feedback can drive up velocity dispersion as well \citep{Efstathiou00,Silk01,Monaco04,Thompson05,Dib06}. This is consistent with the correlation between average velocity dispersion and star formation surface density seen in, for instance, the sample of \citet{Green10}.

\subsection{Dispersion Distributions}
We analyze the distribution of velocity dispersions fit to the individual components in each system. An example histogram of velocity dispersions is shown in Figure 4, with the remaining systems shown in the Appendix. The dispersion distribution is shown as both the fraction of the total number of profiles at a given $\sigma$ and as a fraction of the total F$_{H\alpha}$ in a given velocity bin. All gaussian components detected in at least one emission line are included in the velocity dispersion distribution histograms.

In systems mainly dominated by star formation (e.g. IRAS F16443-2915 N/S, F17222-5953) the majority of the velocity dispersions are consistent with the expected $\sigma$ of a few tens of \kms. As noted in \citet{Rich10,Rich11}, galaxies with a significant contribution from shocks show a significant contribution between~$\sigma$=100-200\kms, consistent with the shock velocities derived from line ratios (e.g. IRAS F01053-1746, F02072-1025, F10257-4339, \citealt{Rich10, Rich11}). Many galaxies also exhibit a significant number of components with $\sigma$ between 40-90\kms, likely consistent with disc turbulence. The galaxies in our sample that show unambiguous AGN signatures also show some contribution at significantly higher velocity dispersions, of 300-600\kms.

\subsection{Line Ratios and Velocity Dispersion}
In their lower spectral resolution IFS survey of U/LIRGs, \citet{Monreal06,Monreal10} established a correlation between velocity dispersion and emission line ratio, which they attributed to shocked gas components. In \citet{Rich11} we see the same correlation for IRAS F01053-1746 and F10257-4339, consistent with the other signatures of shock excitation \citep{Rich11}. 

Nearly all of the galaxies in the WiFeS GOALS sample which have line profiles fit with elevated velocity dispersion components reproduce this enhancement in \NII/\Ha, \SII/\Ha~and \OI/\Ha~with respect to velocity dispersion, regardless of the excitation mechanism. The exceptions to this rule appear to be IRAS F18093-5744 C \& S. IRAS F18093-5744 C in particular shows the opposite trend: a component with elevated velocity dispersion and correspondingly lower \NII/\Ha, \SII/\Ha~and \OI/\Ha~ratios. IRAS F18093-5744 C is a compact Wolf-Rayet galaxy with the emission driven by young, bright stars \citep{Kovo99,Fernandes04}. Our data show two narrow components with an underlying broad component seen in the $\sigma$ vs. ratio diagrams. This broader component is elevated in \OIII/\Hb, and can be seen in the diagnostic diagram in the Appendix, corresponding to lower metallicity gas, consistent with the analysis in \citet{Fernandes04}.

\section{Merger Progress}
All line diagnostic methods described in section 5 have been applied to each galaxy in our sample. We now track how they change, if at all, as a function of merger evolution. We group spaxels from galaxies in each of the four merger stages as outlined in section 2. The four positively identified AGN are placed in a separate group. We now examine the overall shape of the line ratio diagnostic diagrams, velocity dispersion distribution and $\sigma$ vs. line ratio diagrams for each merger stage.

\begin{figure*}[htpb!]
\centering
%{\Large \textbf{IRAS F01053-1746}
{\includegraphics[height=1.2\textwidth]{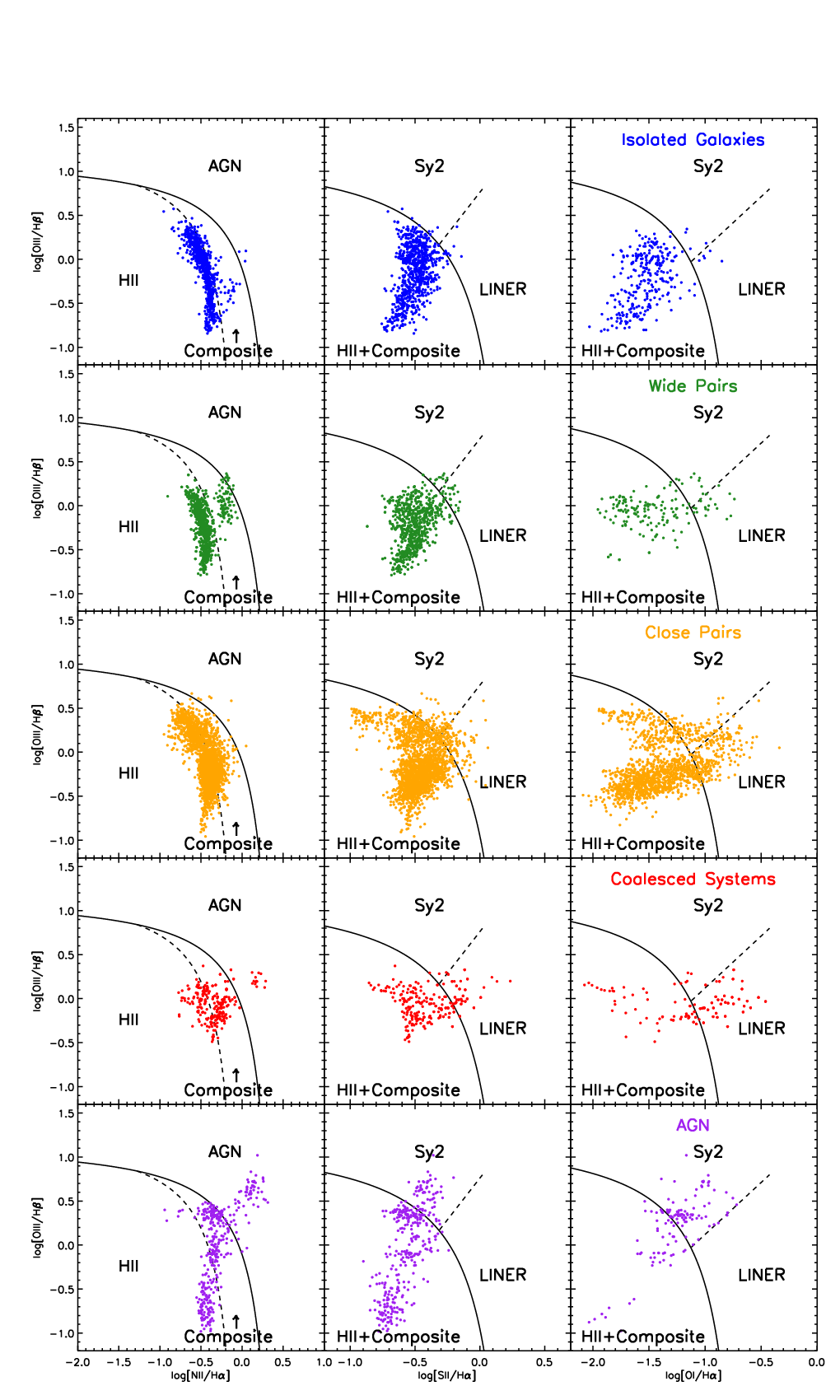}}
\caption{Line ratio diagnostic diagrams created using all spaxels from every galaxy in each merger class. In blue are isolated systems, green are widely separated pairs (``a''), orange are close pairs (``b''), red points are coalesced  galaxies (``cde'') and the bottom diagrams with purple points are the four AGN galaxies. \citet{Kewley06} classifications are shown on each diagram.}
\end{figure*}

\begin{figure*}[htpb!]
\centering
%\vspace{-3.5in}
{\includegraphics[width=\textwidth]{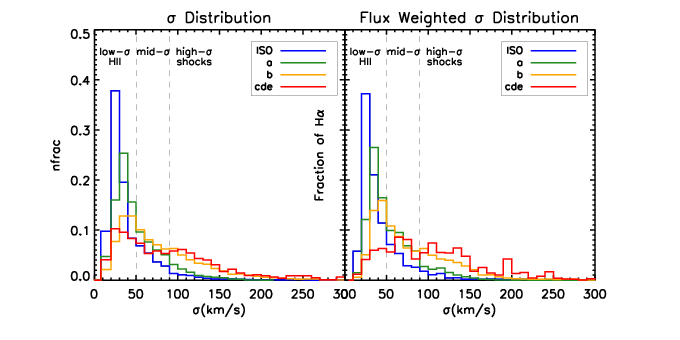}}
\caption{Velocity dispersion distribution histograms for each merger class. As in Figure 4, the fractional number of profiles is shown in the left panel and flux-weighted $\sigma$ distribution in the right panel.}
\end{figure*}
\begin{figure*}[htpb!]
\centering
%{\Large \textbf{IRAS F01053-1746}
%\vspace{+3.5in}
{\includegraphics[width=\textwidth]{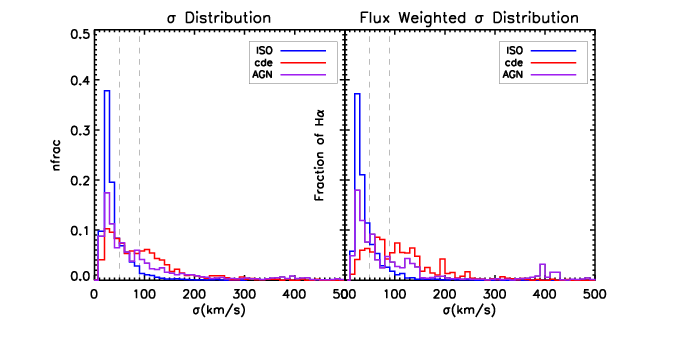}}
\caption{Line ratio vs. $\sigma$ as in Figure 6, but for isolated galaxies, coalesced mergers and the four AGN. Note that this figure is extended to a higher value of $\sigma$ than Figure 6.}
\end{figure*}

\subsection{Diagnostic Diagram Evolution}
Figure 5 shows the three diagnostic diagrams for each merger group and the AGN. The points are generated using the total flux line ratios in each spaxel for every galaxy in a given merger class. In the case of the close pair stage, the data points are somewhat dominated by IRAS F01053-1746 \& F10257-4339, but the other individual galaxies in this merger stage group have line ratios that fall in consistent regions of the diagnostic diagram. The AGN stage is likewise dominated by IRAS F21453-3511 and F23128-5919 which have comparatively spaxels in their diagnostic diagrams, though the relatively fewer points from the other two AGN are again consistent.

The overall shape of the diagnostic diagrams shows an increasing number of spaxels in the 'composite' region of the \NII/\Ha line ratio diagnostic diagram in later merger stages. This result is consistent with the work of  \citet{Yuan10} which shows an increasing fraction of composite behavior with merger stage. 

There is a much narrower range of spaxels in the \HII~region portion of the \NII/\Ha diagram for the ``cde'' galaxies, which is primarily a result of the flattening and dilution of metallicity gradients as mergers progress: we discuss this effect further in \citet{Rich12}. The same effect is seen in individual systems in the ``b'' stage.

\begin{figure*}%[htpb!]
\centering
{\includegraphics[width=\textwidth]{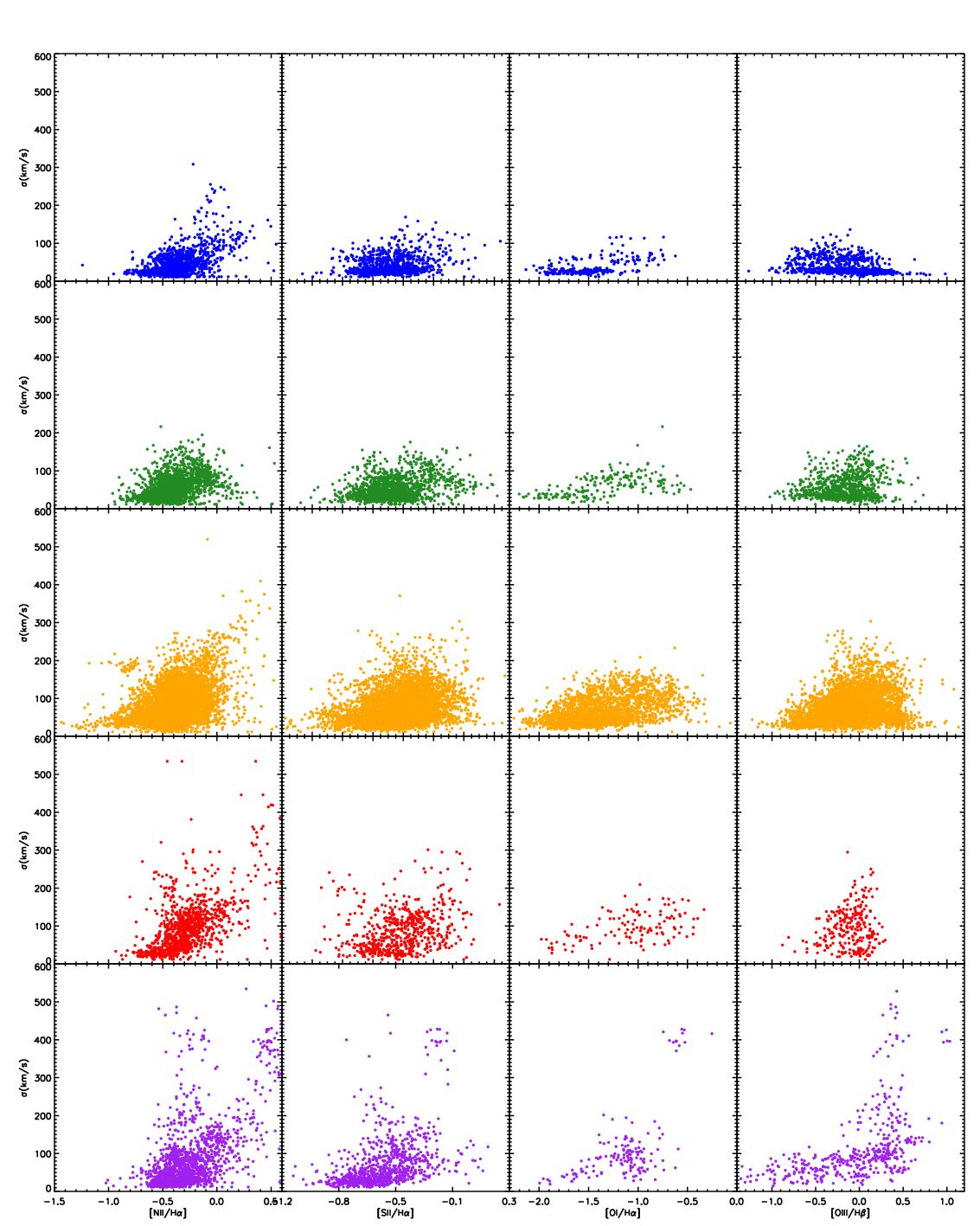}}
\caption{Line ratio vs. $\sigma$ as in Figure 4, for each merger stage, with ``iso, a, b, cde'' and AGN represented by blue, green, orange, red and purple from top to bottom respectively.}
\end{figure*}

\subsection{Velocity Dispersion Evolution}
As in the case of line ratio evolution with merger stage, there is also clear evidence of an corresponding evolution in the observed velocity dispersion. In Figure 7, we show the change in the overall velocity dispersion distribution as a function of merger stage. The isolated systems are dominated primarily by narrow \HII~region-like velocity dispersions, though there is a tail of higher dispersion spaxels corresponding to some turbulence and the LINER emission in IRAS F18341-3413. In the widely separated pairs the peak in both number of profiles and total \Ha flux shifts slightly to the right and the tail of broader spaxels grows, indicating an increase in turbulent star formation. Figure 7 compares the velocity dispersion distribution for the isolated, ``cde'' and AGN systems. The AGN galaxies show a stronger high-$\sigma$ component above 350 km/s that is not seen in the other distributions.

In the closely interacting pairs, the shape of the dispersion distribution begins to change significantly: a strong tail of higher $\sigma$ line profiles is seen in both fractional number and total F$_{Ha}$. This is consistent with the increase in outflows as the merger progresses, as seen in previous IFU studies (\citealt{Rich11, Soto12, Genzel14, Arribas14, Ho14, Wild14}), and in section 5 above. At the latest merger stages, the velocity dispersion distribution is fairly flat from narrow to broad line components, indicating a significantly increased contribution from turbulent star formation and shocks in the emission line gas.

Figure 8 shows the evolution of the line ratio versus $\sigma$ with merger stage. The overall shape of the diagrams tells the same story as the diagnostic diagrams and velocity dispersion distributions taken together: There is an overall trend in the total number of spaxels with enhanced velocity dispersion and emission line ratio as a function of merger stage. 

The AGN show a bimodal distribution in the line ratio vs. $\sigma$, owing mainly to the difference in line ratio values at high velocity dispersion between IRAS F21453-3511 and F23128-5919 which dominate these figures. This is likely due to a difference in metallicity between the AGN-illuminated gas in the two systems.

\section{Discussion}
In the absence of an AGN, a relative increase in the velocity dispersion and shift of the strong emission line ratios in the diagnostic diagram towards the LINER region is indicative of gas excited by slow shocks \citep{Monreal06,Monreal10,Rich10,Rich11}. Moreover, high-$\sigma$ and LINER-like line components are not restricted to the nuclear regions of our non-AGN systems, again consistent with widespread shock excitation caused by merger-driven gas flows as the source of ionization for the observed emission (\citealt{Monreal10,Rich11}, see also individual system notes in appendix). We therefore interpret the overall relative increase in velocity dispersion, particularly in the absence of any detectable influence from an AGN, as an increased relative contribution from slow shock excitation in our sample and discuss the relative strengths of each velocity component in our sample.

\subsection{Shock Fraction and Merger Stage}
With the simple correlations established in the previous sections, $\sigma$ can be used as a proxy for the total contribution from \HII~region emission, turbulent star formation and shocks. The histograms of the distribution of $\sigma$ discussed in section 5 and shown in Figures 6 and 7 allow us to establish approximate cutoffs between the three components. For the discussion here we establish an upper limit of $\sigma$=50 \kms~for pure \HII~region emission and $\sigma>$90 \kms~for spaxels that are primarily dominated by shock excitation in the absence of an AGN.

Figure 9 shows the change in the total fraction of F$_{H\alpha}$ in each component as a function of merger stage. The fraction in each component is the total of the profiles in each merger stage bin. There is a minimal flux-weighted contribution of a few percent from a higher $\sigma$, shock-dominated component in the three isolated galaxies and in the widely separated pairs in our sample. The turbulent, mid-range $\sigma$ component increases as the galaxies begin to interact, but stays roughly flat throughout the merger process. Most interestingly, however, is the rapid increase in a high-$\sigma$ component from wide to close pairs and from close pairs to late-stage mergers. By the ``cde'' stage, just over half of the total F$_{H\alpha}$ we measure lies in high velocity dispersion line emission.

The low-$\sigma$ component is consistent with the velocity dispersions seen in star forming regions, while the mid-$\sigma$ component is likely associated with regions of more turbulent star formation seen in strongly star bursting systems (e.g. \citealt{Green10}). The highest observed velocity dispersions, above 90 \kms, are caused by even more turbulent activity in the ISM of U/LIRGs. The most likely culprit is shocks caused by merger-driven gas flows, such as galactic winds (e.g. \citealt{Armus89,Armus90,Sharp10,Rich10}). These lower velocity shocks exhibit velocity dispersions consistent with the outflow velocity and the observed high-$\sigma$ gas, on the order of 100-200\kms \citep{Rich10,Rich11}. The high-$\sigma$ component observed in our systems is also correlated with elevated emission line ratios, as seen by \citet{Monreal10} and consistent with the radiative spectra of gas ionized by shocks. This combined elevation in emission line ratios and velocity dispersions increases as a function of merger stage, and in the 7 latest merger-stage systems, there appears to be a significant contribution to the optical spectra from merger-induced shocks. 

There is the possibility of contribution from a less luminous AGN (classical LINER), which can similarly increase observed emission line ratios and velocity dispersions. The spatial and spectrally resolved information provided by our IFS observations, however, show not only elevated $\sigma$ and line ratios in the nuclear regions, but at many kpc from the nucleus, more consistent with widespread shocks (see Appendix C). We discuss further possible contamination from AGN in section 7.2.

\begin{figure}%[htpb!]
\centering
{\includegraphics[width=0.45\textwidth]{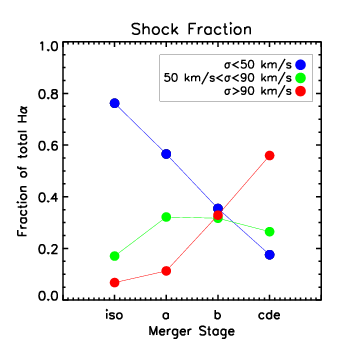}}
\caption{Here the contribution from shocks is traced by the overall velocity dispersion distribution. In each merger stage bin, the total fraction of \Ha flux within each velocity dispersion range is calculated. Cutoffs of $\sigma$=50 and 90 \kms~are chosen to distinguish between \HII~region velocity dispersions, more turbulent dispersions, and shock-dominated dispersions. These values are consistent with the overall distribution of spaxels seen in Figure 6}
\end{figure}

\subsection{AGN Diagnostics and Contamination}
\begin{figure*}%[htpb!]
\centering 
{\includegraphics[scale=0.68]{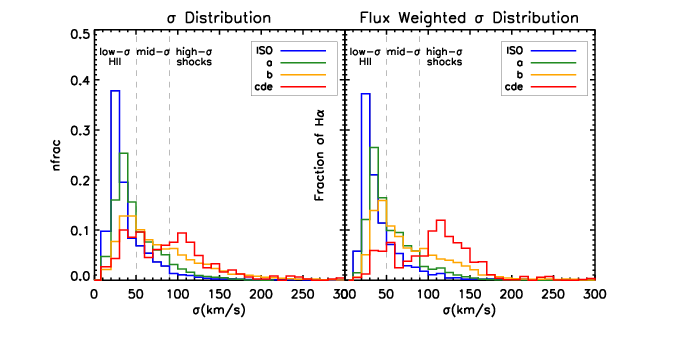}}
\caption{Reproduction of Figure 7 with low 6.2\um~PAH equivalent width galaxies excluded from ``cde''. In this case, only three galaxies remain in the ``cde'' bin, IRAS F12592+0436, F17138-1017 and F22467-4906.}
\end{figure*}

The conclusion that the composite spectra we observe are due to shocks assumes that there is no appreciable contribution to the optical spectra from an AGN. Although we have excluded the four AGN-dominated galaxies from our merger stage discussion, there remains the question of any low-level AGN contribution to the optical spectra in some of our systems. This is of somewhat more of a concern in the latest-stage mergers in our sample, as the optical emission is concentrated primarily in the nuclear regions.   Although none of the galaxies in the merger sample have detectable [Ne V] emission in the mid-IR, this does not necessarily rule out the possibility of a lower-luminosity AGN \citep{Petric11}. Further, not all of the galaxies in our sample have been observed with Chandra, and some systems that have been observed with Chandra but show X-ray emission consistent with intense star formation may still harbor compton-thick AGN \citep{Iwasawa09,Iwasawa11}.

Another measure of AGN activity that can be derived from the mid-IR data is 6.2\um~PAH equivalent width (eqw). In AGN dominated spectra the hot dust continuum depresses the 6.2\um~eqw and the PAHs that generate the feature are apparently destroyed in the EUV photon field \citep{Genzel98,Lutz99,Rigopoulou99,Sturm00,Armus07,Desai07,Wu09,Fu10,Petric11}. \citet{Stierwalt13} determined 6.2\um~PAH eqw values from nuclear spectra for the entire GOALS sample, setting a value eqw$>0.54$\um~for galaxies that are dominated by starburst and eqw$<0.27$ for galaxies dominated by an AGN. Galaxies with eqw values that fall between these values are considered composite systems.

%rise in composites with flat contribution from AGN as a fraction of merger stage.

To make a more conservative estimate of the shock fraction in the late stage mergers galaxies with 6.2\um~PAH eqw $<0.54$\um are removed from the ``cde'' bin (these include IRAS 08355-4944, F10038-3338, F17207-0014 and F20551-4250, with 6.2\um~PAH eqw of 0.19, 0.03, 0.31 and 0.10\um). This leaves 3 coalesced mergers, IRAS F12592+0436, F17138-1017 and F22467-4906 which have eqw of 0.55, 0.68 and 0.45\um respectively. The total number of components with \Ha above the S/N cutoff in these three galaxies is $\sim490$. The new velocity dispersion distribution and shock fraction are shown in Figures 11 and 12. Interestingly, when the potential low-luminosity AGN bearing galaxies are excluded the fraction of emission from broad, slow-shock dominated velocity dispersion components goes up, moving from roughly 55\% to nearly 70\% of the observed \Ha flux. In Fig. 10 IRAS F17138-1017 is classified as a composite system, while F12592+0436 and F22467-4906 host extended LINER-like emission, though in the complete absence of any indication of AGN activity these three systems are clear examples of shock-driven LIER hosts, with LIER emission contaminating and in some cases dominating the observable optical emission line gas.

\subsection{Prevalence of Shocks and Energy Budget}
Our results indicate an overall increase in the total contribution from shocks to the global optical spectra of middle and late stage merging U/LIRGs. The U/LIRGs in the WiFeS GOALS sample contain buried star formation that is so extinguished as to be unobservable at optical wavelengths, which may account in part of the relative increase from non-HII region emission to the total observed emission line flux. 

A prime example of this is IRAS F01053-1746, which has an extended shock-like region in the optical cospatial with a strong IR source indicative of buried star formation \citep{Howell10,Rich11}. Indeed, as noted in \citet{Rich11}, the total luminosity of the shocks in IRAS F01053-1746 is a negligible fraction of the bolometric luminosity, though they can still act as a means of removing energy from the infalling gas over the time scale of the merger process.

Regardless, the \emph{observable} optical emission in our sample shows a combination of HII-region emission and ionization by slow shocks, increasing as a function of merger stage. This combination manifests in some systems as a ``composite'' spectrum, though in this case ``composite'' means starburst+shocks rather than starburst+AGN. Indeed, in \citet{Rich14} we showed that when only nuclear spectra are considered, 75\% of the ``composite'' galaxies in our sample are composite simply due to a sizable contribution from shocks to their emission line spectra.

\begin{figure}[htpb!]
\centering 
{\includegraphics[scale=0.80]{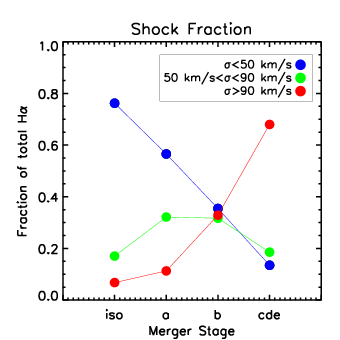}}
\caption{Reproduction of Figure 9 with low 6.2\um~PAH equivalent width galaxies excluded from ``cde'', as in Fig. 12.}
\end{figure}

\section{Summary}
We have analyzed WiFeS optical integral field spectroscopic data cubes of 27 systems from the GOALS sample, 23 of which are a combination of at least two interacting galaxies, ranging from widely separated pairs approaching first pericenter to coalesced systems on their way to becoming elliptical merger remnants. Our analysis focuses on continuum-subtracted emission line gas, in particular through a decomposition of individual emission line profiles into one, two or three distinct kinematic components. We use the results of our emission line fits to create line ratio maps, line diagnostic diagrams and velocity dispersion distributions for each individual galaxy observed. We consider \Ha-flux weighted line component velocity dispersion distributions to avoid over-weighting regions of low-surface brightness emission. The results from each galaxy are placed into bins according to merger stage and the evolution of the emission line ratios and velocity dispersions are tracked as a function of merger progress. Our main conclusions are:

\begin{itemize}

\item Isolated systems and wide pairs are dominated primarily by \HII-region emission ($\sim$75\% and $\sim$55\% of \Ha emission respectively), with line profiles primarily $\sigma$ of a few tens of \kms. 

\item All interacting systems exhibit emission with slightly higher velocity dispersions (50\textless$\sigma$\textless90 \kms), likely associated with turbulent star formation (e.g. \citealt{Genzel08,Green10}. This component makes up roughly 30\% of the total observed \Ha flux throughout the merger process.

\item Closely interacting pairs and coalesced mergers show a strong $\sigma$ component above 100\kms~and an increasingly dominant contribution contribution from composite and LINER-like emission line ratios.  In the absence of an AGN, we assume the broadened $\sigma$ component above 90\kms~is tracing an increasing fractional contribution from shocks, passing 50\% of the total observed \Ha flux in the coalesced mergers. 

\end{itemize}

We consider the IFS data of 4 AGN-dominated systems in our sample separately. The AGN dominant systems are detected in [Ne V] in the mid-IR and show AGN-like X-ray colors and spectra in Chandra and XMM-Newton data \citep{Petric11,Iwasawa11,Franceschini03}. We consider the role that less dominant AGN may play in the latest stage mergers in our system by using the mid-IR diagnostic 6.2\um PAH eqw as a proxy for AGN activity. Our conclusions are:

\begin{itemize}

\item The 4 AGN show a significant component at $\sigma$ of 300-500\kms~and emission line ratios clearly approaching and entering the Seyfert region of the emission line diagnostic diagrams. The AGN also contain shock-like velocity dispersions, driven either by the AGN itself, galactic winds (observed in blue-shifted Na D absorption) or merger-induced gas flows

\item When the low 6.2\um eqw coalesced mergers are removed from the non-AGN sample, the shock-like high-$\sigma$ component increases to 70\% of the observed \Ha emission in the remaining three systems. \citet{Stierwalt13} note that some galaxies show strong evidence for enhanced \hh~emission due to starburst-driven outflows and shocks rather than AGN, with correspondingly low PAH 6.2\um eqw, lending further ambiguity to the picture.

\end{itemize}

Spatially resolved spectroscopy is clearly a powerful tool for distinguishing non-AGN composite emission, particularly in systems where widespread shocked ISM is expected due to a powerful starburst or an ongoing gas-rich interaction. Recent multi-wavelength studies of systems from the GOALS sample and elsewhere have shown promising routes to further study the impact that shocks have on the environment of U/LIRGs (\citealt{Mazzarella12,U13,Inami13}). New surveys of hundreds and thousands of galaxies with IFU spectroscopy, namely CALIFA, SAMI and MaNGA \citet{Sanchez12,Croom12,Bundy15} are revealing a complex mix of forces at work in less luminous galaxies, including LINER excitation in non-mergers caused by shocks as well a warm ionized interstellar medium heated by old stars and UV photons that escape HII regions \citep{Arribas14, Ho14, Belfiore15}. The combination of complementary, resolved multi-wavelength data that probes the multiple sources of activity will be the key to understanding the complex interplay between star formation, shocks, AGN and the ISM in galaxies throughout the universe

\begin{acknowledgements}
The authors acknowledge ARC support under Discovery  project DP0984657. This research has made use of the NASA/IPAC Extragalactic Database (NED) which is operated by  the Jet Propulsion Laboratory, California Institute of Technology, under contract with the National  Aeronautics and Space Administration.  This research has also made use of NASA's Astrophysics Data System, and of SAOImage DS9 \citep{joye03}, developed by the Smithsonian Astrophysical Observatory.
\end{acknowledgements}

%\bibliographystyle{apj}
%\bibliography{ms}

%\pagebreak

\pagebreak

\appendix

\section{Notes on Individual Systems and Maps}
This appendix contains brief notes about each system from our sample and a reproduction of Figures 3 and 4 for every system in the WiFeS GOALS sample. This includes digitized sky survey (DSS) R-band  or \emph{HST} ACS I-band images matching the field of view of the WiFeS GOALS observations combined with emission line ratio maps, emission line ratio diagnostic diagrams, velocity dispersion distribution histograms, velocity dispersion v. emission line ratio diagrams and maps of the distribution of low, mid and high-$\sigma$ velocity components.  We also include a table detailing the observation program for the WiFeS GOALS sample, as well as DSS R-band finder images with individual WiFeS pointings overlaid. Some of the notes on individual systems are adapted from those already published in \citep{Rich12}.

\noindent \textbf{\emph{IRAS F01053-1746}} (IC 1623) This system contains two very closely interacting spiral galaxies. It is kinematically very complex and exhibits evidence of widespread radiative shocks \citep{Rich11}. The eastern system is very intensely star-forming as seen in the infrared (e.g. \citet{Howell10}), but is so enshrouded in dust that our optical spectra do not trace any of the buried star-formation.\\

\noindent \textbf{\emph{IRAS F02072-1025}} (NGC 839) This system is described in great detail in \citet{Rich10}. It is a nearly edge-on starburst galaxy in a compact group (HCG 16). Our emission line maps trace the shocks associated with the starburst-driven outflow and we see evidence of a post-starburst E+A spectrum in portions of the galaxy.\\

\noindent \textbf{\emph{IRAS F06076-2139}} \citet{Arribas08} describe this ULIRG in great detail in their IFS study of ULIRGs. They determined that the system consists of two galaxies that may never merge given their significantly different velocities, though they are interacting as evidenced by a ring of \Ha~emission \citep{Elmegreen06,Arribas08}. Our velocity dispersion and line ratio data imply the northern galaxy is dominated by star formation, with some amount of composite ratios and $\sigma$ in the southern galaxy. The southern galaxy in this closely interacting pair has a fairly low 6.2\um~PAH eqw, which also corresponds to enhanced \NII/\Ha~in our maps.\\

\noindent \textbf{\emph{IRAS 08355-4944}} 
While \emph{HST} I-band images show remnant tidal tails extending nearly 20 kpc, only the central 5 kpc or so appears to harbor the entirety of the intense ongoing star formation in this system. Our spectra are dominated by \HII~region emission, with evidence in some portions of the system of a blue-shifted component with low-velocity shock-dominated line ratios which could be associated with a galactic wind. This is in agreement with the IFS observations of \citet{Monreal10}.\\

\noindent \textbf{\emph{IRAS F10038-3338}} (ESO 374-IG032)
This post-merger exhibits significant ongoing star formation in its southwestern tidal arm unlike the two other coalesced systems in our sample. The total line emission in this region is much weaker than in the nucleus, inducing a large uncertainty in the extinction map and \OII~lines, creating the discrepant values seen in the metallicity gradients for this galaxy \citep{Rich12}. This system hosts an OH megamaser and has soft x-ray emission, all consistent with the advanced stage of merging and increasingly intense nuclear starburst \citep{Henkel90,Staveley-Smith92,Darling02,Iwasawa09}. Our spectra also show evidence for extended off-nuclear shock emission dominating in areas where there is little to no evidence of ongoing star formation, consistent with the IFU observations of \citet{Monreal10}.\\

\noindent \textbf{\emph{IRAS F10257-4339}} (NGC 3256) This advanced merger is the nearest galaxy in our sample (z$\sim0.0094$) and is well studied. As with IRAS F01053-1746, one of the galaxies in IRAS F10257-4339 is very extinguished. The second, buried system and its nucleus are revealed at longer wavelengths, south of the main optical nucleus \citep{Alonso02,Rothberg10}. The gas and tidal tails in this system extend 10's of kpc from the nuclear regions (e.g. \citealt{Rothberg10}). Our IFU mosaic covers only the central 6 kpc, though this appears to be the physical extent of most of the ongoing star formation in this system. NGC 3256 also shows evidence for widespread shocks in our data and is discussed alongside IC 1623 in \citet{Rich11}.\\

\noindent \textbf{\emph{IRAS F12043-3140}} (ESO 440-IG058)
This pair of closely interacting galaxies is poorly studied and there exists no higher spatial resolution imagery to examine. The DSS image, though, shows tidal features indicative of the ongoing interaction. Our IFS data indicate a combination of ongoing star formation and shocked gas, seen in the diagnostic diagrams as \HII-region + composite spaxels and with a tail out to $\sigma \sim$200\kms in the velocity dispersion distribution. Our map shows the enhanced ratios seem to form a cone extending to the north and south of the southern galaxy, and agrees with the IFU observation of \citep{Monreal10}. The overlap in the two galaxies and the lower S/N of our data make it difficult to discern if this is indeed associated with an outflow from the southern galaxy.\\

\noindent \textbf{\emph{IRAS F12592+0436}} (CGCG 043-099)
This late-stage merger is very dusty and subsequently the blue data are mostly too low S/N to measure any emission in single spaxels. The red data indicate low line ratio and $\sigma$ gas in a tidal tail indicative of star formation and higher line ratio and $\sigma$ data toward the nucleus. Our nuclear spectrum indicates composite behavior dominated by shocks. \citet{Poggianti00} found balmer emission and absorption indicative of star formation, with no indication of Seyfert behavior.  X-ray and IR data indicate a galaxy with ongoing star formation and no indication of an AGN, consistent with our results \citep{Rush96,Petric11,Stierwalt13}.\\

\noindent \textbf{\emph{IRAS 13120-5453}}
This nearby ULIRG is dominated by AGN emission at all wavelengths. X-ray data indicate an AGN signature \citet{Iwasawa09,Iwasawa11} and mid-IR PAH values and [Ne V] are consistent with a galaxy dominated by an AGN \citep{Farrah07,Pereira10b,Petric11}. Our optical IFS data are quite consistent with this picture: the strongest emission we detect is dominated by Seyfert-like ratios in the diagnostic diagrams, and the velocity dispersion distribution shows a contribution a high-$\sigma$ component in the nuclear regions of 300-400\kms in addition to the lower $\sim$100\kms component.\\

\noindent \textbf{\emph{IRAS F13373+0105 W}} (NGC 5257) Although NGC 5257 and its equal mass partner NGC 5258 are still widely separated and retain much of their structure, they exhibit interaction features including tidal tails and a bridge between the two galaxies. The spiral arms show regions of strong star formation with some signatures of post-starburst populations away from the nucleus. The nuclear regions of NGC 5257 are dominated by an older stellar population as the intense nuclear starburst associated with the later stages of major mergers has not yet begun. The extinction is higher in the nuclear regions and the measurable line-ratios place the nucleus in the composite region of the standard diagnostic diagrams, indicating possible LINER activity-previous nuclear observations and integrated spectrophotometry are consistent with the higher nuclear extinction and overall line ratios we observe \citep{Veilleux95,Kewley01b,Moustakas06}.\\

\noindent \textbf{\emph{IRAS F13373+0105 E}} (NGC 5258) NGC 5258 is a near-twin to NGC 5257 in mass and luminosity. Our spectra indicate higher extinction and a flatter gradient, though the extent of measurable \HII-region metallicities is smaller than in NGC 5257. Again similar to NGC 5257 the nuclear region of NGC 5258 is dominated by older stars and very little line-emission and there is some evidence of younger post-starburst populations away from the nucleus. The strongest line emission is associated with the knot of star formation to the southwest of the nuclear region.\\

\noindent \textbf{\emph{IRAS F15107+0724}} (CGCG 049-057)
This galaxy bears a low-luminosity OH megamaser \citep{Martin88}, and was classified as an \HII-region galaxy by \citep{Kim95,Veilleux95}. Our IFS data are dominated primarily by an older stellar population in the blue, with the emission lines primarily indicative of ongoing star formation plus some composite contribution, likely shocks given the velocity dispersion distribution and relatively high 6.2\um~PAH eqw value of 0.51\um~\citep{Stierwalt13}. An \emph{HST} NICMOS H-band image is the highest resolution data available, showing what appears to be a relatively undisturbed spiral disc \citep{Haan11b}. This is in keeping with it's status as a isolated galaxy \citep{Yuan10}.\\

\noindent \textbf{\emph{IRAS F16164-0746}}
This coalesced merger is AGN dominated in the X-ray and mid-IR \citep{Iwasawa09,Iwasawa11,Petric11}. Our optical spectra show rotation and a narrow and a broad component, consistent with star formation and shocks, though our line ratio map shows a composite spectrum. Our data do not have sufficient S/N to detect \OIII/\Hb where the other line ratios increase towards the nucleus, where more LINER-like ratios dominate \citep{Veilleux95}. Our nuclear spectrum is somewhat ambiguous, indicating composite, SB and Seyfert-like spectra with \OIII/\Hb near unity.\\

\noindent \textbf{\emph{IRAS F16399-0937}} 
This close-merger pair shows \HII-region+composite emission and a velocity dispersion distribution dominated by star formation. The emission line ratio map indicates a ring of star forming gas surrounding the northern nucleus. The northern nucleus has much stronger emission line ratios, though very few spaxels are detected in both \OIII and \Hb. The composite nuclear spectrum of the northern nucleus has \OIII/\Hb near unity, indicating LINER-like emission. This may be associated with shocks, possibly caused by an outflow. There is some indication of broad blue wings in the Na D profile possibly associated with such an outflow (e.g. \citealt{Rupke05}), though no fit was performed.

\noindent \textbf{\emph{IRAS F16443-2915 N\&S}} (ESO 453-G005)
This widely separated pair of galaxies appear to be completely dominated by star formation. The emission line ratio maps are dominated by \HII~emission and the velocity dispersions in both systems are primarily a few tens of \kms. The southern galaxy does show a larger turbulent component approaching $\sigma \sim$100\kms, but neither indicates any shock component. These galaxies show no apparent signs of interaction, though the available imaging data are low resolution.\\

\noindent \textbf{\emph{IRAS F17138-1017}}
Our observations of this coalesced merger are consistent with the IFS observations of \citet{Monreal10}. The central regions of the galaxy are dominated by star formation, with an underlying shock-like velocity dispersion component. In the outer portions of the galaxy, our line ratios are also consistent with a dominant contribution from shocks both in line ratio and velocity dispersion. There is no indication of any AGN activity at any wavelength, with a very high 6.2\um~PAH eqw of 0.68\um~\citep{Stierwalt13}.\\

\noindent \textbf{\emph{IRAS F17207-0014}}
This ULIRG is the second brightest in the GOALS sample, behind only MRK 231. Despite its high L$_{IR}$, the X-ray data from Chandra and XMM are apparently dominated by the starburst \citep{Franceschini03,Iwasawa11}. The 6.2\um~PAH eqw is low (0.3\um), though there is no detected [Ne V] emission \citep{Petric11, Stierwalt13}. \citet{Arribas03} carried out early IFS observations of this system and suggest that this ULIRG is not evolving into a QSO. Our data show composite-like line ratios,though the overall velocity dispersion distribution is quite high, with $\sigma$ exceeding 100\kms in nearly all measurable spaxels. \citet{Rupke05,Rupke05b} detect blue shifted Na D absorption consistent with a galactic wind, which we also see in our data. In a recent study, \citet{Medling15} analyzed the nuclear region in great detail with AO-aided near-IR IFU data which also shows evidence of outflow-driven shocks. Taken all together, the data indicate an intense starburst with a galactic wind driving low velocity shocks.\\

\noindent \textbf{\emph{IRAS F17222-5953}} (ESO 138-G027).\textemdash This system is more akin to a typical, non-interacting, strongly starbursting spiral galaxy in our sample. It is in the vicinity of a few other galaxies, including the similarly bright ESO 138-G026, but even the nearest galaxy is at a projected distance of over 100 kpc and IRAS F17222-5953 is not yet interacting with any of these systems. Our IFU data applied to the \NII/\Ha~v \OIII/\Hb~ BPT diagram show a clean curve following the shape of the SDSS sequence of local star forming galaxies (e.g. \citet{Kewley06}), but with a slight apparent shift in total \NII/\Ha. We interpret this shift as an overall nitrogen enhancement ~\citep{Perez-Montero09}.\\

\noindent \textbf{\emph{IRAS 17578-0400}}
There is very little information available about this system in the literature. Our data are very low S/N in the blue, though the binned nuclear spectrum is indicative of pure star formation. This is consistent with the \NII~and \SII/\Ha~line ratios and the velocity dispersion distribution, which shows little to no indication of any appreciable non-\HII~region component and a rotation-dominated velocity field.\\

\noindent \textbf{\emph{IRAS F18093-5744 N}} (IC 4687) We classify this galaxy as a close merger; in fact it is a member of a triplet. IC 4687 is undergoing a close merger with the less massive starburst IC 4686, classed as a Wolf-Rayet galaxy by \citet{Kovo99,Fernandes04}. IC 4687 itself is a is also in a wide merger with the equally massive spiral IC 4689. The archived \emph{hst} images of IC 4687 show a complex morphology tangled up with IC 4686: gas and dust from IC 4687 appear to be obscuring the less massive system. Our IFU data cover the entirety of IC 4686/4687 and the metallicities we measure are consistent with the expected metallicties in the outskirts of IC 4687 as extrapolated from the gradient we present in this paper as well as a low-metallicity, flattened gradient in IC 4686. The kinematic information from our data also indicates that we are indeed seeing gas from both systems.\\

\noindent \textbf{\emph{IRAS F18093-5744 S}} (IC 4689) This spiral galaxy is slightly less massive and luminous than IC 4687 and less well-observed, though it is still intensely star forming ~\citep{Howell10}. It is less morphologically disturbed than the other two interacting galaxies IC 4686/4687, though its gradient is quite flattened already according to our observations ~\citep{Rich12}. Although it is widely separated from IC 4687, we include it as part of the closely interacting system of IRAS F18093-5744.\\

\noindent \textbf{\emph{IRAS F18093-5744 C}} (IC 4686)
IC 4686 is a compact strongly star-bursting galaxy with Wolf-Rayet features in its spectrum \citep{Kovo99,Fernandes04}. As noted in \citet{Rich12}, this galaxy shows a strong moderate $\sigma$ component with lower line ratios. This may be due to the very compact nature, causing an overlap in the kinematics that may require a higher-component line fit or higher resolution spectroscopy than have been considered in this thesis.\\

\noindent \textbf{\emph{IRAS F18293-3413}}
The IR-bright source in this galaxy is the northern, highly extinguished galaxy. K-band data from NIRC and \emph{HST} NICMOS show very buried spiral structure \citep{Vaisanen08b,Haan11b}. The southern galaxy appears elliptical, with our IFS spectra having no detectable emission line gas and a continuum indicative of an old stellar population. The northern galaxy is dominated by star formation with some contribution from shocks in the outskirts of the galaxy. The study of Chandra observations by \citet{Iwasawa09,Iwasawa11} shows resolved soft X-ray emission possibly associated with an outflow, but with no indication of an AGN.\\

\noindent \textbf{\emph{IRAS F18341-5732}} (IC 4734) Like IRAS F17222-5953, this isolated galaxy is in the vicinity of a few other luminous galaxies, but is not yet undergoing any interactions. Our nuclear spectra are dominated by a LINER combined with an aging stellar population. The strongest sites of star formation are where the bar in this galaxy meet the spiral arms, evidenced by the two strong clumps of \HII-region like spaxels seen in the line-ratio and metallicity. \Ha~imaging by \citet{Dopita02} shows further knots of star formation along the spiral arms and our nuclear spectra also show signs of an aging stellar population in the nucleus of IC 4734.\\

\noindent \textbf{\emph{IRAS F19115-2124}} (ESO 593-IG 008) \citet{Vaisanen08} analyze the morphology of this close pair in detail using Adaptive Optics K-band images and suggest the possibility of a triple-galaxy interaction. Kinematic modelling can be satisfied by an interacting pair, however, and indicates that though this system is classified as a 'close pair', it is actually just entering first pericenter (Josh Barnes, private communication). The metallicity gradient in this system, however, has already flattened considerably \citep{Rich12}.\\

\noindent \textbf{\emph{IRAS F20551-4250}} (ESO 286-IG019)
This late stage merger shows ambiguous AGN signatures. The Chandra X-ray data show extended soft emission inconsistent with AGN \citep{Ptak03,Grimes05,Iwasawa09,Iwasawa11}, but \citep{Franceschini03} classify the galaxy as an AGN based on the XMM data. There is no [Ne V] detection in the mid-IR, though the 6.2 \um~PAH eqw is very low (0.1\um) more consistent with AGN emission \citep{Petric11,Stierwalt13}. \citet{Imanishi10,Sani08,Nardini08} also claim signs of a heavily obscured AGN with IR data. Our IFS data are dominated by a post-starburst spectrum and composite LINER emission, with the emission line velocity dispersion distribution centered around 100\kms and trailing to 300\kms. The emission line maps show the lowest emission line ratios near the nucleus with increasing ratios further away, more consistent with shock excitation in the optical rather than a low-luminosity AGN. The Na D absorption line profile in this galaxy also appears to have broad blue wings consistent with an outflow (e.g. \citealt{Rupke05}), though no fit to the complex was performed. Our observations are in agreement with previous optical spectra of the nucleus, classified as HII-dominated \citep{Veilleux95,Kewley01b}, coupled with extended shock excitation.\\

\noindent \textbf{\emph{IRAS F21330-3846}} (ESO 343-IG013)
This closely interacting pair shows tidal tails in the available DSS image. Our IFS data show both discs are dominated primarily by \HII~emission. The line ratio diagnostic diagrams show a composite component, which corresponds primarily to the outskirts of the outskirts of the northern galaxy. The ionized emission between the two systems is also higher line ratio and velocity dispersion than the discs, consistent with shock excitation possibly caused directly by the interaction, as seen in systems like Stephan's Quintet and the Taffy Galaxies \citep{Cluver10,Guillard10,Peterson12} as opposed to outflow-driven shocks.\\

\noindent \textbf{\emph{IRAS F21453-3511}} (NGC 7130)
NGC 7130 is an isolated face-on spiral galaxy with a previously observed Seyfert 2 nucleus \citep{Phillips83}. The AGN dominates the mid-IR spectra in the nucleus and is detected in x-ray as well \citep{Petric11,Iwasawa11}. Because of it is nearby and nearly uninclined, our IFS spectroscopy resolve well the AGN and starburst component in this galaxy showing spiral arms dominated by star formation and a strong Seyfert component in the nucleus. The distribution of spaxels in the diagnostic diagram and the line ratio maps show this well, and agree with \citet{Monreal10} who also observed this system. Na D absorption shows a blue-shifted component which may be due to an AGN-driven outflow. Further analysis of the WiFeS data for NGC 7130 is presented in \citet{Leslie14}.\\

\noindent \textbf{\emph{IRAS F22467-4906}} (ESO 239-IG002)
The nuclear regions of this compact late-stage merger have very high \NII/\Ha and \SII/\Ha line ratios, with moderate \OIII/\Hb ratios indicating composite/LINER behavior. The nucleus also shows a post-starburst spectrum in the blue, with older stellar features in the tidal debris to the southeast of the nucleus. There is some detection of low-level \HII-region like emission in the tidal arm to the SE as well, likely similar to the star formation in IRAS F10038-3338, likely related to the shock-induced star formation in NGC 7252 described by \citet{Chien10}. The Chandra data show a point-like source detected in the hard band, while the soft X-rays extend to the N and S and the overall hardness of the X-ray data are not consistent with an AGN \citep{Iwasawa11}. There is a very broad blue shifted Na D absorption feature indicating an outflow which may be driving shocks in the is galaxy. The velocity dispersion distribution seems consistent with low velocity shocks with some high-velocity components, but the limited spatial resolution of our data in this system make this more difficult to discern.\\

\noindent \textbf{\emph{IRAS F23128-5919}} (ESO 148-IG002)
The nuclei in this closely interacting ULIRG are separated by a projected distance of about 4 kpc and show distinctly different properties in our spectra. The northern nucleus has line ratio and velocity dispersion values consistent with star formation. The southern nucleus is about three times brighter in at 24\um and hosts an X-ray and IR detected AGN \citep{Franceschini03,Petric11,Iwasawa11}, which are mirrored in the enhanced composite/Seyfert line ratios and broadened velocity dispersions seen in our IFS diagnostic diagrams and emission line maps. The emission line components associated with the composite/Seyfert emission show broad velocity dispersions of 200-400\kms, though there is still a significant contribution below 100\kms, likely due to the ongoing star formation in both nuclei. The broad component shows signs of rotation while the underlying narrow component seems to show none, indicating most of the narrow emission may be coming from the non-AGN galaxy. We do not detect a blue-shifted component in fits to the Na D component. Further analysis of the WiFeS data for ESO 148-IG002 is presented in \citet{Leslie14}.\\

%\pagebreak

\setcounter{figure}{-1}

\begin{figure*}[htpb!]
\centering
{\includegraphics[scale=0.95,angle=90]{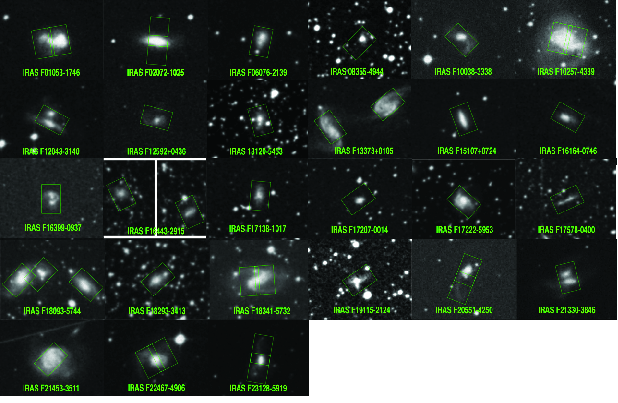}}
\caption{DSS R-band images with individual WiFeS pointings overlaid in green. Coordinates for the pointings are given in Table A.1}
\end{figure*}

\clearpage

\begin{table*}
\centering
\caption{WiFeS Observations \label{table_observations}}
\hspace{-0.5cm}
\begin{tabular}{lllccr}
IRAS Name & \multicolumn{1}{c}{Optical ID} & \multicolumn{1}{c}{Observation Date} & \multicolumn{1}{c}{Pointing center (J2000)} & \multicolumn{1}{c}{PA\degr} & \multicolumn{1}{c}{Exp.} \\

\multicolumn{1}{c}{(a)} & \multicolumn{1}{c}{(b)} & \multicolumn{1}{c}{(c)} & \multicolumn{1}{c}{(d)} & \multicolumn{1}{c}{(e)} & \multicolumn{1}{c}{(f)} \\
\hline\extravspace
%  Name      & Other name    & Date(s) Observed   & PTcent                    & PA  & exp time     \\         
F19115-2124 & ESO 593-IG008 & 2009 Jul 26       & 19h14m30.8s~~$-$21d19m05s & 305 & 120m \\
F16443-2915 & ESO 453-G005  & 2009 Jul 28       & 16h47m31.1s~~$-$29d21m22s & 115 & 67m  \\
            &               & 2009 Jul 28       & 16h47m29.5s~~$-$29d19m07s & 115 & 67m  \\
F23128-5919 & ESO 148-IG002 & 2009 Jul 28       & 23h15m47.1s~~$-$59d03m25s & 173 & 67m  \\
            &               & 2009 Aug 14, 15   & 23h15m46.7s~~$-$59d03m00s & 173 & 59m  \\
F18293-3413 & -             & 2009 Aug 14       & 18h32m41.1s~~$-$34d11m26s & 130 & 84m  \\ 
F20551-4250 & ESO 286-IG019 & 2009 Aug 15       & 20h58m26.7s~~$-$42d39m00s & 160 & 100m \\ 
            &               & 2009 Aug 15       & 20h58m27.5s~~$-$42d39m23s & 160 & 100m \\ 
F18341-5732 & IC 4734       & 2009 Aug 18       & 18h38m24.5s~~$-$57d29m27s & 10  & 60m  \\ 
            &               & 2009 Aug 18       & 18h38m27.0s~~$-$57d29m28s & 10  & 60m  \\ 
F22467-4906 & ESO 239-IG002 & 2009 Aug 18       & 22h49m40.6s~~$-$48d51m00s & 30  & 40m  \\ 
            &               & 2009 Aug 18       & 22h39m39.0s~~$-$48d50m51s & 30  & 40m  \\ 
F02072-1025 & NGC 839       & 2009 Aug 18       & 02h09m42.6s~~$-$10d10m49s & 350 & 40m  \\ 
            &               & 2009 Aug 18       & 02h09m42.8s~~$-$10d11m13s & 350 & 40m  \\ 
F21330-3846 & ESO 343-IG013 & 2009 Aug 20       & 21h36m10.7s~~$-$38d32m38s & 15  & 80m  \\
F01053-1746 & IC 1623       & 2009 Aug 20       & 01h07m46.8s~~$-$17d30m25s & 10  & 40m  \\ 
            &               & 2009 Aug 20       & 01h07m48.3s~~$-$17d30m29s & 10  & 40m  \\ 
F21453-3511 & NGC 7130      & 2009 Sep 18       & 21h48m19.6s~~$-$34d57m08s & 130 & 60m  \\
F06076-2139 & -             & 2010 Mar 12, 13   & 06h09m45.9s~~$-$21d40m30s & 170 & 60m  \\
08355-4944  & -             & 2010 Mar 14, 15   & 08h37m02.2s~~$-$49d54m36s & 135 & 80m  \\
13120-5453  & -             & 2010 Mar 14, 16   & 13h15m06.3s~~$-$55d09m26s & 15  & 95m  \\
F10257-4339 & NGC 3256      & 2010 Mar 15       & 10h27m51.9s~~$-$43d54m10s & 165 & 60m  \\
            &               & 2010 Mar 15       & 10h27m50.1s~~$-$43d54m15s & 165 & 60m  \\
F13373+0105 & Arp 240       & 2010 Mar 15, 16   & 13h39m52.8s~~$+$00d50m26s & 35  & 60m  \\
            &               & 2010 May 7        & 13h39m57.8s~~$+$00d49m56s & 130 & 50m  \\
F10038-3338 & ESO 374-IG032 & 2010 Mar 16       & 10h06m04.5s~~$-$33d53m10s & 45  & 60m  \\
F12592+0436 & CGCG 043-099   & 2010 Mar 16       & 13h01m50.3s~~$+$04d20m04s & 75  & 40m  \\
F12043-3140 & ESO 440-IG058 & 2010 May 7, 8     & 12h06m51.7s~~$-$31d56m55s & 60  & 100m \\
F16164-0746 & -             & 2010 May 7        & 16h19m11.6s~~$-$07d54m03s & 60  & 75m  \\
F18093-5744 & IC 4687       & 2010 May 7, 13    & 18h13m40.1s~~$-$57d43m28s & 50  & 116m \\
            &               & 2010 May 9        & 18h13m40.2s~~$-$57d44m56s & 140 & 109m \\
            &               & 2010 May 8, 9, 15 & 18h13m38.9s~~$-$57d43m56s & 50  & 60m  \\
F16399-0937 & -             & 2010 May 10       & 16h42m40.3s~~$-$09d43m15s & 0   & 90m  \\ 
F17138-1017 & -             & 2010 May 12, 17   & 17h16m35.8s~~$-$10d20m40s & 175 & 175m \\
F15107+0724 & CGCG 049-057  & 2010 May 12, 13   & 15h13m13.0s~~$+$07d13m31s & 20  & 75m  \\ 
17578-0400  & -             & 2010 May 13       & 18h00m32.1s~~$-$04d00m55s & 115 & 75m  \\
F17222-5953 & ESO 138-G027  & 2010 May 14       & 17h26m43.2s~~$-$59d55m57s & 45  & 75m  \\
F17207-0014 & -             & 2010 May 15       & 17h23m22.1s~~$+$00d16m59s & 315 & 75m  \\
\hline
\end{tabular}
\begin{quote}
(a) IRAS identifier from \citet{Sanders03,Armus09} (b) Optical counterpart from \citet{Armus09} (c) Observation Dates (d) center of pointing (e) Position Angle, degrees E of N (f) Total pointing exposure time in minutes
\end{quote}
\end{table*}

\clearpage

%\onecolumn

\pagebreak

\begin{sidewaysfigure*}%[htb]
\vspace{0.00in}
\centering
{\Large \textbf{IRAS F01053-1746}}
{\includegraphics[width=\textheight]{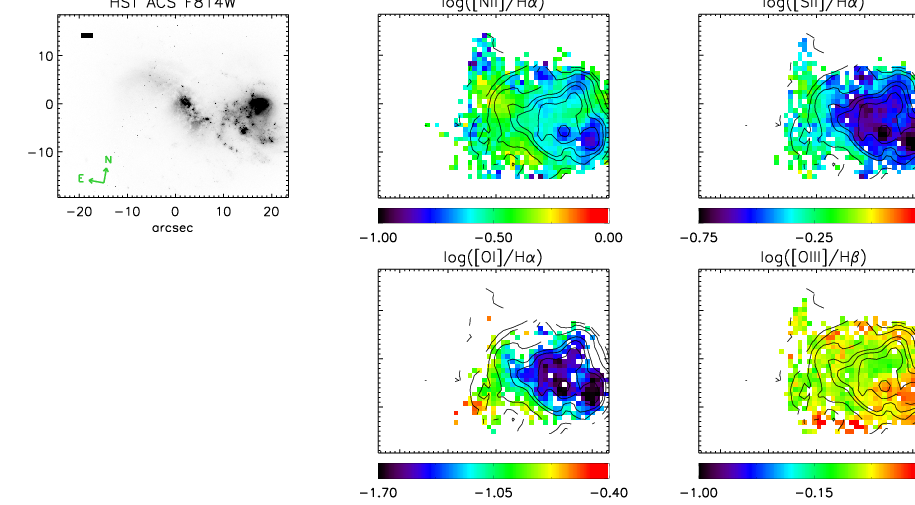}}
\caption[]{}
\end{sidewaysfigure*}
\pagebreak
\begin{sidewaysfigure*}%[htbp!]
\ContinuedFloat
\vspace{0.000in}
{\includegraphics[width=\textheight]{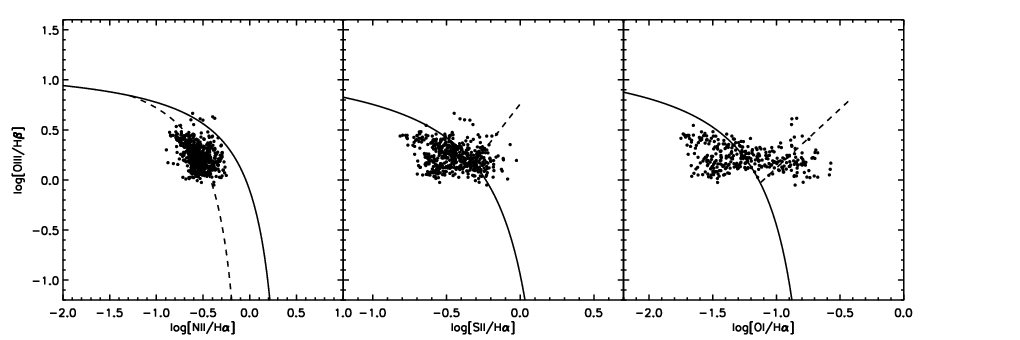}}
{\includegraphics[width=\textheight]{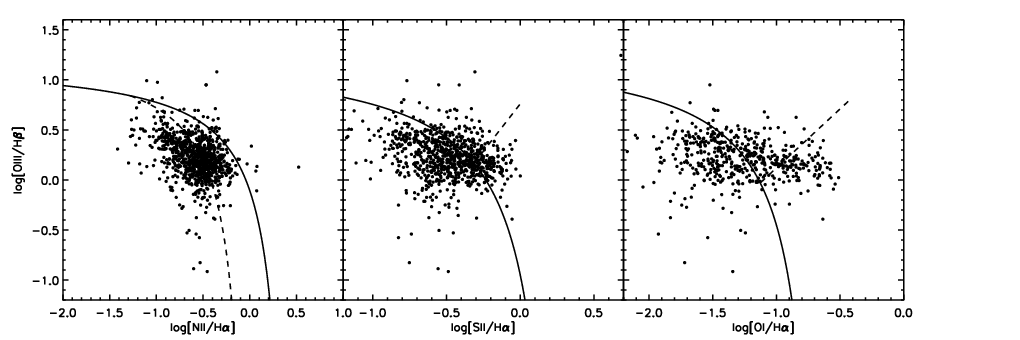}}
\caption{The top left panel shows an \emph{HST} I-band image of IRAS F01053-1746 aligned with the total WiFeS field of view with a 1-kpc bar. The top four panels to the right show emission line ratio maps for \NII/\Ha, \SII/\Ha, \OI/\Ha~and \OIII/\Hb~ with \Ha~contours measured from our data overlaid.
The six bottom panels are standard diagnostic diagrams (\citealt{Kewley06}). Points in the top three correspond to the total flux ratios in a single spaxel. Points in the bottom three correspond to line ratios from the individual components above S/N of 3 in each line. The same diagrams for the remaining portion of the sample are in Appendix 1.}
\end{sidewaysfigure*}

%\pagebreak

%\pagebreak

\begin{sidewaysfigure*}%[htb]
\vspace{0.00in}
\centering
{\Large \textbf{IRAS F02072-1025}}
{\includegraphics[width=\textwidth]{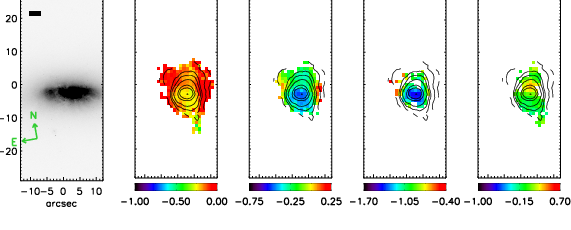}}
%\caption[]{}
\end{sidewaysfigure*}
\begin{sidewaysfigure*}%[htb]
\ContinuedFloat
\vspace{0.000in}
{\includegraphics[width=\textheight]{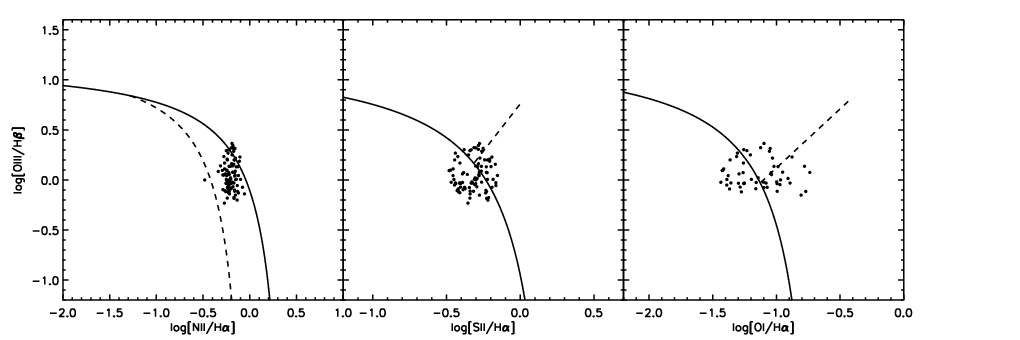}}
{\includegraphics[width=\textheight]{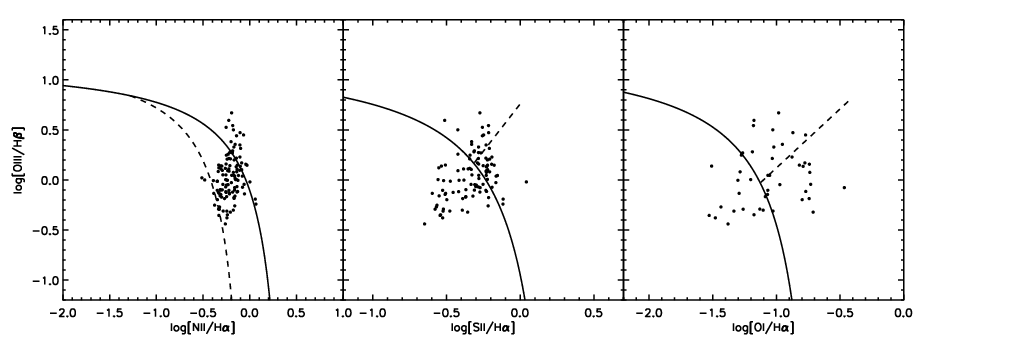}}
\caption{Same as figure A1, but for IRAS F02072-1025.}
\end{sidewaysfigure*}

\begin{sidewaysfigure*}%[htb]
\vspace{0.00in}
\centering
{\Large \textbf{IRAS F06076-2139}}
{\includegraphics[width=0.80\textwidth]{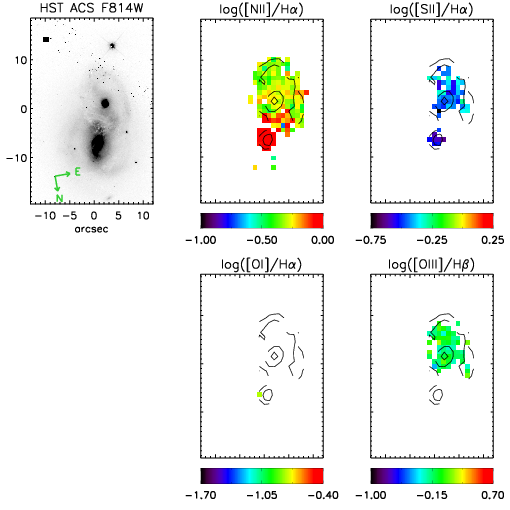}}
%\caption[]{}
\end{sidewaysfigure*}
\begin{sidewaysfigure*}%[htb]
\ContinuedFloat
\vspace{0.000in}
{\includegraphics[width=\textheight]{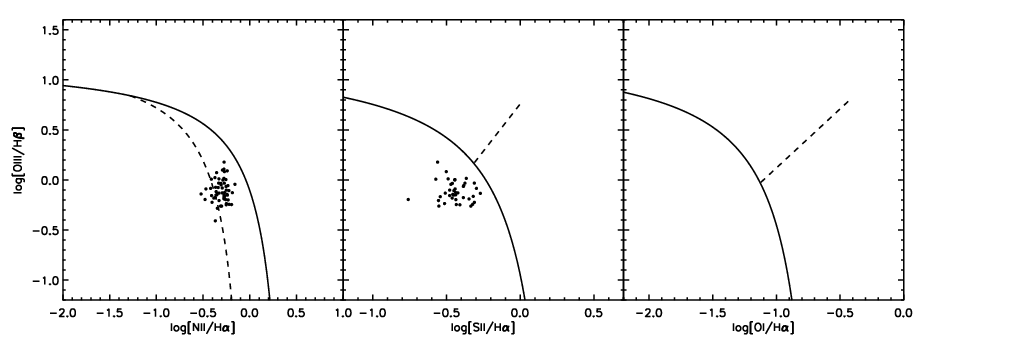}}
{\includegraphics[width=\textheight]{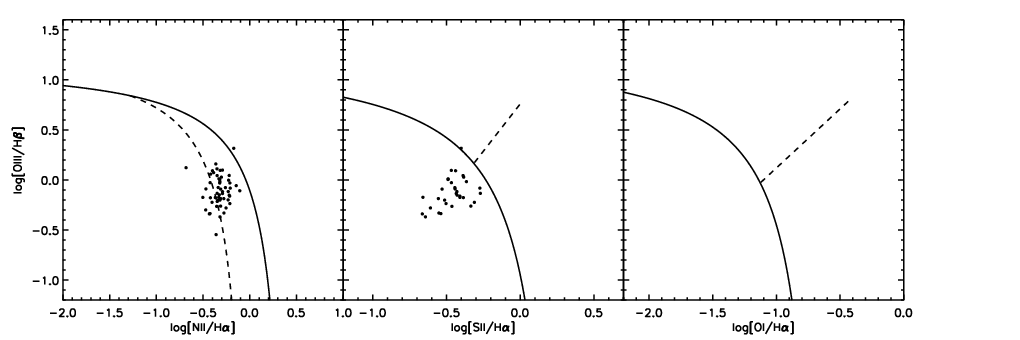}}
\caption{Same as figure A1, but for IRAS F06076-2139.}
\end{sidewaysfigure*}

\begin{sidewaysfigure*}%[htb]
\vspace{0.00in}
\centering
{\Large \textbf{IRAS F08355-4944}}
{\includegraphics[width=0.80\textwidth]{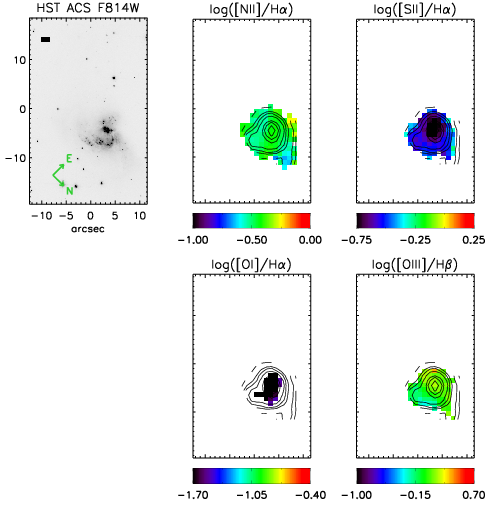}}
%\caption[]{}
\end{sidewaysfigure*}
\begin{sidewaysfigure*}%[htb]
\ContinuedFloat
\vspace{0.000in}
{\includegraphics[width=\textheight]{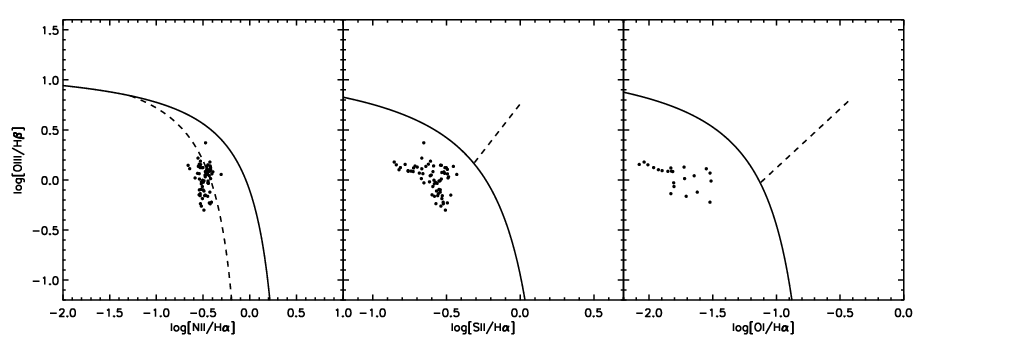}}
{\includegraphics[width=\textheight]{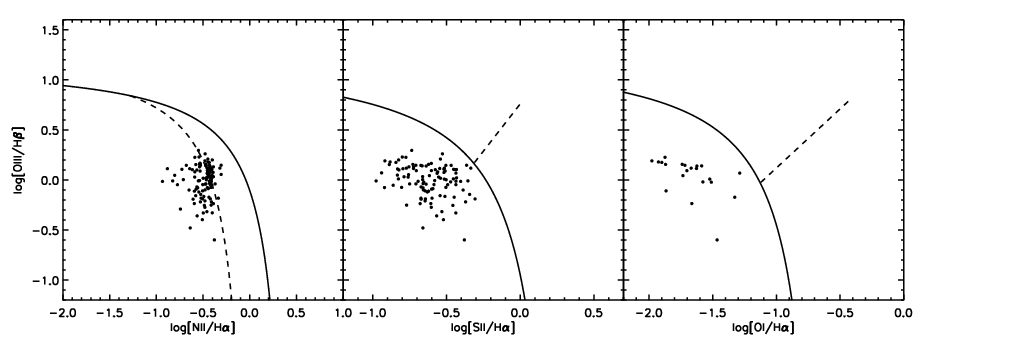}}
\caption{Same as figure A1, but for IRAS 08355-4944.}
\end{sidewaysfigure*}

\begin{sidewaysfigure*}%[htb]
\vspace{0.00in}
\centering
{\Large \textbf{IRAS F10038-3338}}
{\includegraphics[width=0.80\textwidth]{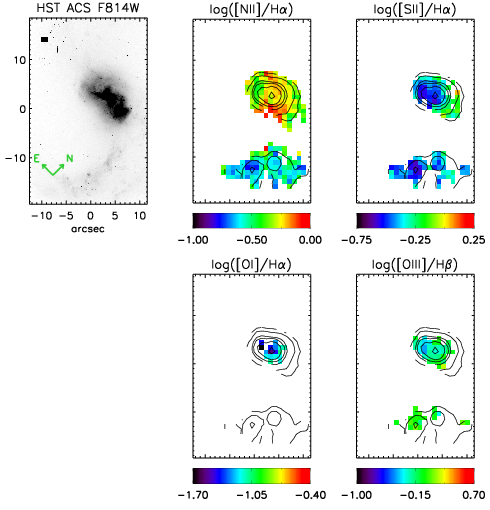}}
%\caption[]{}
\end{sidewaysfigure*}
\begin{sidewaysfigure*}%[htb]
\ContinuedFloat
\vspace{0.000in}
{\includegraphics[width=\textheight]{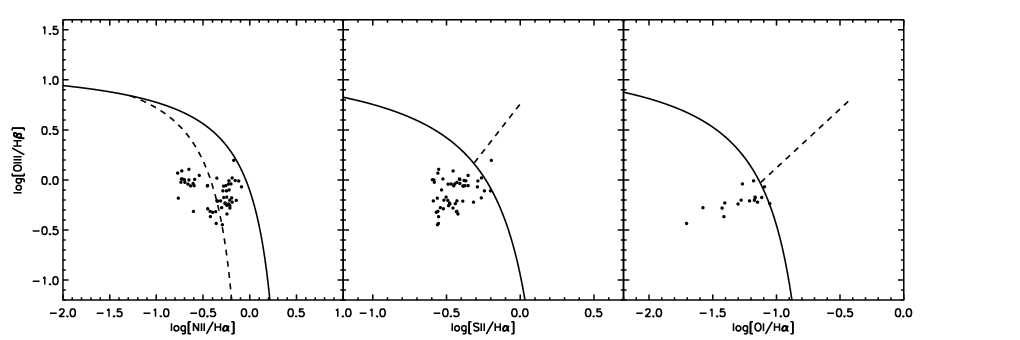}}
{\includegraphics[width=\textheight]{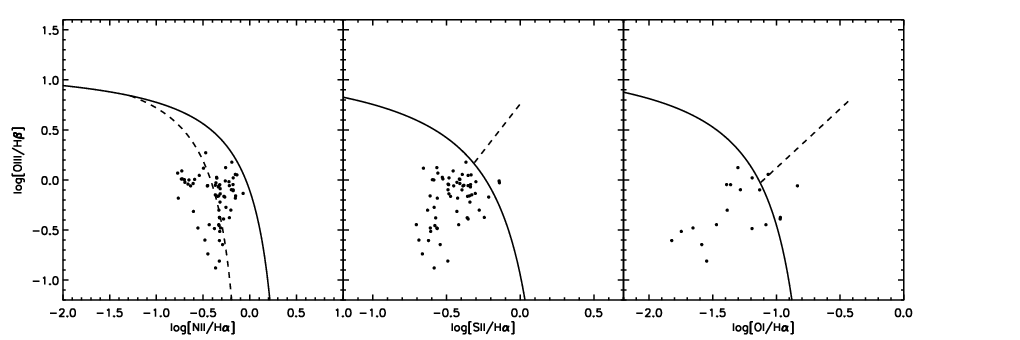}}
\caption{Same as figure A1, but for IRAS F10038-3338.}
\end{sidewaysfigure*}

\begin{sidewaysfigure*}%[htb]
\vspace{0.00in}
\centering
{\Large \textbf{IRAS F10257-4339}}
{\includegraphics[width=\textwidth]{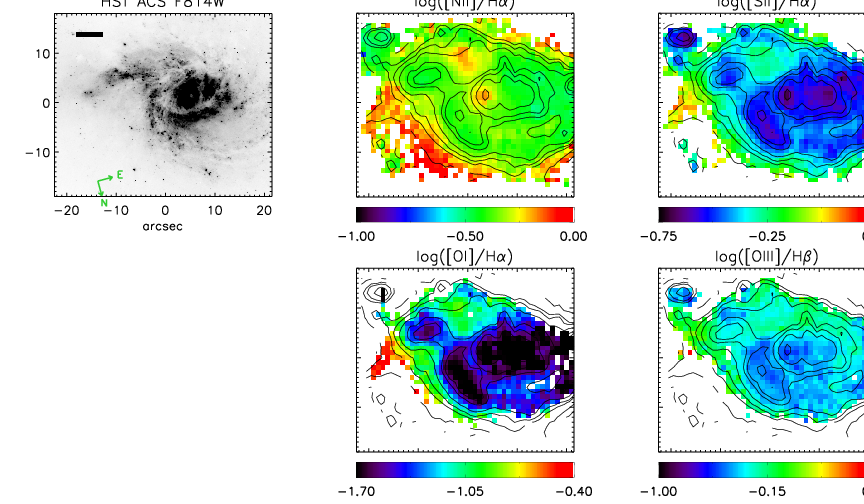}}
%\caption[]{}
\end{sidewaysfigure*}
\begin{sidewaysfigure*}%[htb]
\ContinuedFloat
\vspace{0.000in}
{\includegraphics[width=\textheight]{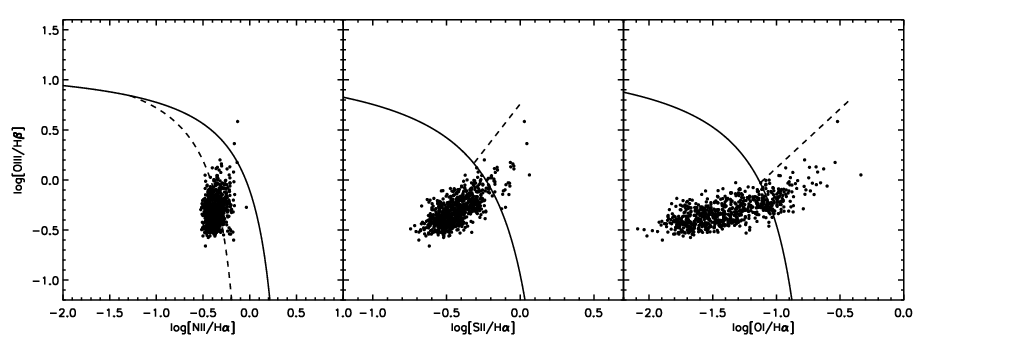}}
{\includegraphics[width=\textheight]{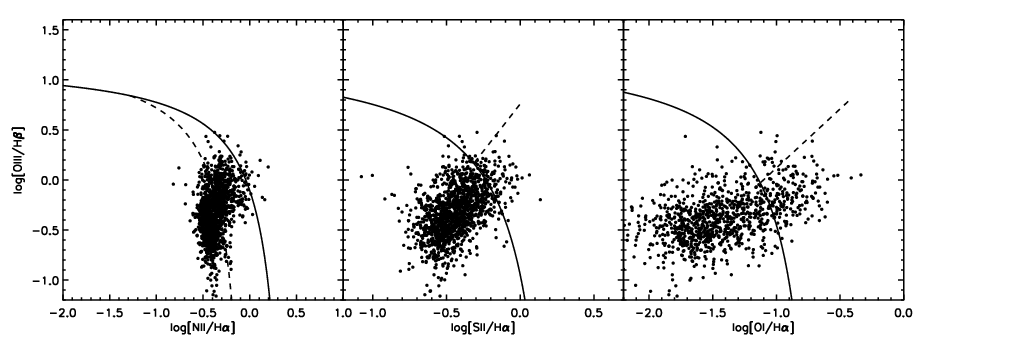}}
\caption{Same as figure A1, but for IRAS F10257-4339.}
\end{sidewaysfigure*}

\begin{sidewaysfigure*}%[htb]
\vspace{0.00in}
\centering
{\Large \textbf{IRAS F12043-3140}}
{\includegraphics[width=0.80\textwidth]{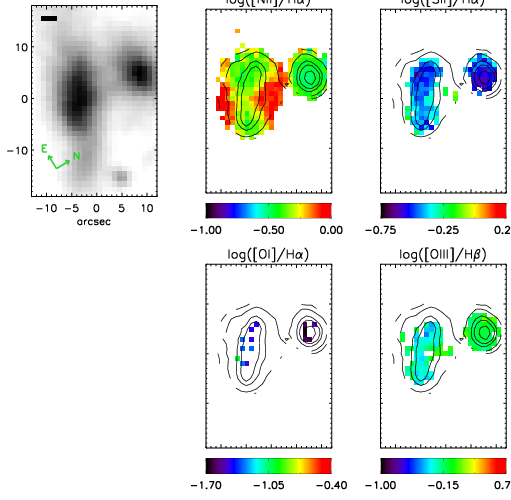}}
%\caption[]{}
\end{sidewaysfigure*}
\begin{sidewaysfigure*}%[htb]
\ContinuedFloat
\vspace{0.000in}
{\includegraphics[width=\textheight]{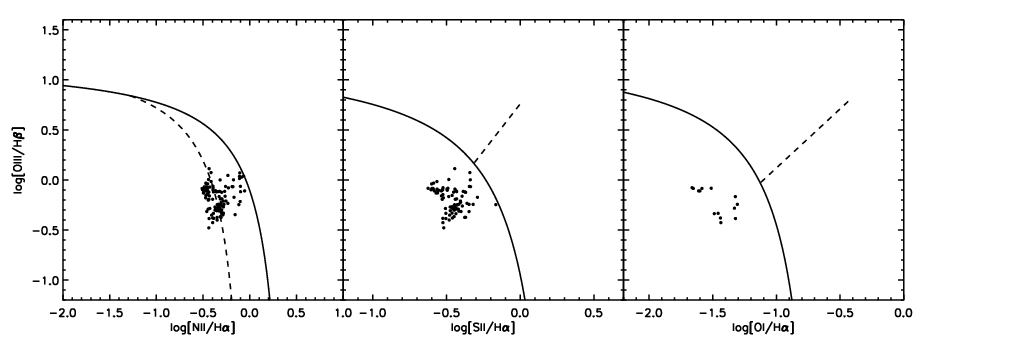}}
{\includegraphics[width=\textheight]{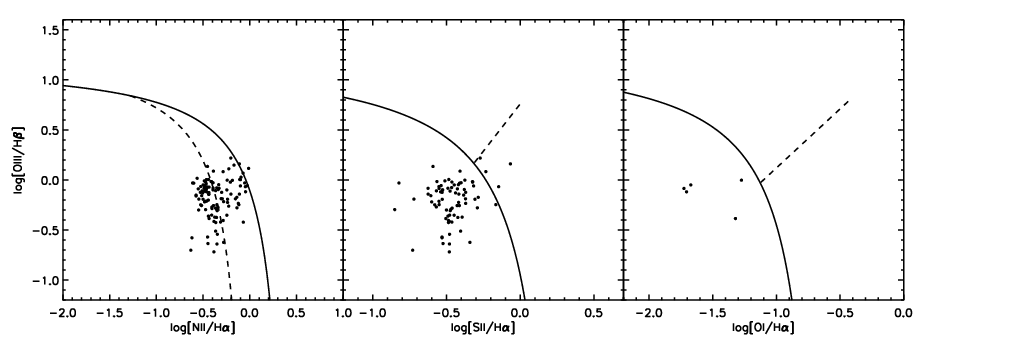}}
\caption{Same as figure A1, but for IRAS F12043-3140 and a DSS R-band image.}
\end{sidewaysfigure*}

\begin{sidewaysfigure*}%[htb]
\vspace{0.00in}
\centering
{\Large \textbf{IRAS F12592+0436}}
{\includegraphics[width=0.80\textwidth]{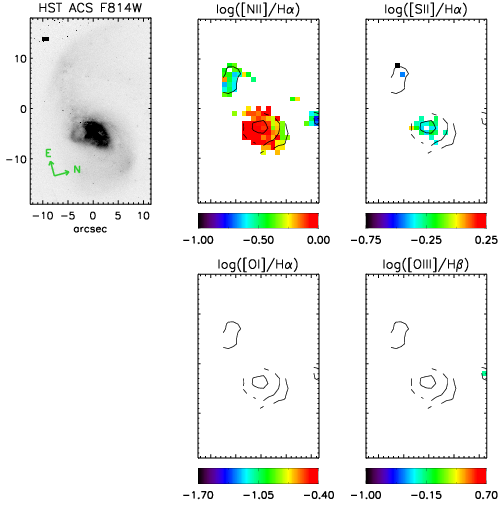}}
%\caption[]{}
\end{sidewaysfigure*}
\begin{sidewaysfigure*}%[htb]
\ContinuedFloat
\vspace{0.000in}
{\includegraphics[width=\textheight]{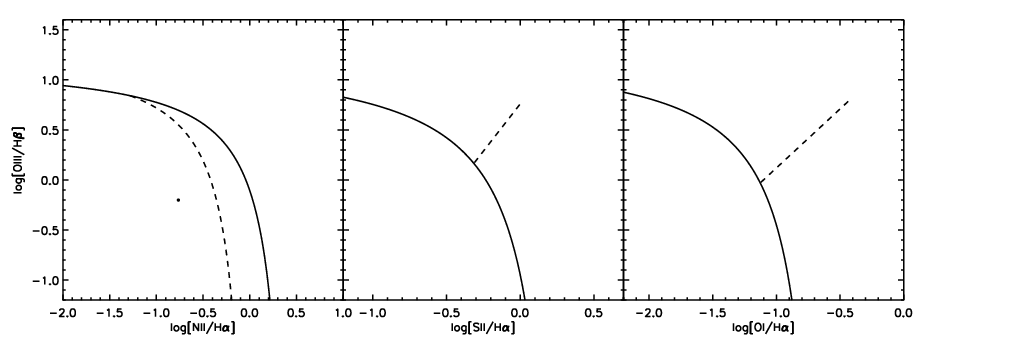}}
{\includegraphics[width=\textheight]{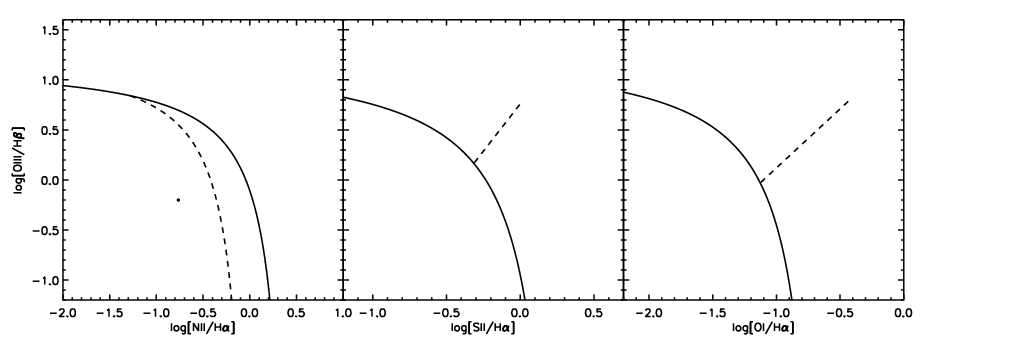}}
\caption{Same as figure A1, but for IRAS F12592+0436.}
\end{sidewaysfigure*}

\begin{sidewaysfigure*}%[htb]
\vspace{0.00in}
\centering
{\Large \textbf{IRAS 13120-5453}}
{\includegraphics[width=0.80\textwidth]{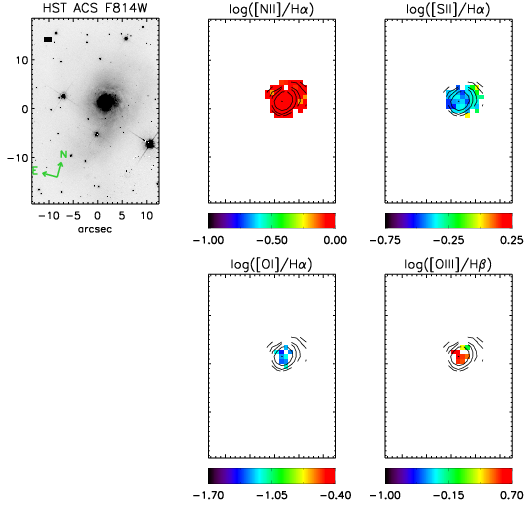}}
%\caption[]{}
\end{sidewaysfigure*}
\begin{sidewaysfigure*}%[htb]
\ContinuedFloat
\vspace{0.000in}
{\includegraphics[width=\textheight]{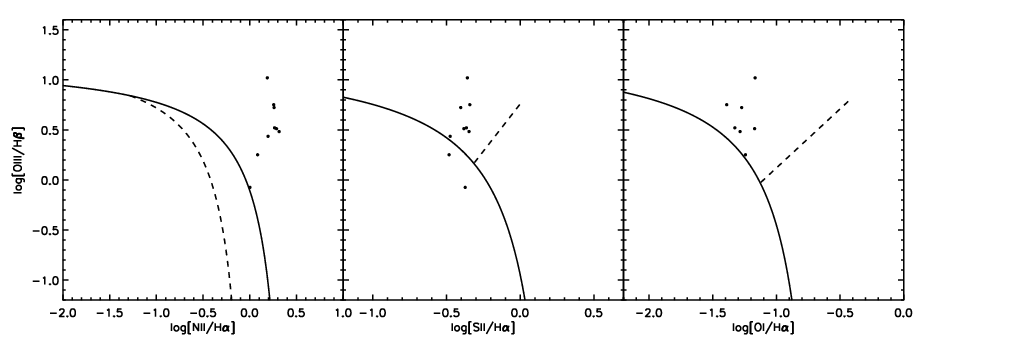}}
{\includegraphics[width=\textheight]{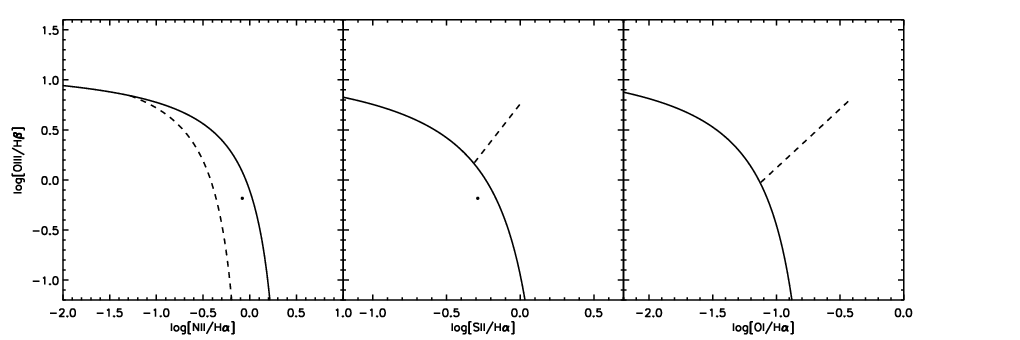}}
\caption{Same as figure A1, but for IRAS 13120-5453.}
\end{sidewaysfigure*}
\clearpage

\begin{sidewaysfigure*}%[htb]
\vspace{0.00in}
\centering
{\Large \textbf{IRAS F13373+0105 W}}
{\includegraphics[width=0.80\textwidth]{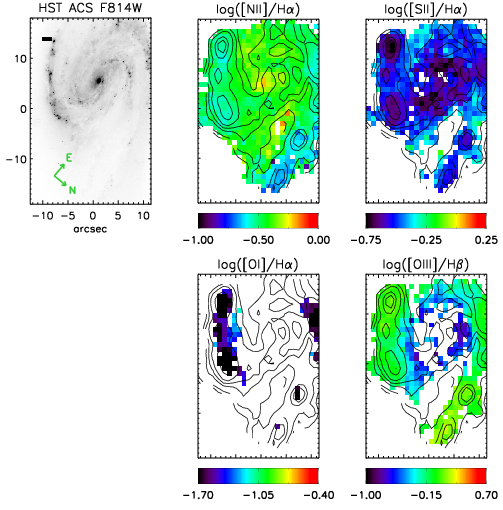}}
%\caption[]{}
\end{sidewaysfigure*}
\begin{sidewaysfigure*}%[htb]
\ContinuedFloat
\vspace{0.000in}
{\includegraphics[width=\textheight]{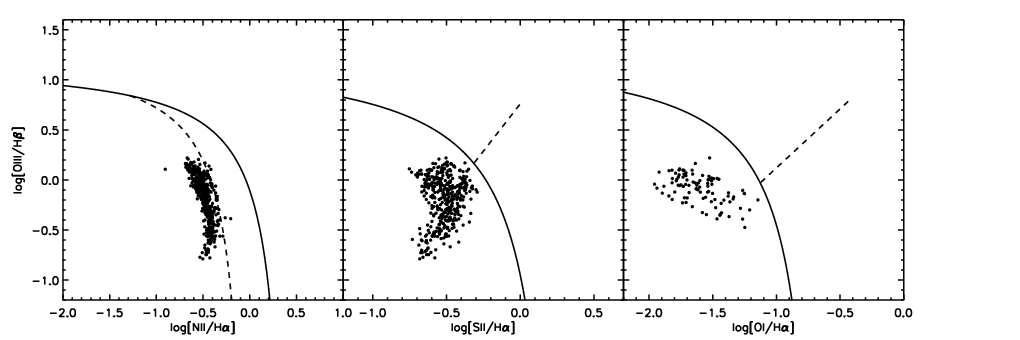}}
{\includegraphics[width=\textheight]{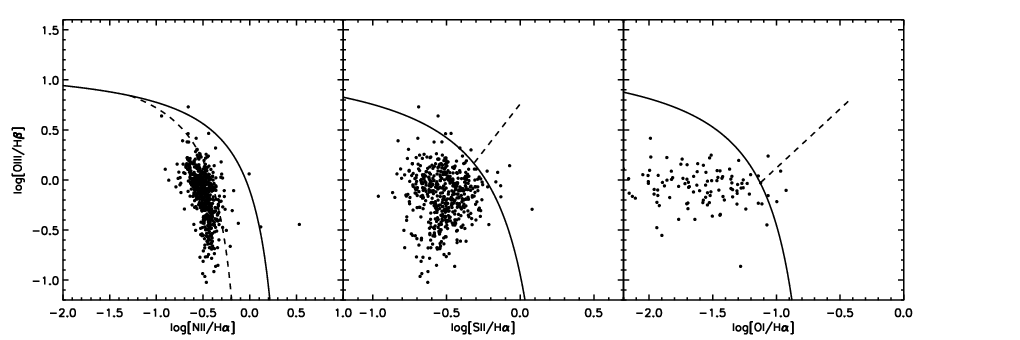}}
\caption{Same as figure A1, but for IRAS F13373+0105 W.}
\end{sidewaysfigure*}

\begin{sidewaysfigure*}%[htb]
\vspace{0.00in}
\centering
{\Large \textbf{IRAS F13373+0105 E}}
{\includegraphics[width=0.80\textwidth]{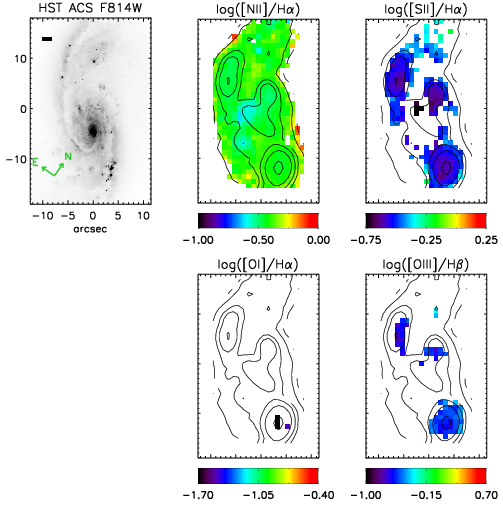}}
%\caption[]{}
\end{sidewaysfigure*}
\begin{sidewaysfigure*}%[htb]
\ContinuedFloat
\vspace{0.000in}
{\includegraphics[width=\textheight]{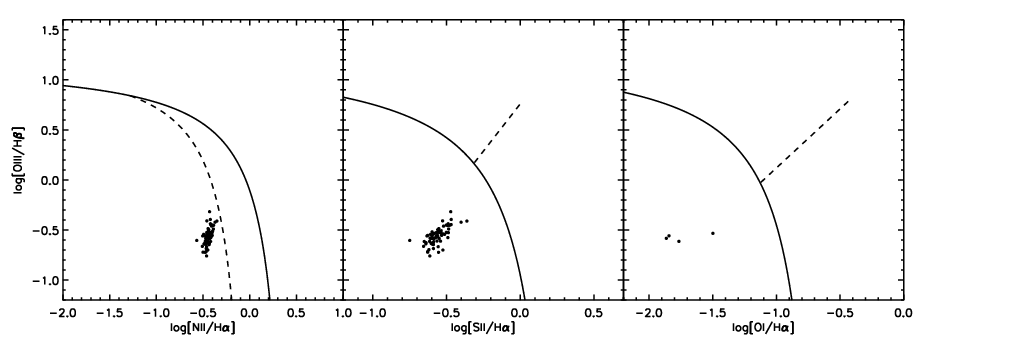}}
{\includegraphics[width=\textheight]{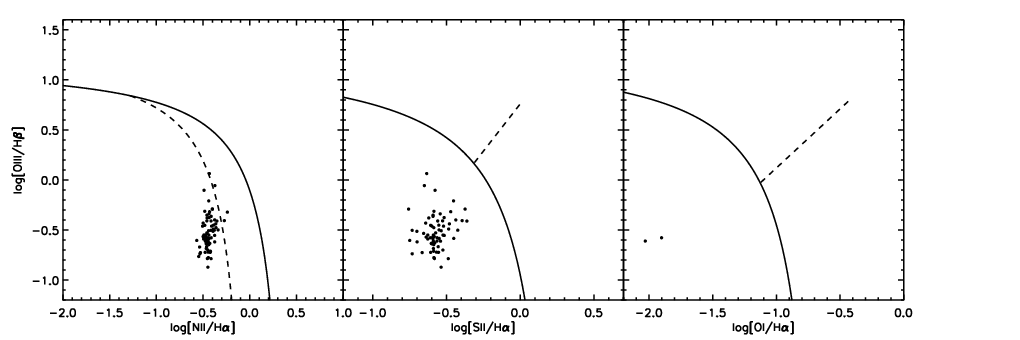}}
\caption{Same as figure A1, but for IRAS F13373+0105 E.}
\end{sidewaysfigure*}

\begin{sidewaysfigure*}%[htb]
\vspace{0.00in}
\centering
{\Large \textbf{IRAS F15107+0724}}
{\includegraphics[width=0.80\textwidth]{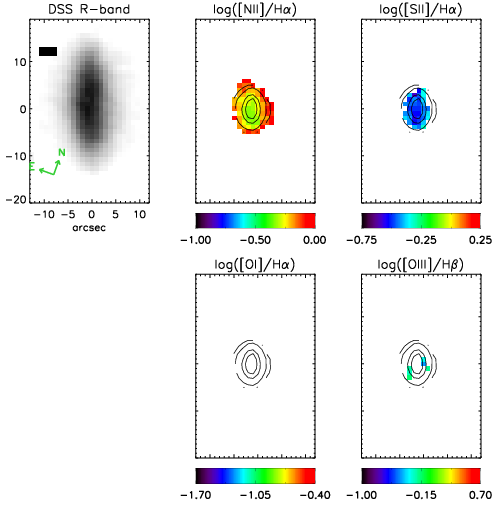}}
%\caption[]{}
\end{sidewaysfigure*}
\begin{sidewaysfigure*}%[htb]
\ContinuedFloat
\vspace{0.000in}
{\includegraphics[width=\textheight]{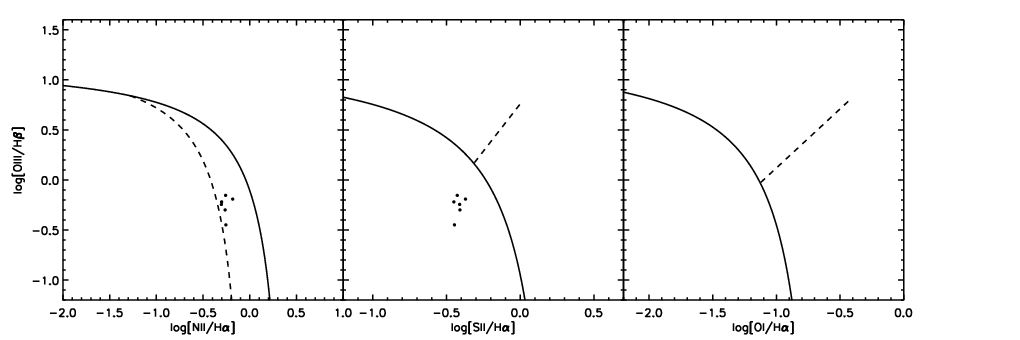}}
{\includegraphics[width=\textheight]{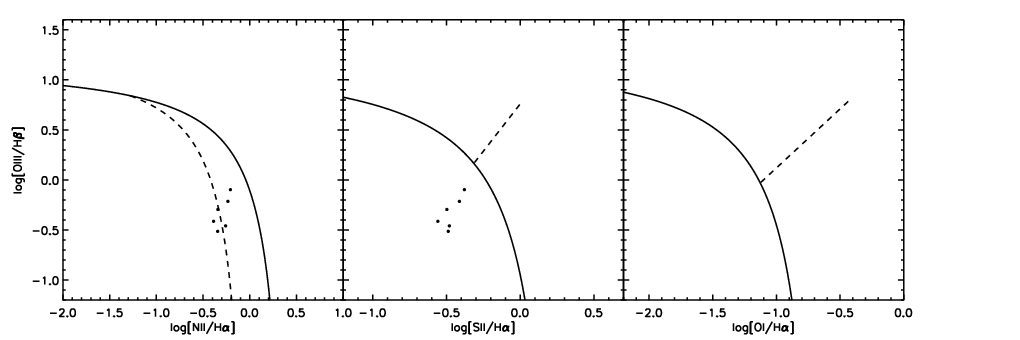}}
\caption{Same as figure A1, but for IRAS F15107+0724 and a DSS R-band image.}
\end{sidewaysfigure*}

\begin{sidewaysfigure*}%[htb]
\vspace{0.00in}
\centering
{\Large \textbf{IRAS F16164-0746}}
{\includegraphics[width=0.80\textwidth]{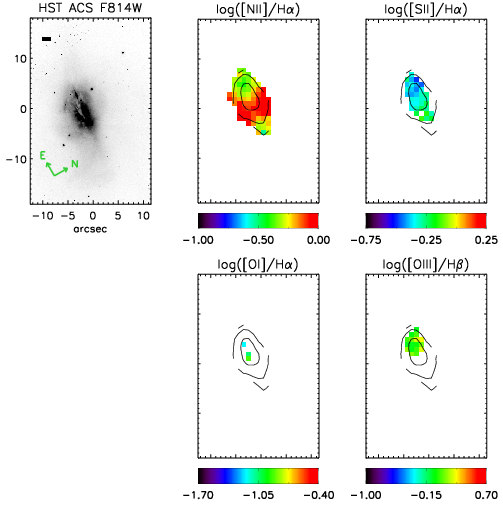}}
%\caption[]{}
\end{sidewaysfigure*}
\begin{sidewaysfigure*}%[htb]
\ContinuedFloat
\vspace{0.000in}
{\includegraphics[width=\textheight]{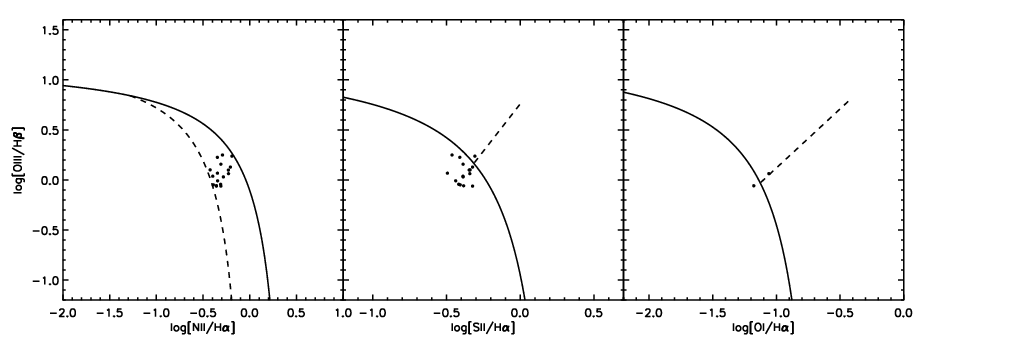}}
{\includegraphics[width=\textheight]{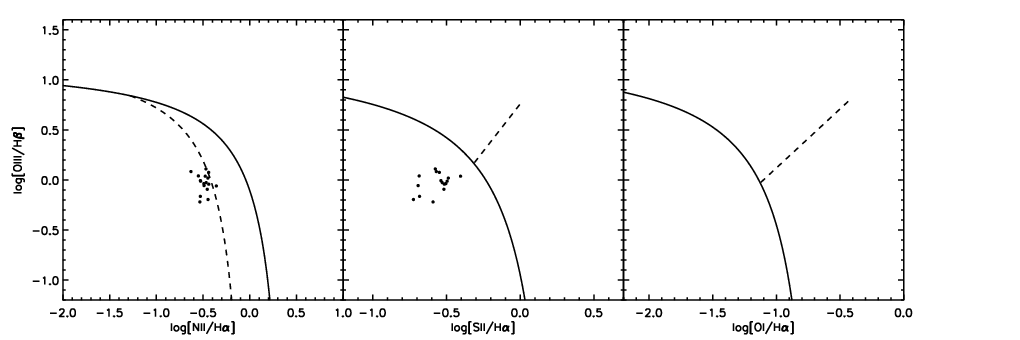}}
\caption{Same as figure A1, but for IRAS F16164-0746.}
\end{sidewaysfigure*}

\begin{sidewaysfigure*}%[htb]
\vspace{0.00in}
\centering
{\Large \textbf{IRAS F16399-0937}}
{\includegraphics[width=0.80\textwidth]{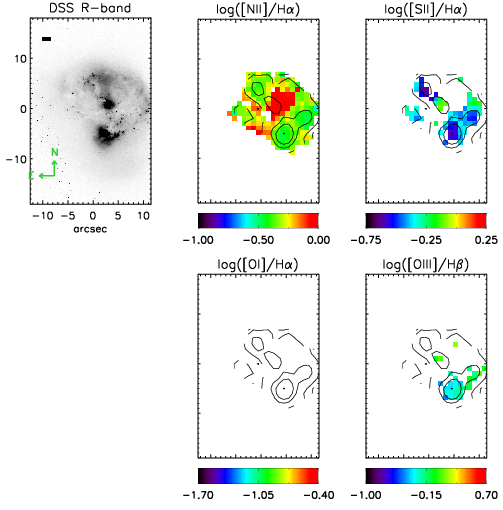}}
%\caption[]{}
\end{sidewaysfigure*}
\begin{sidewaysfigure*}%[htb]
\ContinuedFloat
\vspace{0.000in}
{\includegraphics[width=\textheight]{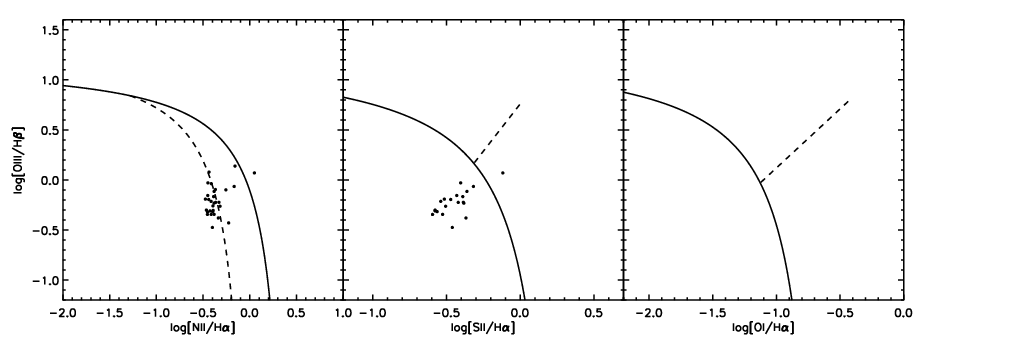}}
{\includegraphics[width=\textheight]{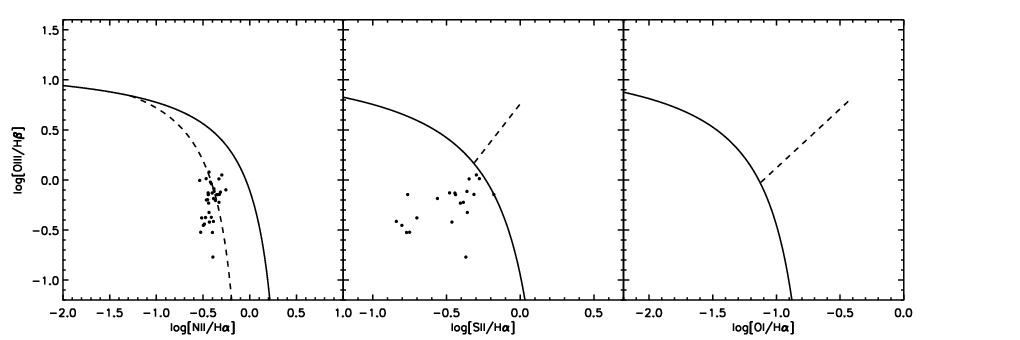}}
\caption{Same as figure A1, but for IRAS F16399-0937.}
\end{sidewaysfigure*}

\clearpage

\begin{sidewaysfigure*}%[htb]
\vspace{0.00in}
\centering
{\Large \textbf{IRAS F16443-2915 N}}
{\includegraphics[width=0.80\textwidth]{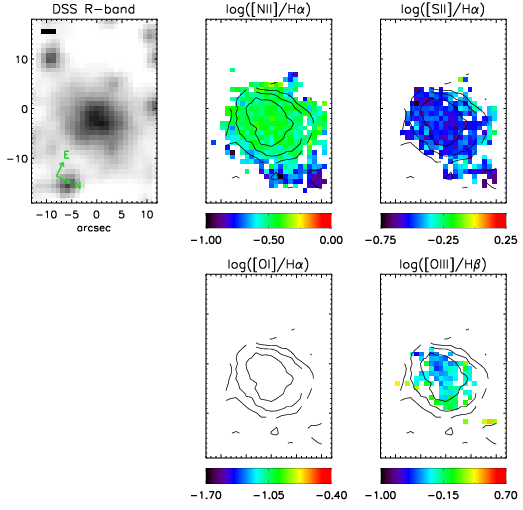}}
%\caption[]{}
\end{sidewaysfigure*}
\begin{sidewaysfigure*}%[htb]
\ContinuedFloat
\vspace{0.000in}
{\includegraphics[width=\textheight]{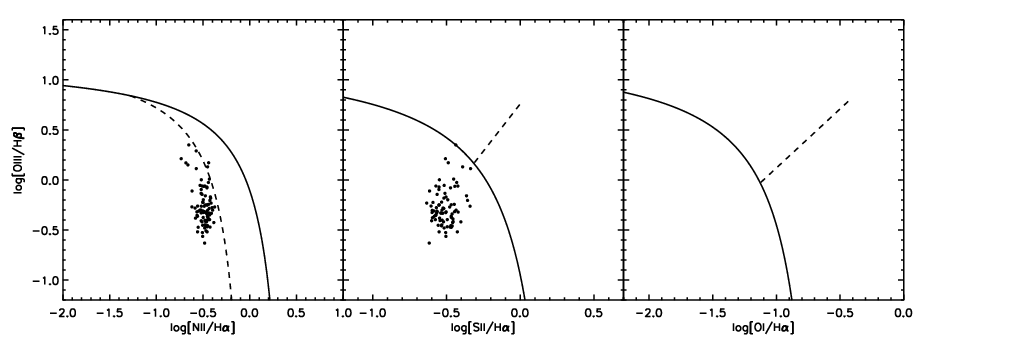}}
{\includegraphics[width=\textheight]{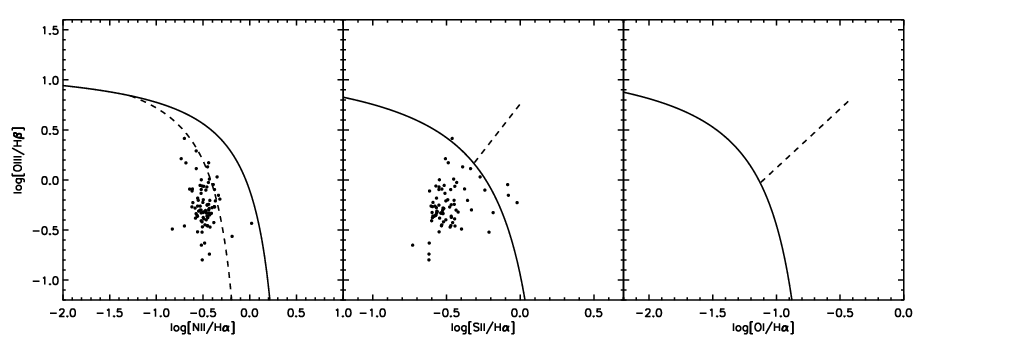}}
\caption{Same as figure A1, but for IRAS F16443-2915 N and a DSS R-band image.}
\end{sidewaysfigure*}

\clearpage

\begin{sidewaysfigure*}%[htb]
\vspace{0.00in}
\centering
{\Large \textbf{IRAS F16443-2915 S}}
{\includegraphics[width=0.80\textwidth]{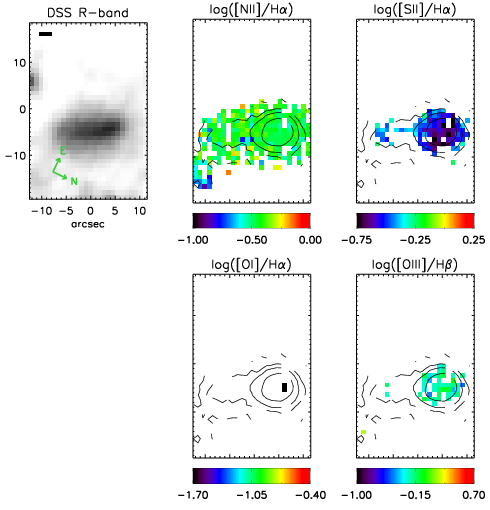}}
%\caption[]{}
\end{sidewaysfigure*}
\begin{sidewaysfigure*}%[htb]
\ContinuedFloat
\vspace{0.000in}
{\includegraphics[width=\textheight]{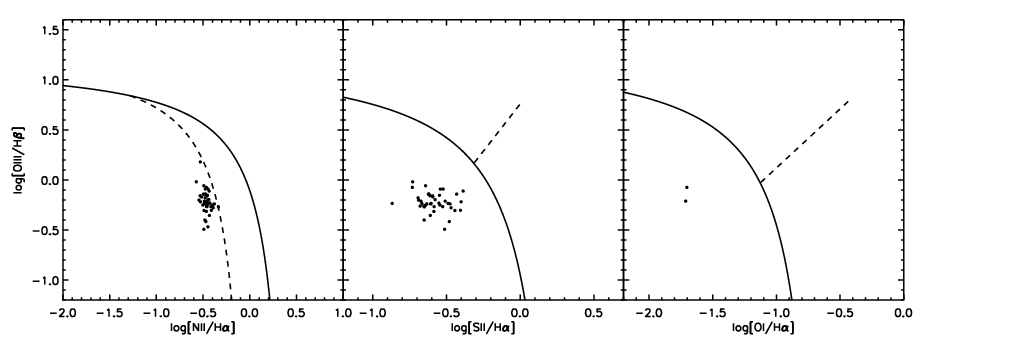}}
{\includegraphics[width=\textheight]{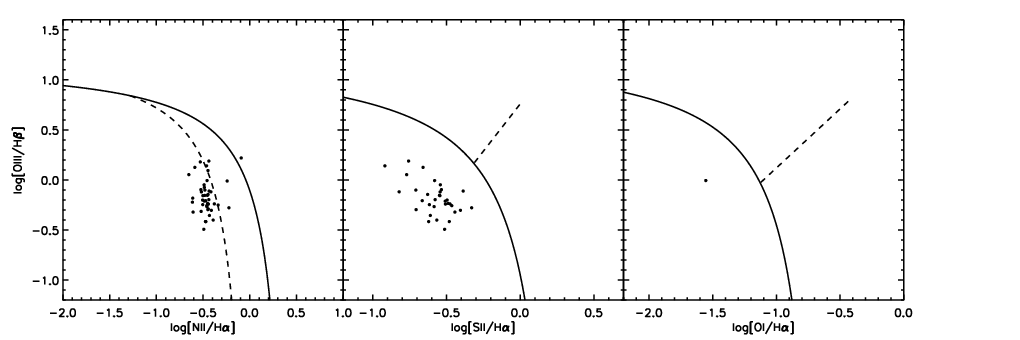}}
\caption{Same as figure A1, but for IRAS F16443-2915 S and a DSS R-band image.}
\end{sidewaysfigure*}

\begin{sidewaysfigure*}%[htb]
\vspace{0.00in}
\centering
{\Large \textbf{IRAS F17138-1017}}
{\includegraphics[width=0.80\textwidth]{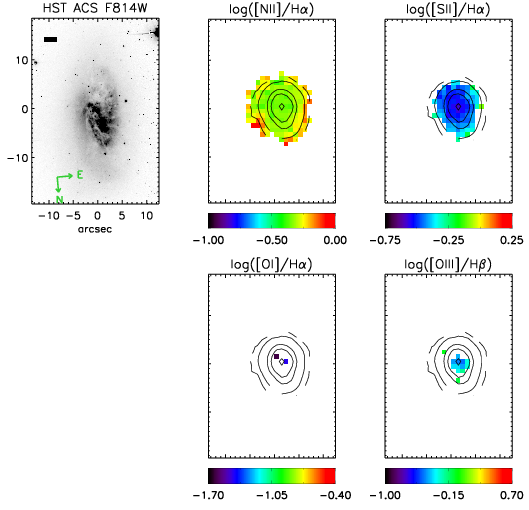}}
%\caption[]{}
\end{sidewaysfigure*}
\begin{sidewaysfigure*}%[htb]
\ContinuedFloat
\vspace{0.000in}
{\includegraphics[width=\textheight]{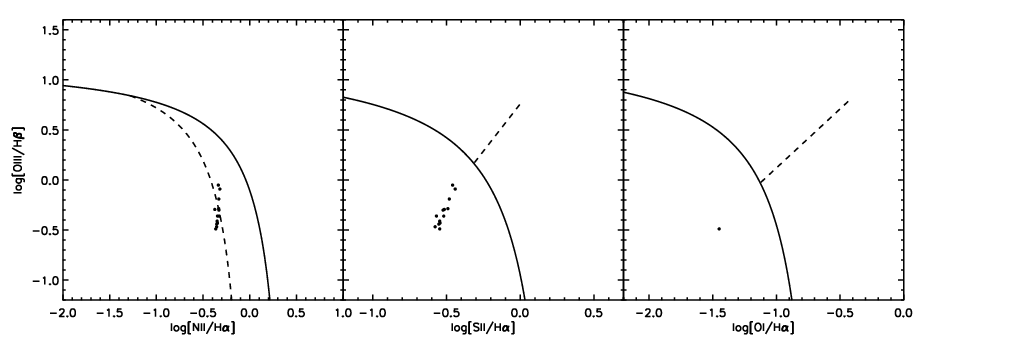}}
{\includegraphics[width=\textheight]{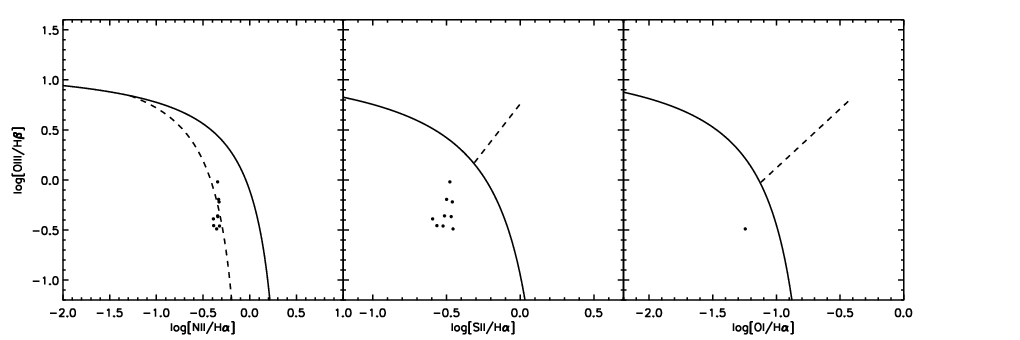}}
\caption{Same as figure A1, but for IRAS F17138-1017.}
\end{sidewaysfigure*}

\begin{sidewaysfigure*}%[htb]
\vspace{0.00in}
\centering
{\Large \textbf{IRAS F17207-0014}}
{\includegraphics[width=0.80\textwidth]{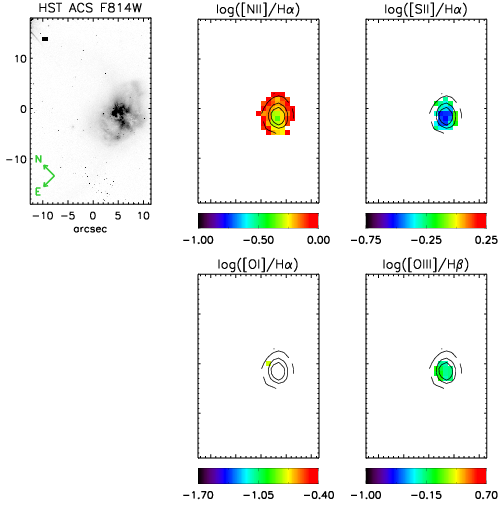}}
%\caption[]{}
\end{sidewaysfigure*}
\begin{sidewaysfigure*}%[htb]
\ContinuedFloat
\vspace{0.000in}
{\includegraphics[width=\textheight]{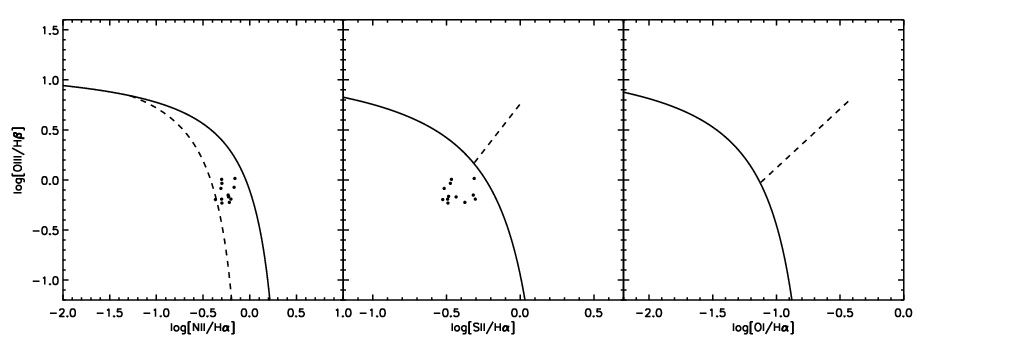}}
{\includegraphics[width=\textheight]{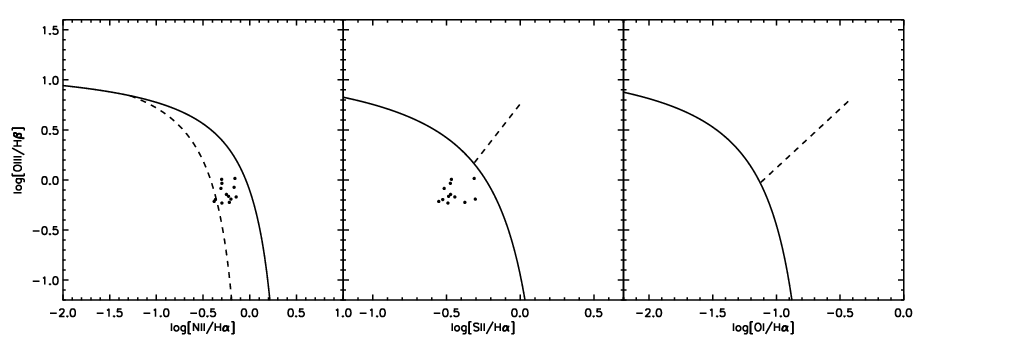}}
\caption{Same as figure A1, but for IRAS F17207-0014.}
\end{sidewaysfigure*}

\clearpage

\begin{sidewaysfigure*}%[htb]
\vspace{0.00in}
\centering
{\Large \textbf{IRAS F17222-5953}}
{\includegraphics[width=0.80\textwidth]{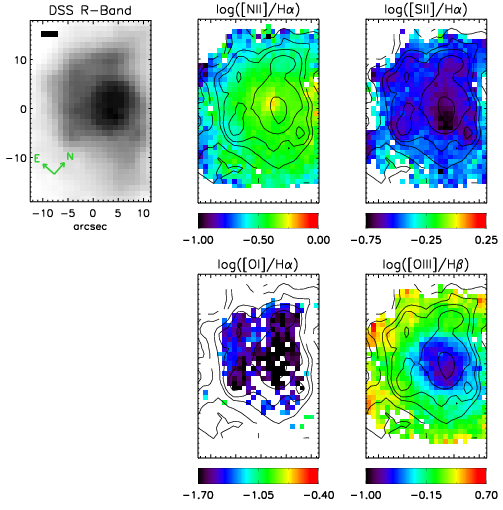}}
%\caption[]{}
\end{sidewaysfigure*}
\begin{sidewaysfigure*}%[htb]
\ContinuedFloat
\vspace{0.000in}
{\includegraphics[width=\textheight]{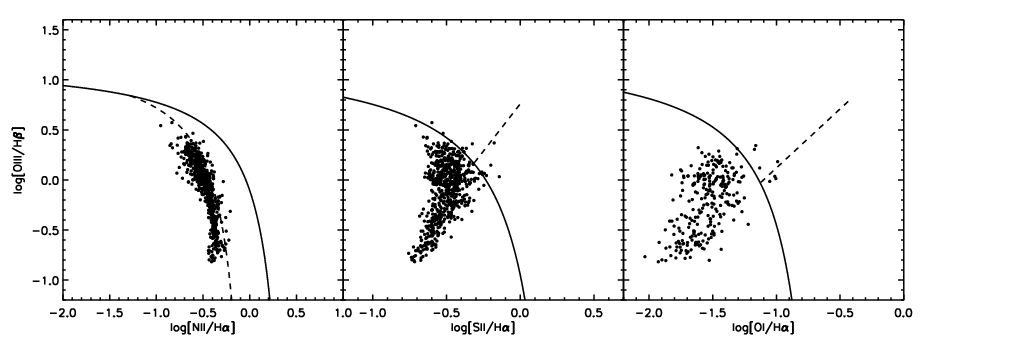}}
{\includegraphics[width=\textheight]{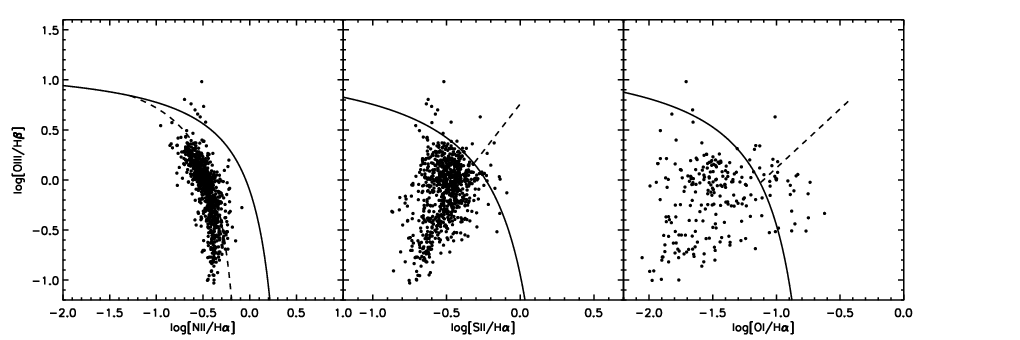}}
\caption{Same as figure A1, but for IRAS F17222-5953 and a DSS R-band image.}
\end{sidewaysfigure*}

\begin{sidewaysfigure*}%[htb]
\vspace{0.00in}
\centering
{\Large \textbf{IRAS 17578-0400}}
{\includegraphics[width=0.80\textwidth]{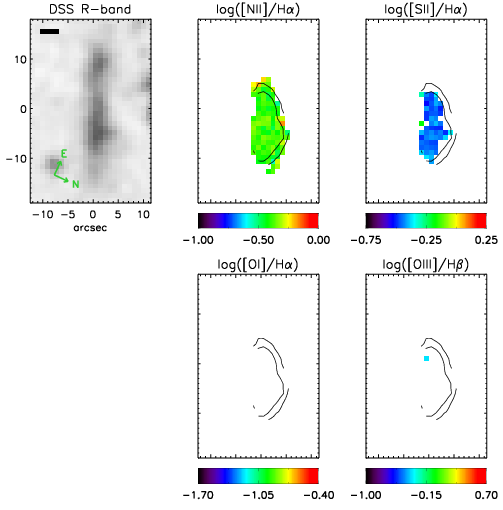}}
%\caption[]{}
\end{sidewaysfigure*}
\begin{sidewaysfigure*}%[htb]
\ContinuedFloat
\vspace{0.000in}
{\includegraphics[width=\textheight]{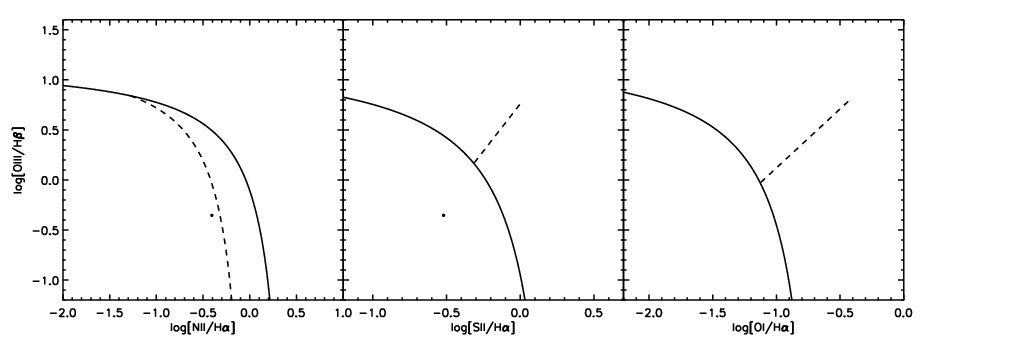}}
{\includegraphics[width=\textheight]{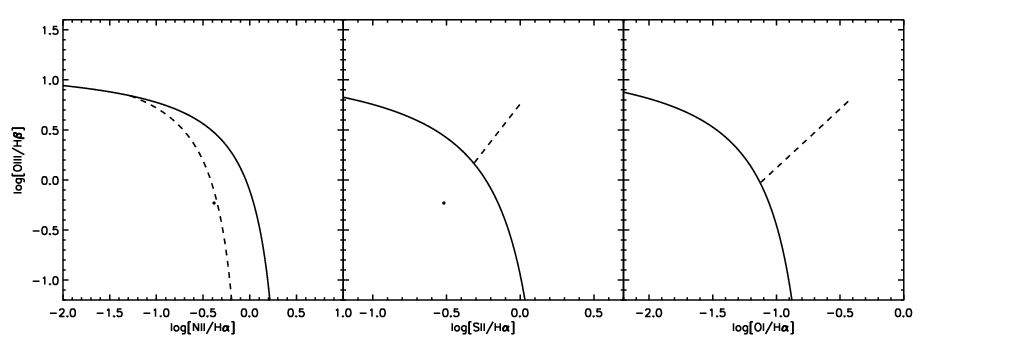}}
\caption{Same as figure A1, but for IRAS 17578-0400 and a DSS R-band image.}
\end{sidewaysfigure*}

\begin{sidewaysfigure*}%[htb]
\vspace{0.00in}
\centering
{\Large \textbf{IRAS F18093-5744 N}}
{\includegraphics[width=0.80\textwidth]{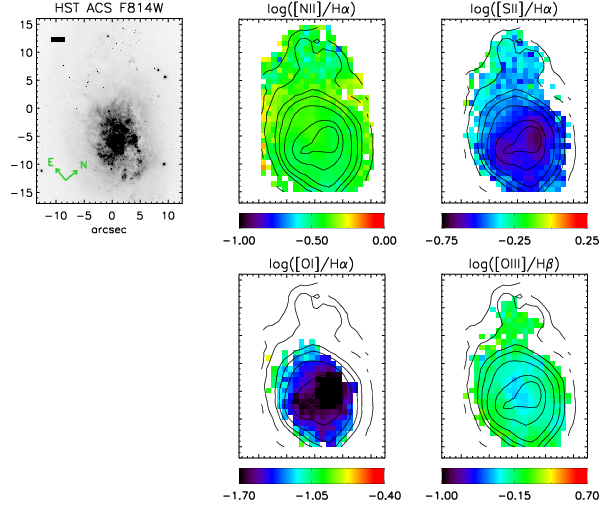}}
%\caption[]{}
\end{sidewaysfigure*}
\begin{sidewaysfigure*}%[htb]
\ContinuedFloat
\vspace{0.000in}
{\includegraphics[width=\textheight]{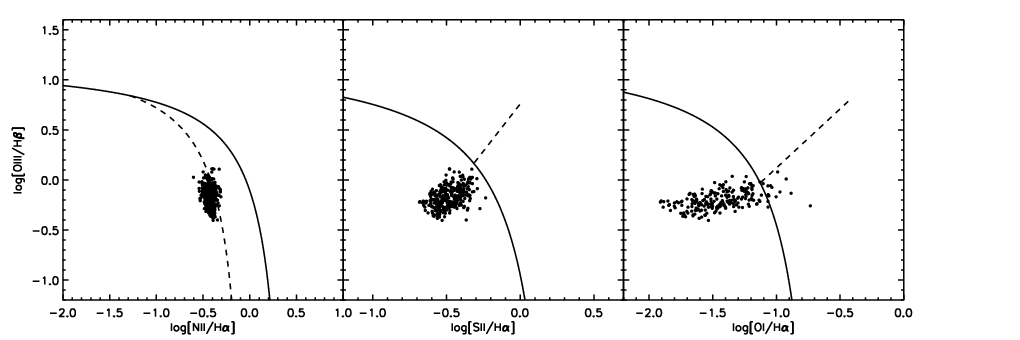}}
{\includegraphics[width=\textheight]{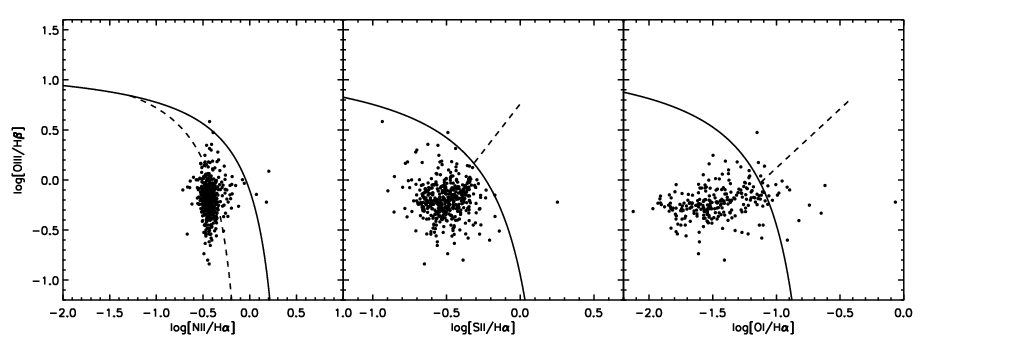}}
\caption{Same as figure A1, but for IRAS F18093-5744 N.}
\end{sidewaysfigure*}

\begin{sidewaysfigure*}%[htb]
\vspace{0.00in}
\centering
{\Large \textbf{IRAS F18093-5744 S}}
{\includegraphics[width=0.80\textwidth]{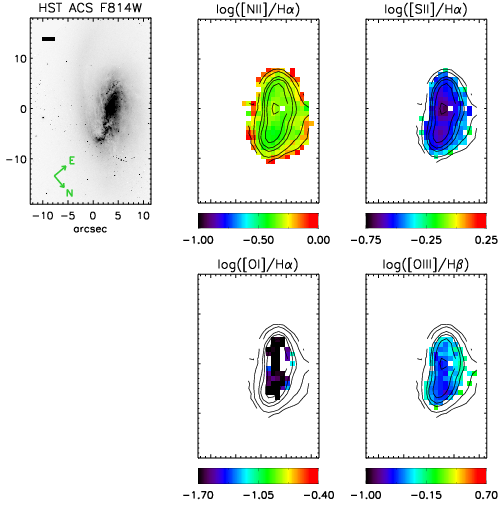}}
%\caption[]{}
\end{sidewaysfigure*}
\begin{sidewaysfigure*}%[htb]
\ContinuedFloat
\vspace{0.000in}
{\includegraphics[width=\textheight]{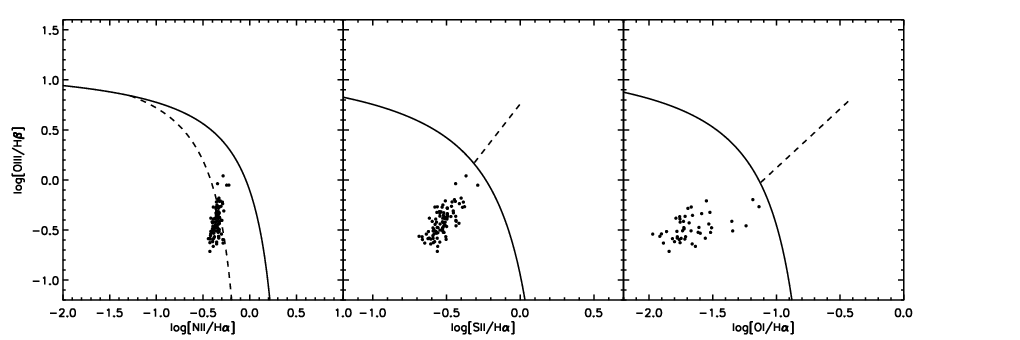}}
{\includegraphics[width=\textheight]{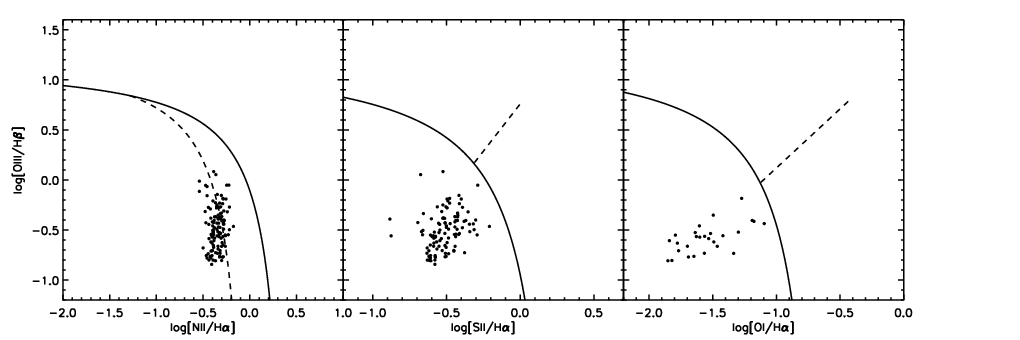}}
\caption{Same as figure A1, but for IRAS F18093-5744 S.}
\end{sidewaysfigure*}

\begin{sidewaysfigure*}%[htb]
\vspace{0.00in}
\centering
{\Large \textbf{IRAS F18093-5744 C}}
{\includegraphics[width=\textwidth]{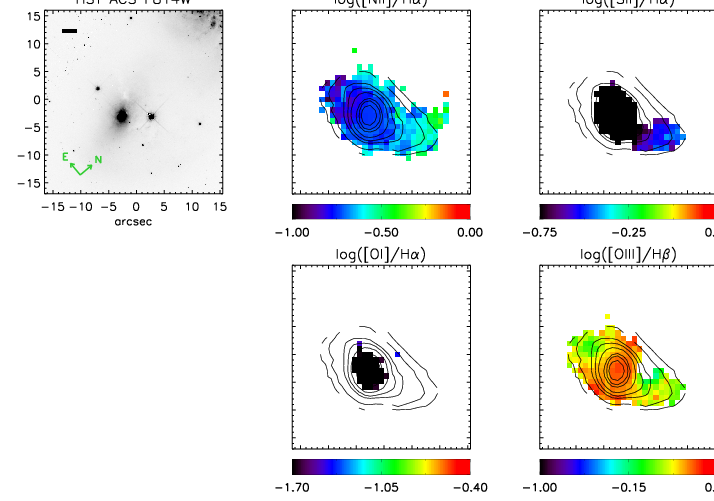}}
%\caption[]{}
\end{sidewaysfigure*}
\begin{sidewaysfigure*}%[htb]
\ContinuedFloat
\vspace{0.000in}
{\includegraphics[width=\textheight]{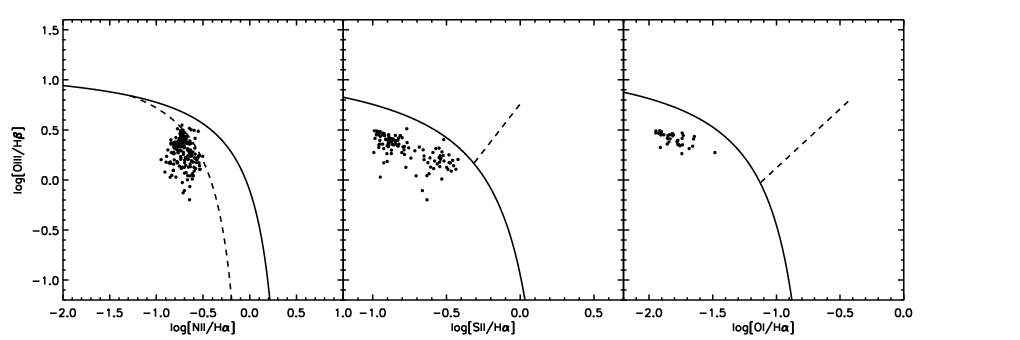}}
{\includegraphics[width=\textheight]{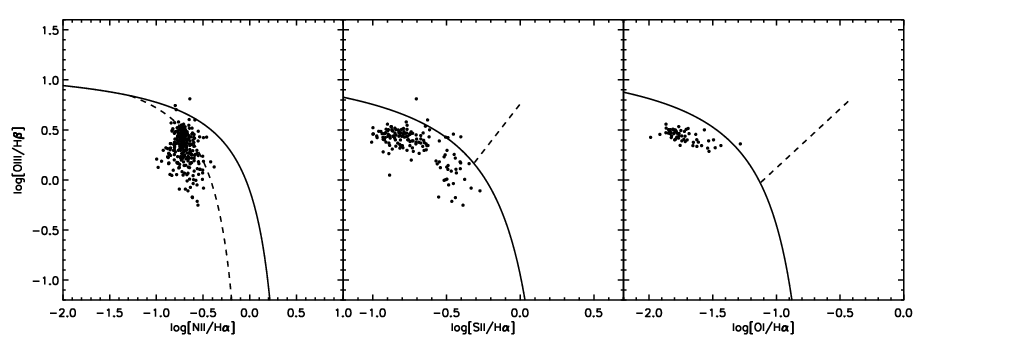}}
\caption{Same as figure A1, but for IRAS F18093-5744 C.}
\end{sidewaysfigure*}

\begin{sidewaysfigure*}%[htb]
\vspace{0.00in}
\centering
{\Large \textbf{IRAS F18293-3413}}
{\includegraphics[width=0.80\textwidth]{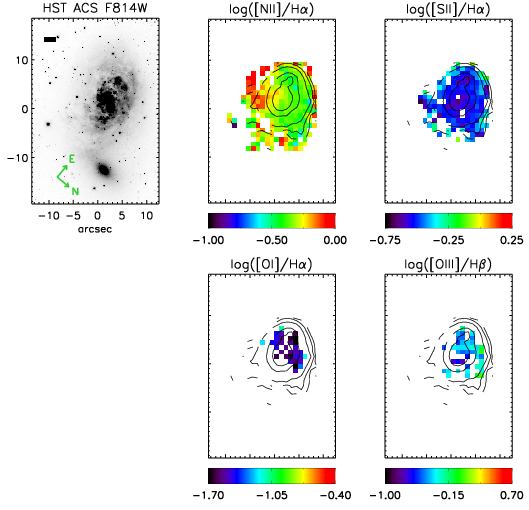}}
%\caption[]{}
\end{sidewaysfigure*}
\begin{sidewaysfigure*}%[htb]
\ContinuedFloat
\vspace{0.000in}
{\includegraphics[width=\textheight]{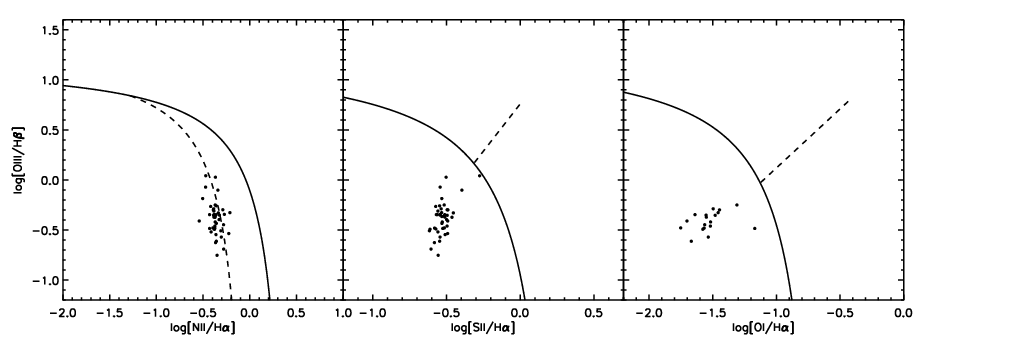}}
{\includegraphics[width=\textheight]{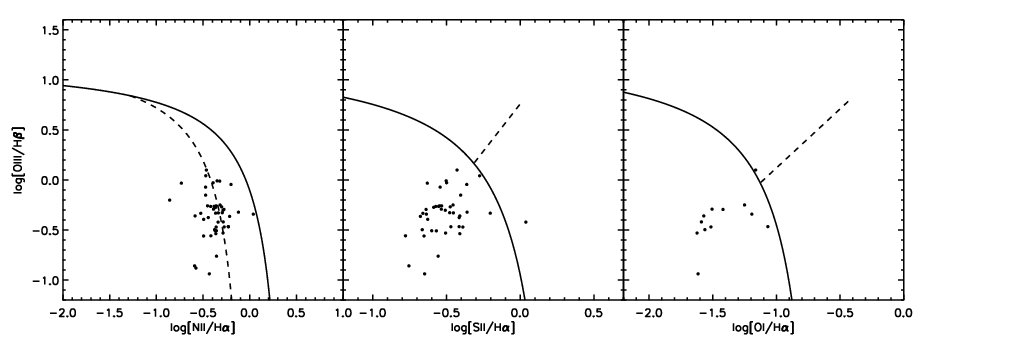}}
\caption{Same as figure A1, but for IRAS F18293-3413.}
\end{sidewaysfigure*}

\begin{sidewaysfigure*}%[htb]
\vspace{0.00in}
\centering
{\Large \textbf{IRAS F18341-5732}}
{\includegraphics[width=\textwidth]{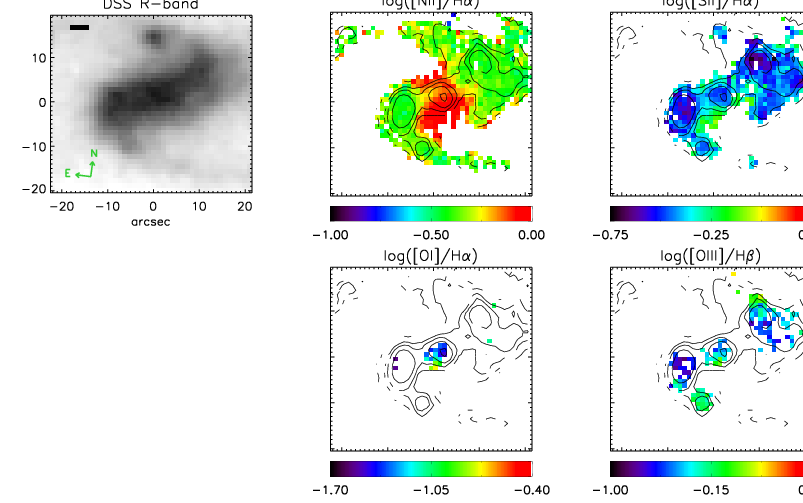}}
%\caption[]{}
\end{sidewaysfigure*}
\begin{sidewaysfigure*}%[htb]
\ContinuedFloat
\vspace{0.000in}
{\includegraphics[width=\textheight]{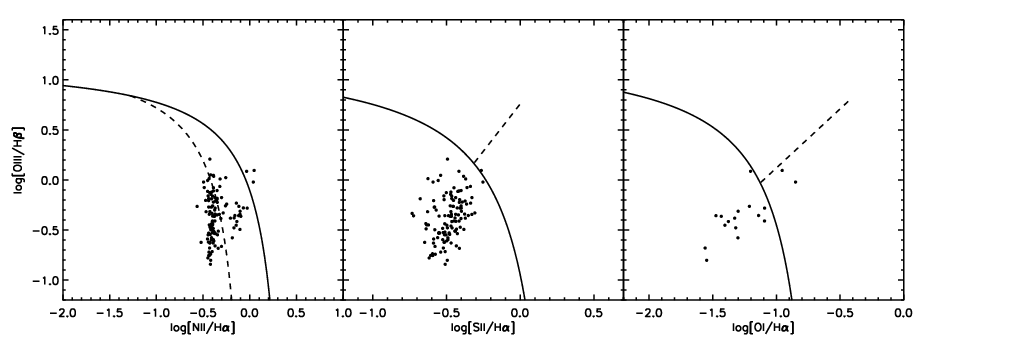}}
{\includegraphics[width=\textheight]{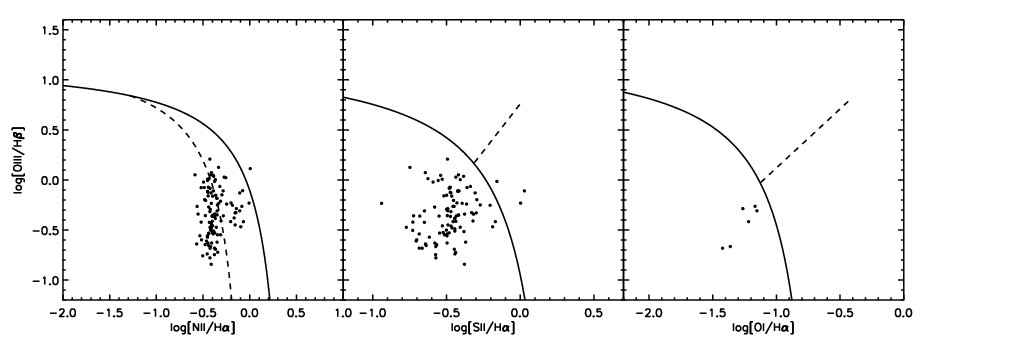}}
\caption{Same as figure A1, but for IRAS F18341-5732 and a DSS R-band image.}
\end{sidewaysfigure*}

\begin{sidewaysfigure*}%[htb]
\vspace{0.00in}
\centering
{\Large \textbf{IRAS F19115-2124}}
{\includegraphics[width=0.80\textwidth]{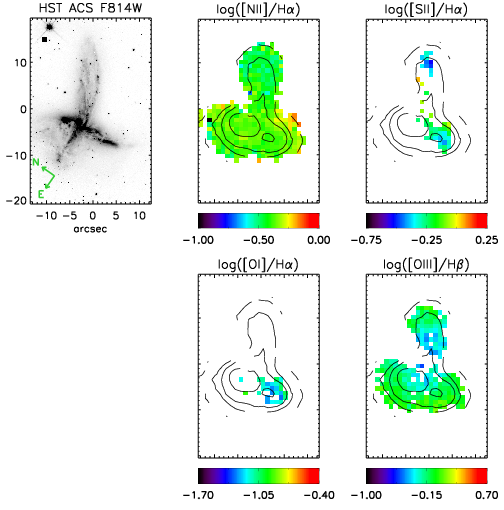}}
%\caption[]{}
\end{sidewaysfigure*}
\begin{sidewaysfigure*}%[htb]
\ContinuedFloat
\vspace{0.000in}
{\includegraphics[width=\textheight]{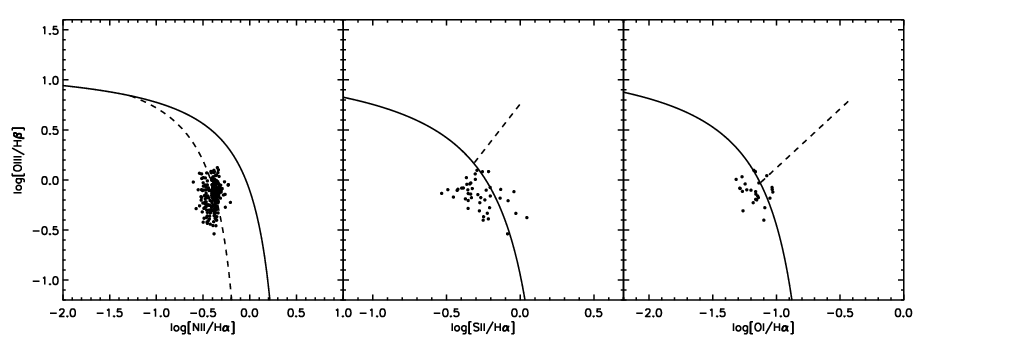}}
{\includegraphics[width=\textheight]{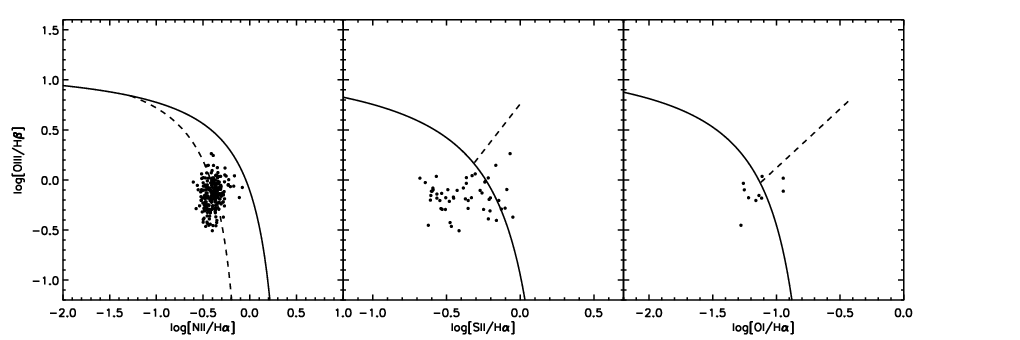}}
\caption{Same as figure A1, but for IRAS F19115-2124.}
\end{sidewaysfigure*}

\begin{sidewaysfigure*}%[htb]
\vspace{0.00in}
\centering
{\Large \textbf{IRAS F20551-4250}}
{\includegraphics[width=0.80\textwidth]{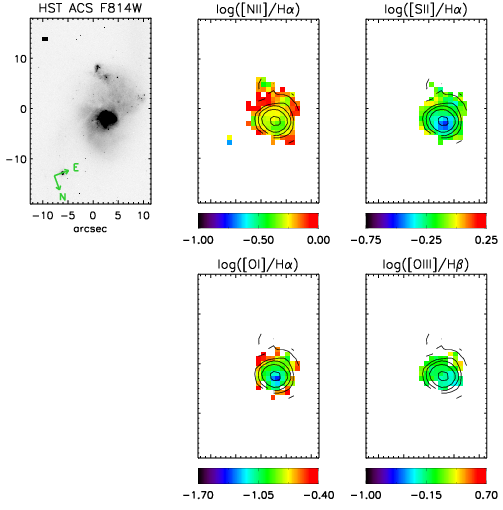}}
%\caption[]{}
\end{sidewaysfigure*}
\begin{sidewaysfigure*}%[htb]
\ContinuedFloat
\vspace{0.000in}
{\includegraphics[width=\textheight]{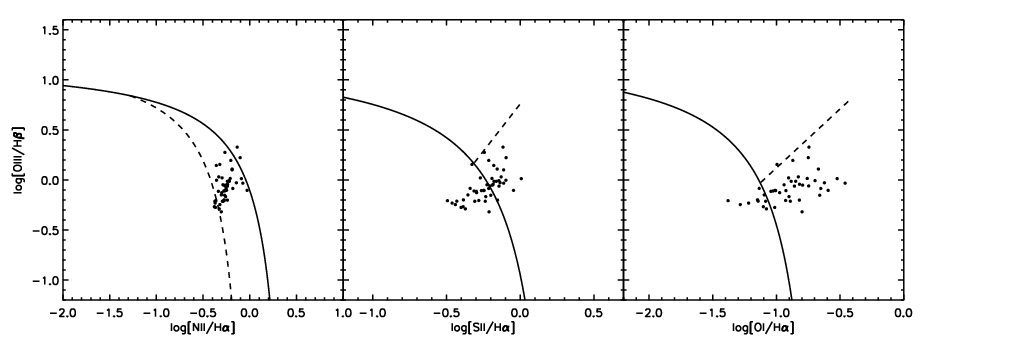}}
{\includegraphics[width=\textheight]{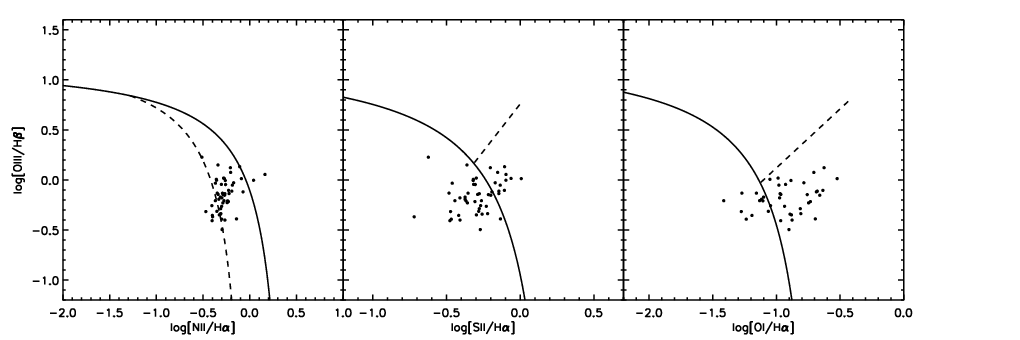}}
\caption{Same as figure A1, but for IRAS F20551-4250.}
\end{sidewaysfigure*}

\begin{sidewaysfigure*}%[htb]
\vspace{0.00in}
\centering
{\Large \textbf{IRAS F21330-3846}}
{\includegraphics[width=0.80\textwidth]{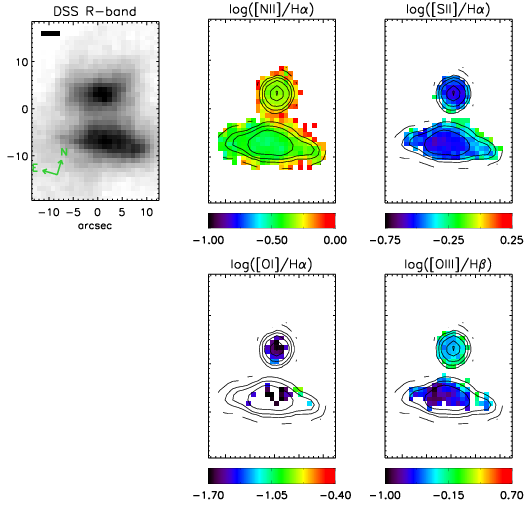}}
%\caption[]{}
\end{sidewaysfigure*}
\begin{sidewaysfigure*}%[htb]
\ContinuedFloat
\vspace{0.000in}
{\includegraphics[width=\textheight]{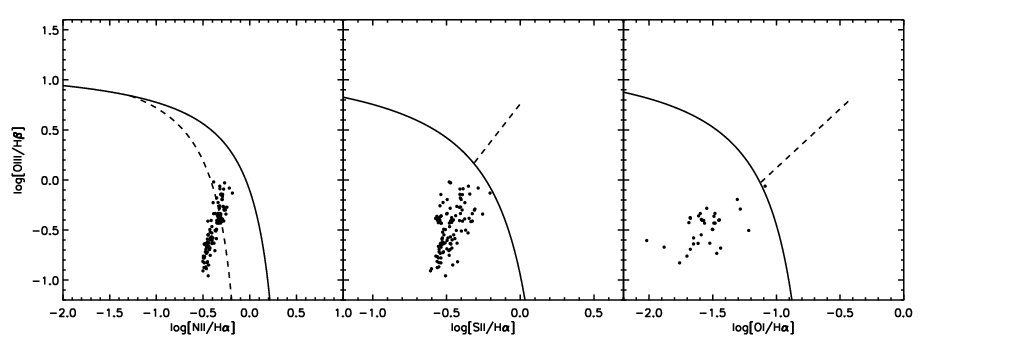}}
{\includegraphics[width=\textheight]{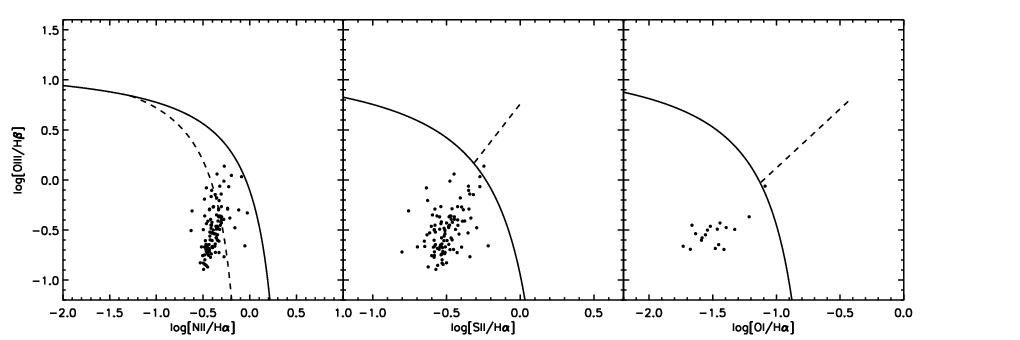}}
\caption{Same as figure A1, but for IRAS F21330-3846 and a DSS R-band image.}
\end{sidewaysfigure*}

\begin{sidewaysfigure*}%[htb]
\vspace{0.00in}
\centering
{\Large \textbf{IRAS F21453-3511}}
{\includegraphics[width=0.80\textwidth]{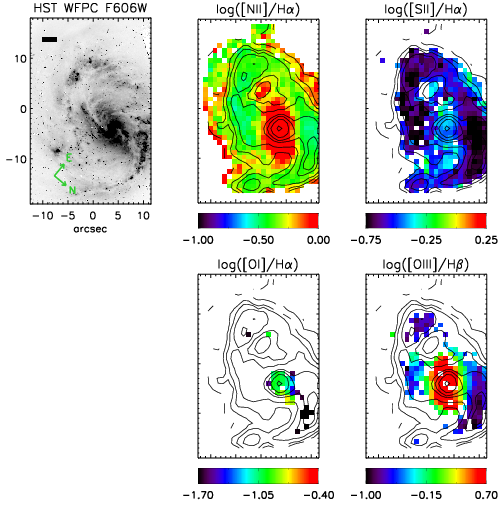}}
%\caption[]{}
\end{sidewaysfigure*}
\begin{sidewaysfigure*}%[htb]
\ContinuedFloat
\vspace{0.000in}
{\includegraphics[width=\textheight]{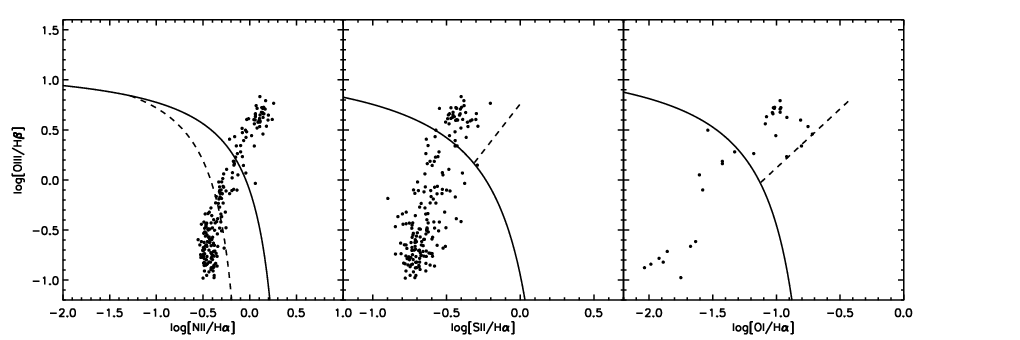}}
{\includegraphics[width=\textheight]{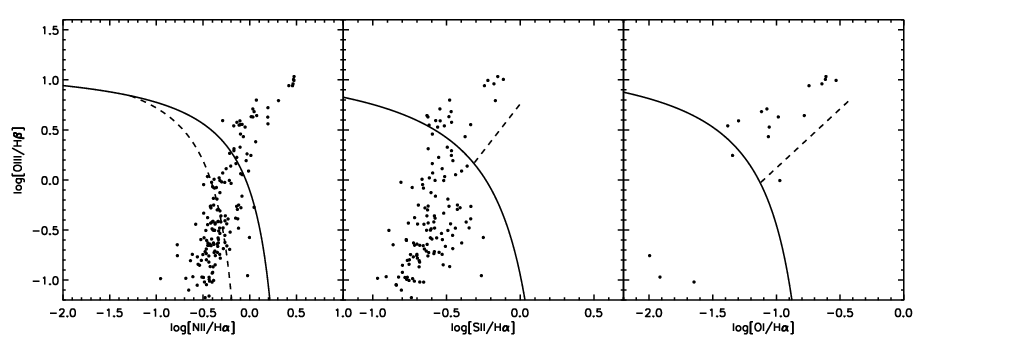}}
\caption{Same as figure A1, but for IRAS F21453-3511 and an \emph{HST} G-band image.}
\end{sidewaysfigure*}

\clearpage

\begin{sidewaysfigure*}%[htb]
\vspace{0.00in}
\centering
{\Large \textbf{IRAS F22467-4906}}
{\includegraphics[width=\textwidth]{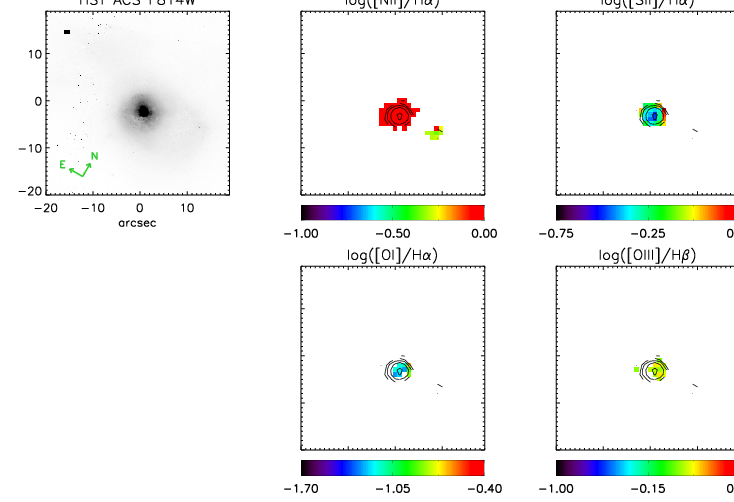}}
%\caption[]{}
\end{sidewaysfigure*}
\begin{sidewaysfigure*}%[htb]
\ContinuedFloat
\vspace{0.000in}
{\includegraphics[width=\textheight]{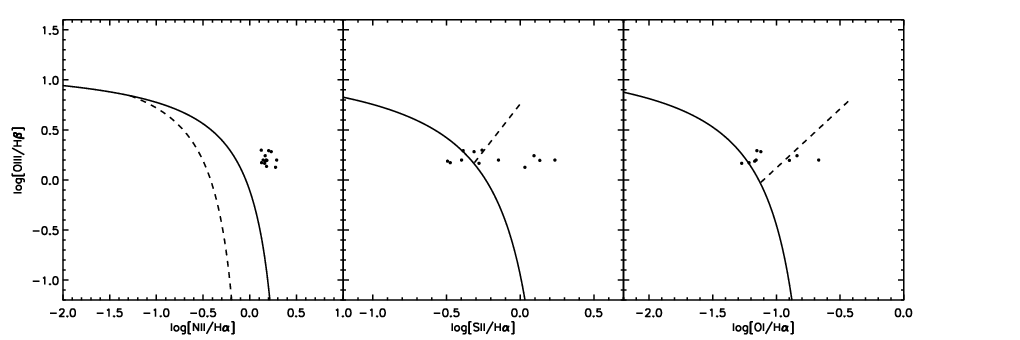}}
{\includegraphics[width=\textheight]{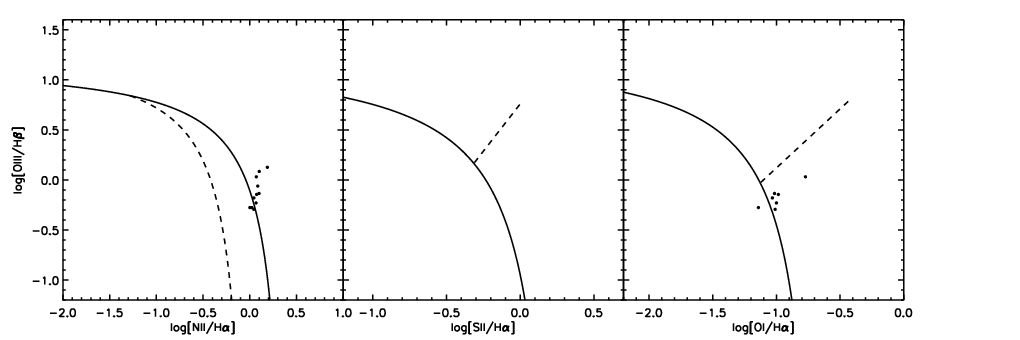}}
\caption{Same as figure A1, but for IRAS F22467-4906.}
\end{sidewaysfigure*}

\begin{sidewaysfigure*}%[htb]
\vspace{0.00in}
\centering
{\Large \textbf{IRAS F23128-5919}}
{\includegraphics[width=1.0\textwidth]{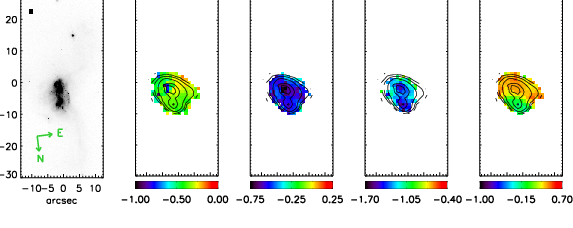}}
%\caption[]{}
\end{sidewaysfigure*}
\begin{sidewaysfigure*}%[htb]
\ContinuedFloat
\vspace{0.000in}
{\includegraphics[width=\textheight]{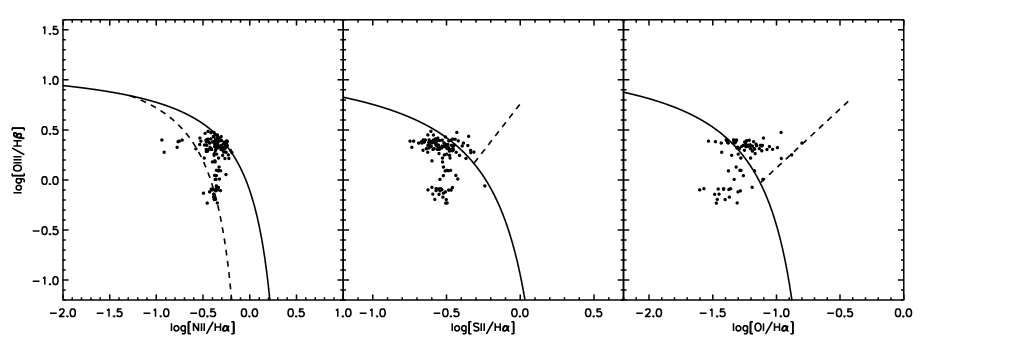}}
{\includegraphics[width=\textheight]{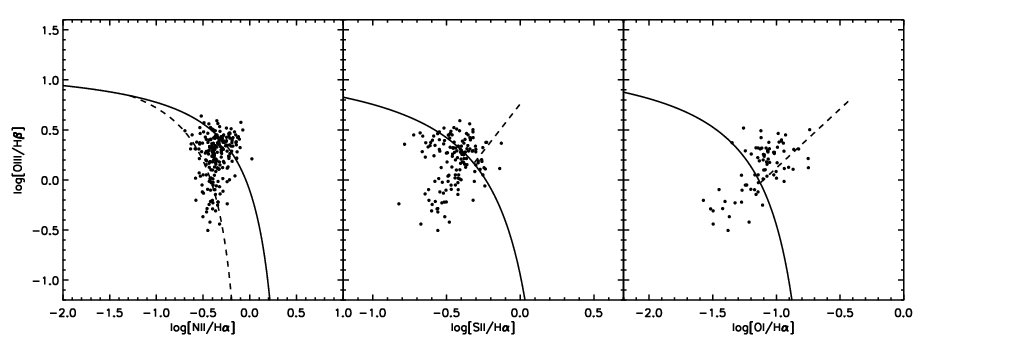}}
\caption{Same as figure A1, but for IRAS F23128-5919.}
\end{sidewaysfigure*}

\clearpage

%\onecolumn

\section{Velocity Dispersion Distributions}
This section contains the velocity dispersion distributions and comparison to line ratios for the entire WiFeS GOALS sample, generated as described in section 5.

\begin{figure*}[htpb!]
\centering
{\Large \textbf{IRAS F01053-1746}}
{\includegraphics[width=\textwidth]{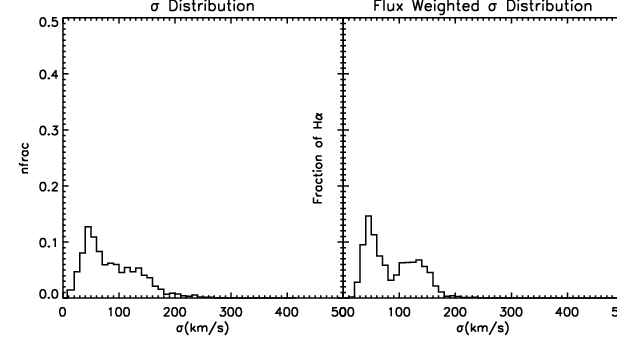}}
{\includegraphics[width=\textwidth]{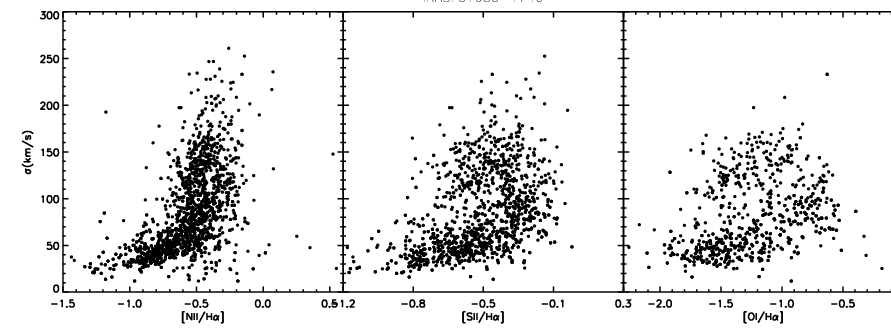}}
%{\includegraphics[width=\textwidth]{01053figBb_ind.png}}
\caption{Velocity dispersion distribution and comparison with line ratios. The top left panel shows the histogram for the fraction of the total number of components for a given velocity dispersion The top right panel shows a similar histogram, but with the fraction of the total \Ha~flux for a given velocity dispersion. Systems with a stronger contribution from shocks show a larger fraction above $\sigma>90km~s^{-1}$.
The bottom three panels show various diagnostic emission line ratios vs. velocity dispersion. Lower emission line ratios are, on average, dominated by lower-$\sigma$ star forming regions.}
\end{figure*}

\begin{figure*}%[htpb!]
\centering
{\Large \textbf{IRAS F02072-1025}}
{\includegraphics[width=\textwidth]{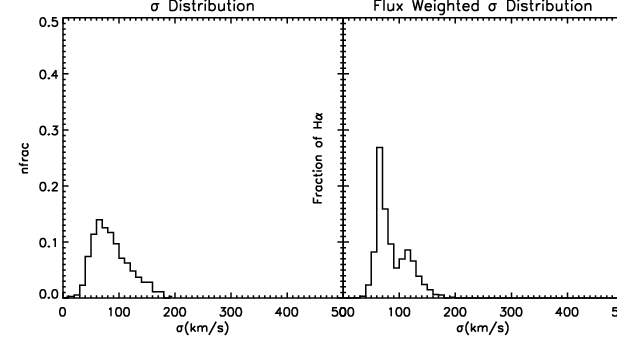}}
{\includegraphics[width=\textwidth]{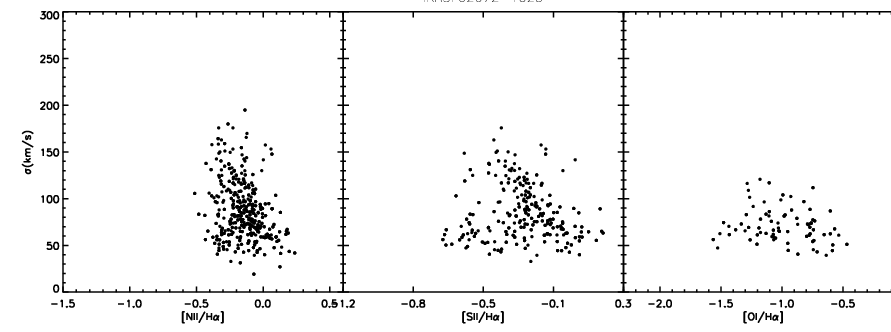}}
%{\includegraphics[width=\textwidth]{02072figBb_ind.png}}
\caption{Same as Figure B1, but for IRAS F02072-1025.}
\end{figure*}

\begin{figure*}%[htpb!]
\centering
{\Large \textbf{IRAS F06076-2139}}

{\includegraphics[width=\textwidth]{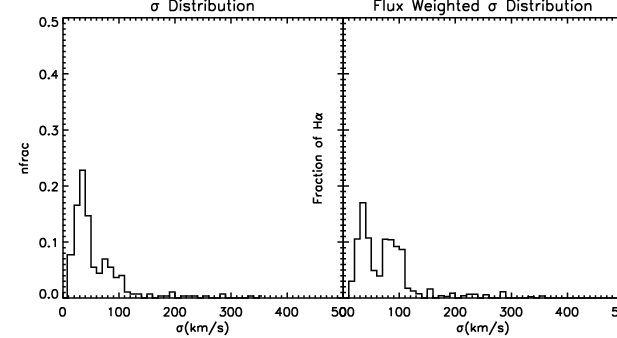}}
{\includegraphics[width=\textwidth]{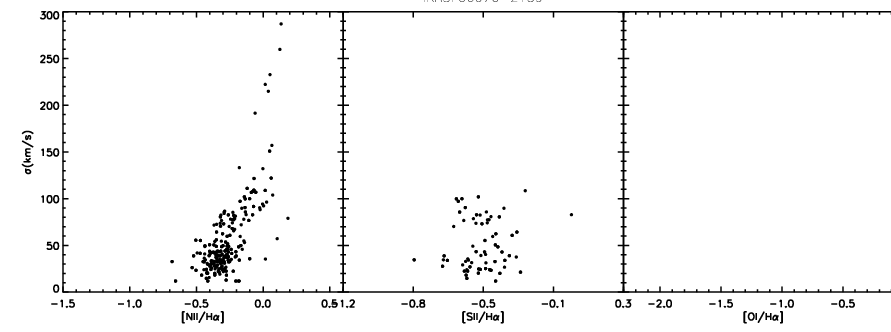}}
%{\includegraphics[width=\textwidth]{06076figBb_ind.png}}
\caption{Same as Figure B1, but for IRAS F06076-2139.}
\end{figure*}

\begin{figure*}%[htpb!]
\centering
{\Large \textbf{IRAS F08355-4944}}

{\includegraphics[width=\textwidth]{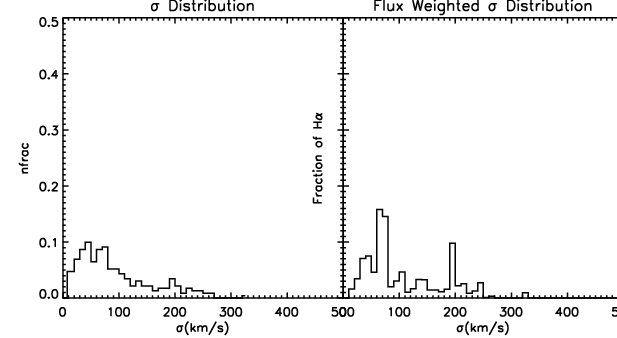}}
{\includegraphics[width=\textwidth]{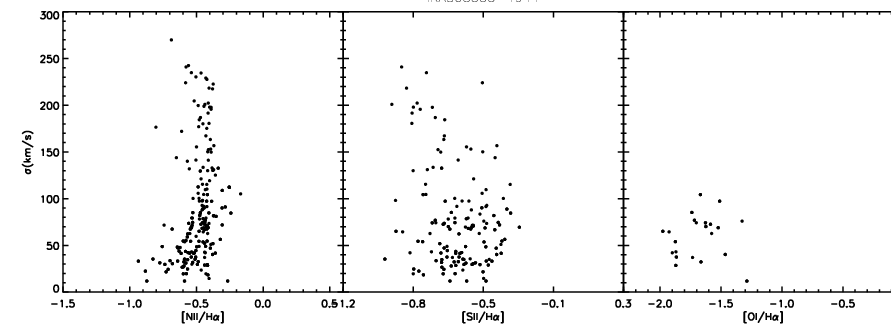}}
%{\includegraphics[width=\textwidth]{08355figBb_ind.png}}
\caption{Same as Figure B1, but for IRAS 08355-4944.}
\end{figure*}

\begin{figure*}%[htpb!]
\centering
{\Large \textbf{IRAS F10038-3338}}

{\includegraphics[width=\textwidth]{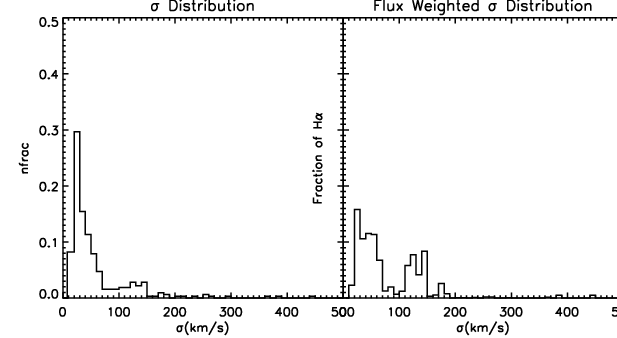}}
{\includegraphics[width=\textwidth]{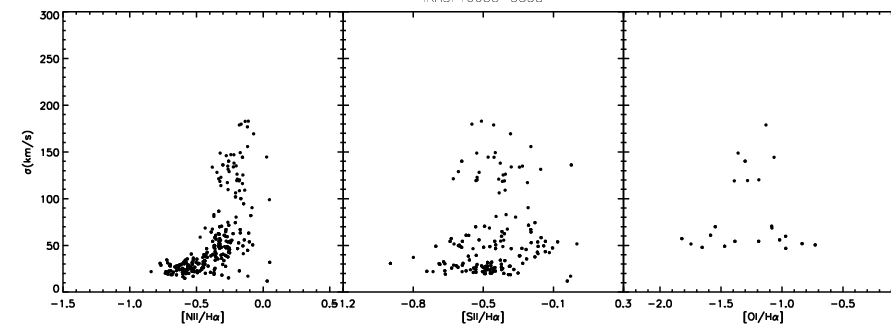}}
%{\includegraphics[width=\textwidth]{10038figBb_ind.png}}
\caption{Same as Figure B1, but for IRAS F10038-3338.}
\end{figure*}

\begin{figure*}%[htpb!]
\centering
{\Large \textbf{IRAS F10257-4339}}

{\includegraphics[width=\textwidth]{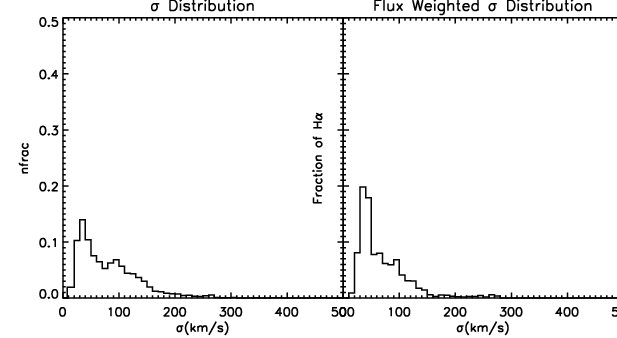}}
{\includegraphics[width=\textwidth]{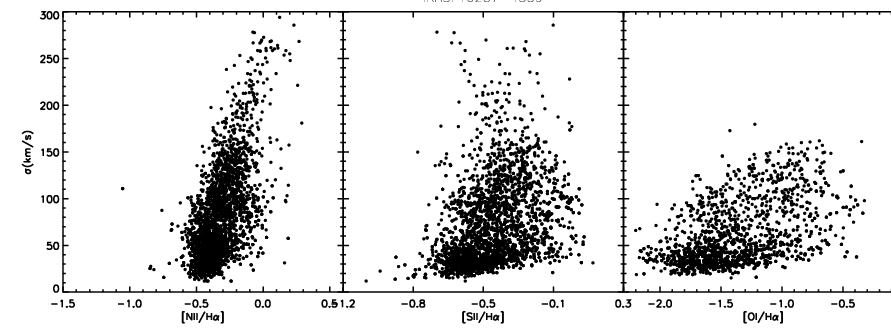}}
%{\includegraphics[width=\textwidth]{10257figBb_ind.png}}
\caption{Same as Figure B1, but for IRAS F10257-4339.}
\end{figure*}

\begin{figure*}%[htpb!]
\centering
{\Large \textbf{IRAS F12043-3140}}

{\includegraphics[width=\textwidth]{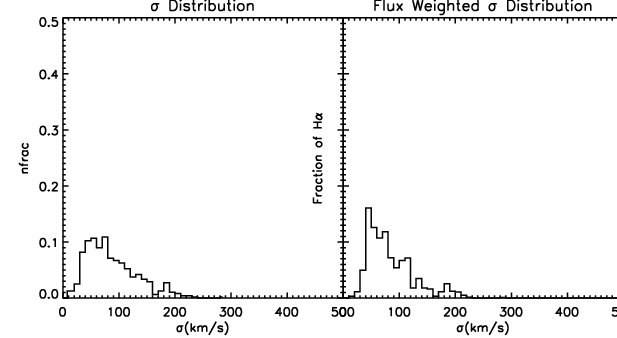}}
{\includegraphics[width=\textwidth]{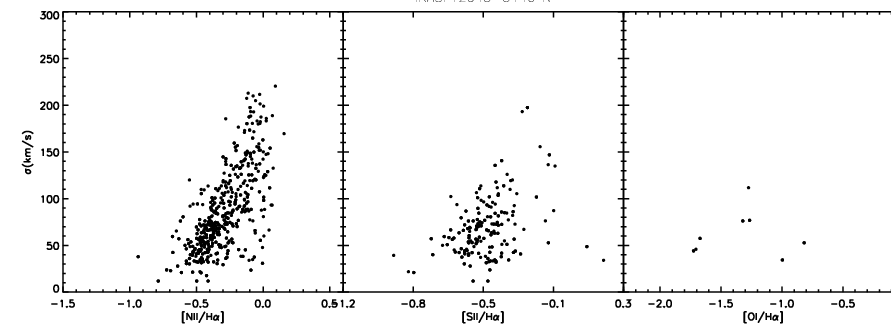}}
%{\includegraphics[width=\textwidth]{12043_nfigBb_ind.png}}
\caption{Same as Figure B1, but for IRAS F12043-3140.}
\end{figure*}

\begin{figure*}%[htpb!]
\centering
{\Large \textbf{IRAS F12592+0436}}

{\includegraphics[width=\textwidth]{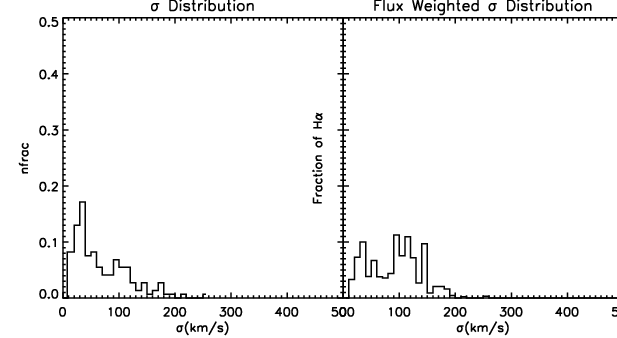}}
{\includegraphics[width=\textwidth]{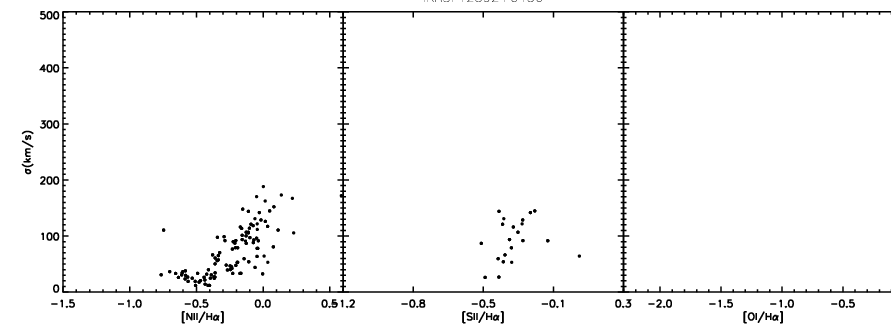}}
%{\includegraphics[width=\textwidth]{12592figBb_ind.png}}
\caption{Same as Figure B1, but for IRAS F12592+0436.}
\end{figure*}

\begin{figure*}%[htpb!]
\centering
{\Large \textbf{IRAS 13120-5453}}

{\includegraphics[width=\textwidth]{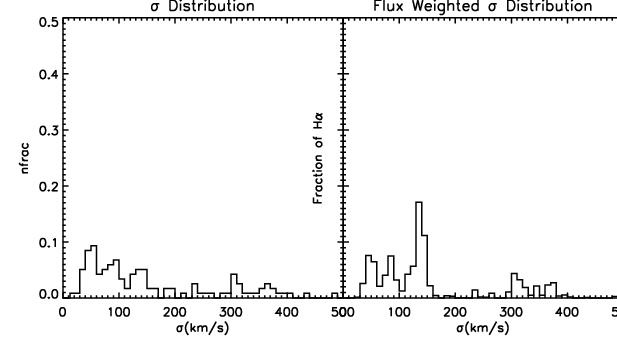}}
{\includegraphics[width=\textwidth]{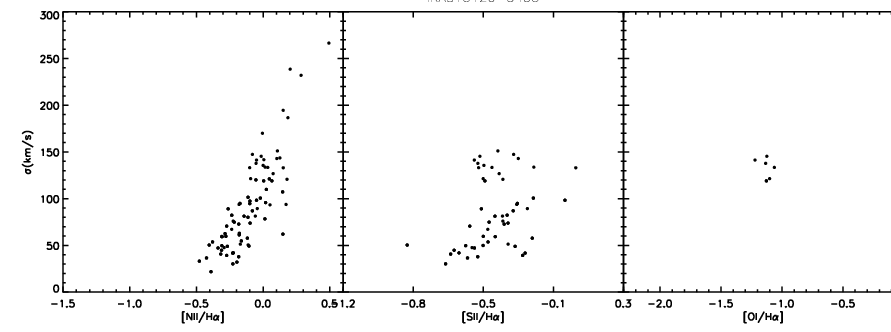}}
%{\includegraphics[width=\textwidth]{13120figBb_ind.png}}
\caption{Same as Figure B1, but for IRAS 13120-5453.}
\end{figure*}

\begin{figure*}%[htpb!]
\centering
{\Large \textbf{IRAS F13373+0105 W}}

{\includegraphics[width=\textwidth]{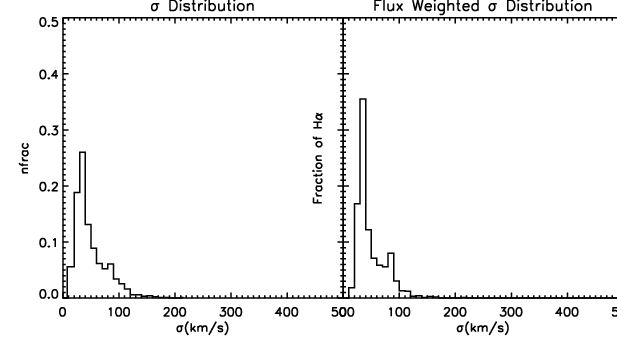}}
{\includegraphics[width=\textwidth]{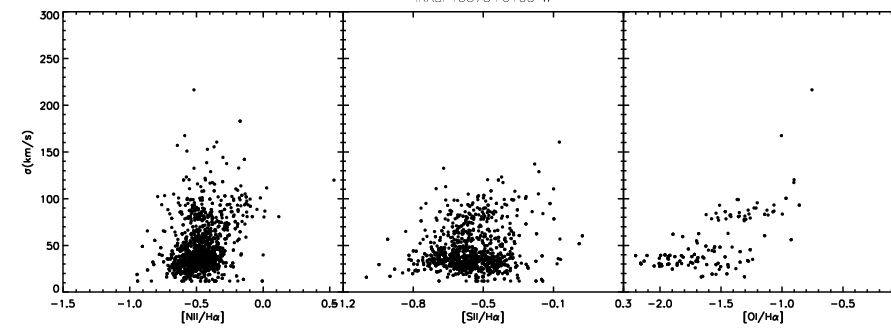}}
%{\includegraphics[width=\textwidth]{13373_wfigBb_ind.png}}
\caption{Same as Figure B1, but for IRAS F13373+0105 W.}
\end{figure*}

\begin{figure*}%[htpb!]
\centering
{\Large \textbf{IRAS F13373+0105 E}}

{\includegraphics[width=\textwidth]{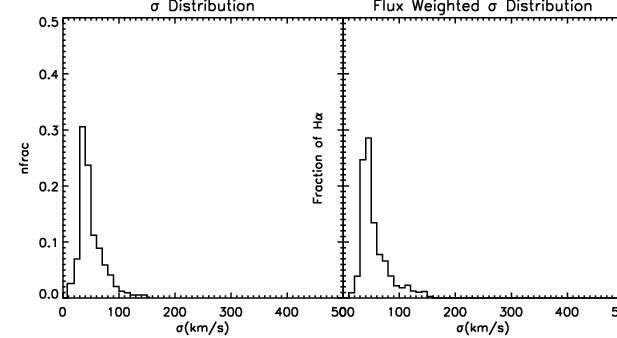}}
{\includegraphics[width=\textwidth]{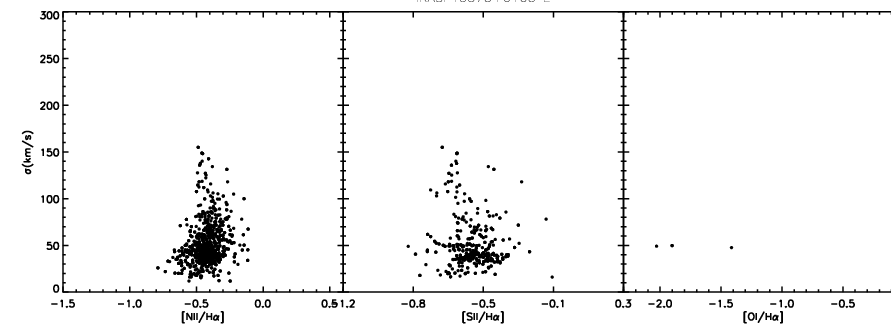}}
%{\includegraphics[width=\textwidth]{13373_efigBb_ind.png}}
\caption{Same as Figure B1, but for IRAS F13373+0105 E.}
\end{figure*}

\begin{figure*}%[htpb!]
\centering
{\Large \textbf{IRAS F15107+0724}}

{\includegraphics[width=\textwidth]{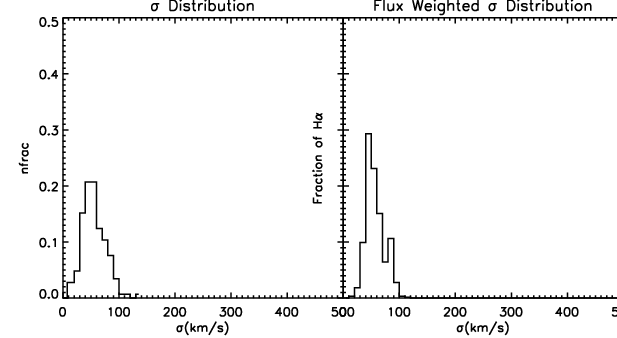}}
{\includegraphics[width=\textwidth]{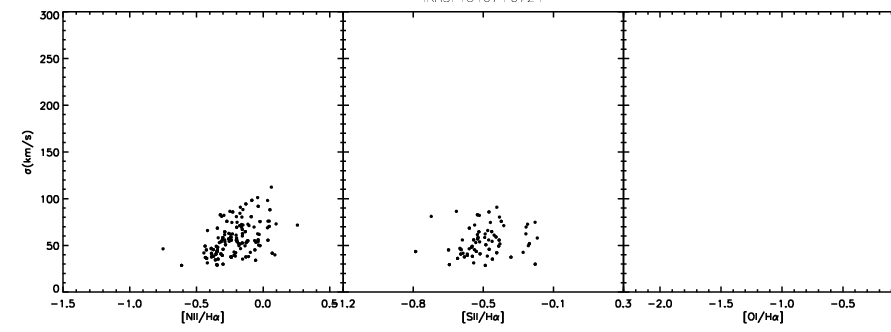}}
%{\includegraphics[width=\textwidth]{15107figBb_ind.png}}
\caption{Same as Figure B1, but for IRAS F15107+0724.}
\end{figure*}

\begin{figure*}%[htpb!]
\centering
{\Large \textbf{IRAS F16164-0746}}

{\includegraphics[width=\textwidth]{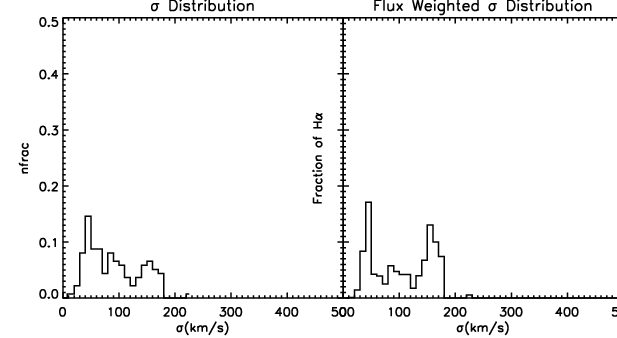}}
{\includegraphics[width=\textwidth]{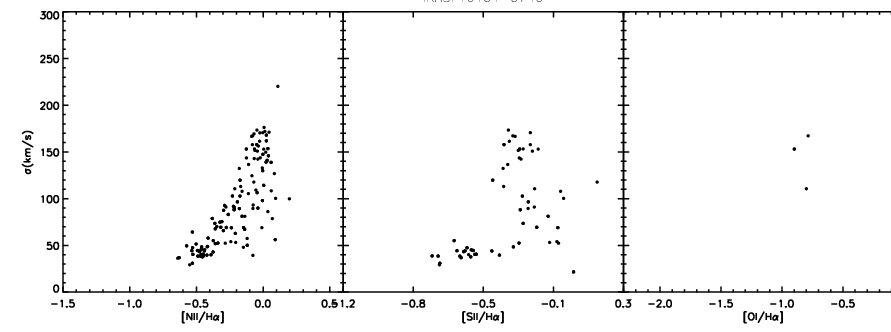}}
%{\includegraphics[width=\textwidth]{16164figBb_ind.png}}
\caption{Same as Figure B1, but for IRAS F16164-0746.}
\end{figure*}

\begin{figure*}%[htpb!]
\centering
{\Large \textbf{IRAS F16399-0937}}

{\includegraphics[width=\textwidth]{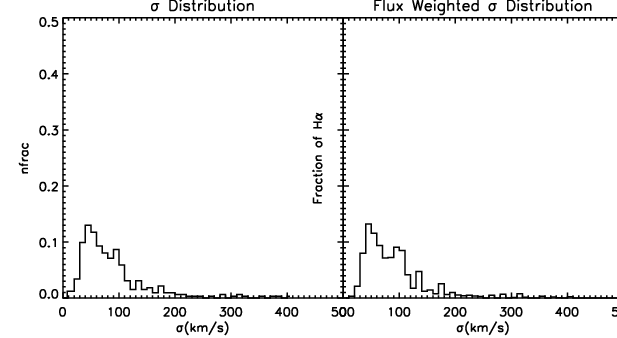}}
{\includegraphics[width=\textwidth]{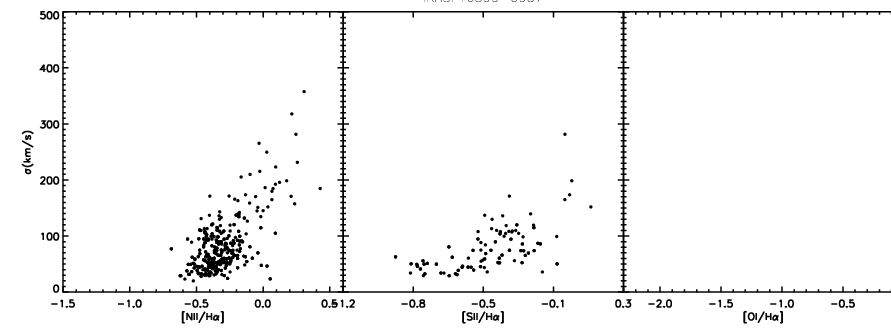}}
%{\includegraphics[width=\textwidth]{16399figBb_ind.png}}
\caption{Same as Figure B1, but for IRAS F16399-0937.}
\end{figure*}

\begin{figure*}%[htpb!]
\centering
{\Large \textbf{IRAS F16443-2915 N}}

{\includegraphics[width=\textwidth]{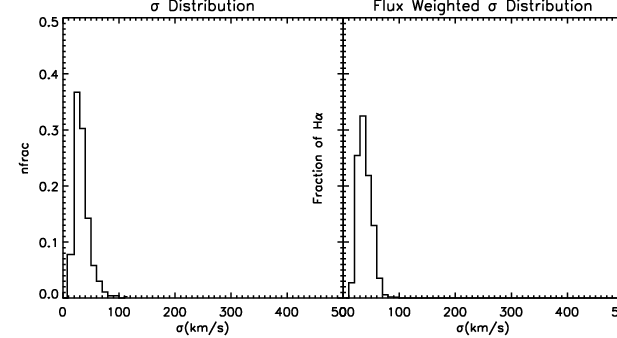}}
{\includegraphics[width=\textwidth]{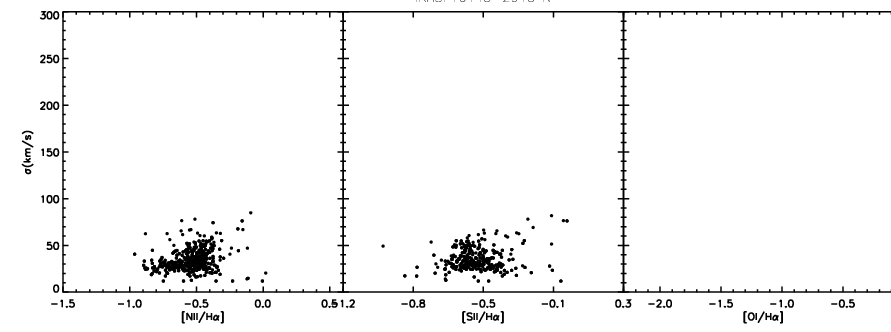}}
%{\includegraphics[width=\textwidth]{16443_nfigBb_ind.png}}
\caption{Same as Figure B1, but for IRAS F16443-2915 N.}
\end{figure*}

\begin{figure*}%[htpb!]
\centering
{\Large \textbf{IRAS F16443-2915 S}}

{\includegraphics[width=\textwidth]{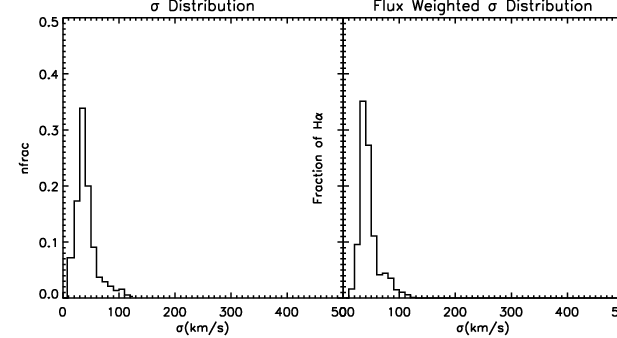}}
{\includegraphics[width=\textwidth]{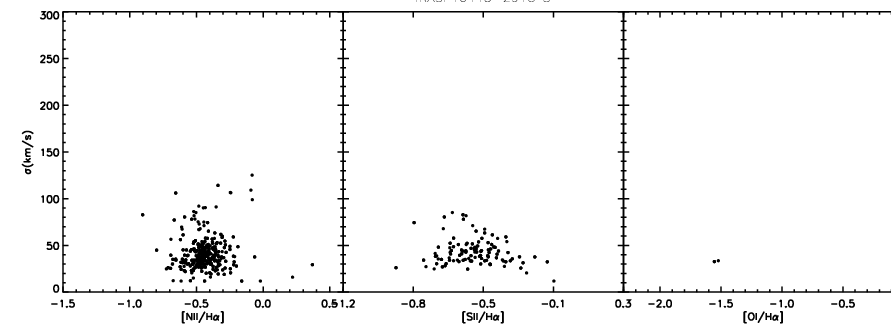}}
%{\includegraphics[width=\textwidth]{16443_sfigBb_ind.png}}
\caption{Same as Figure B1, but for IRAS F16443-2915 S.}
\end{figure*}

\begin{figure*}%[htpb!]
\centering
{\Large \textbf{IRAS F17138-1017}}

{\includegraphics[width=\textwidth]{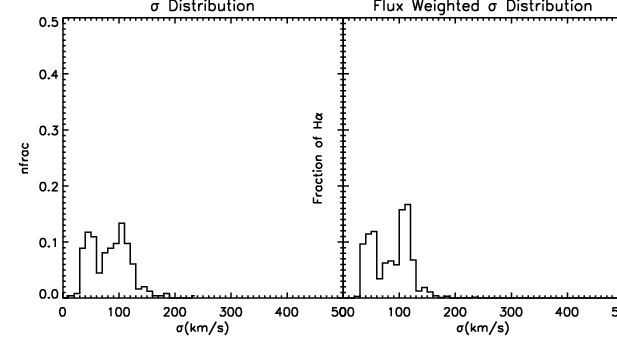}}
{\includegraphics[width=\textwidth]{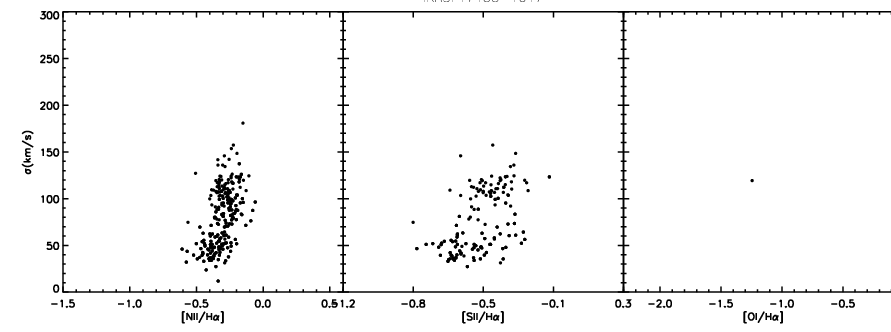}}
%{\includegraphics[width=\textwidth]{17138figBb_ind.png}}
\caption{Same as Figure B1, but for IRAS F17138-1017.}
\end{figure*}

\begin{figure*}%[htpb!]
\centering
{\Large \textbf{IRAS F17207-0014}}

{\includegraphics[width=\textwidth]{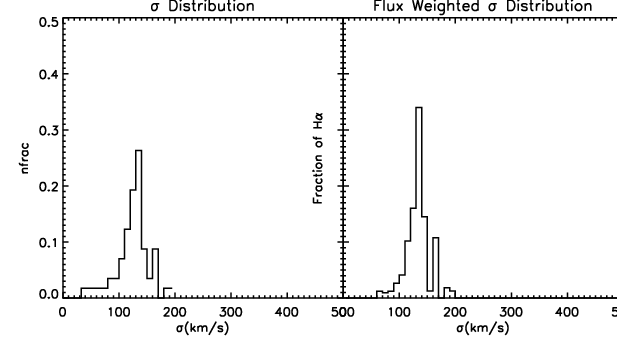}}
{\includegraphics[width=\textwidth]{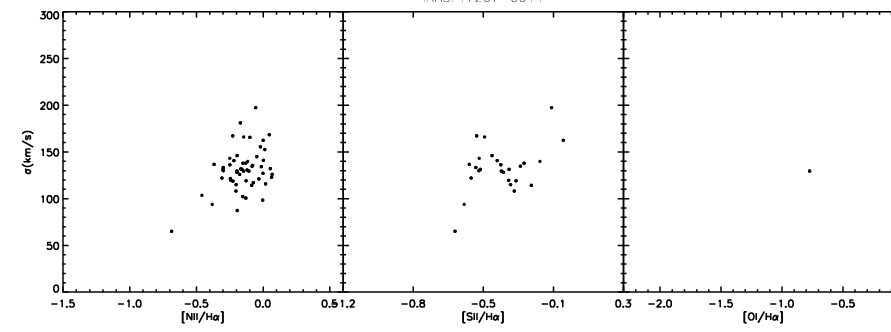}}
%{\includegraphics[width=\textwidth]{17207figBb_ind.png}}
\caption{Same as Figure B1, but for IRAS F17207-0014.}
\end{figure*}

\clearpage

\begin{figure*}%[htpb!]
\centering
{\Large \textbf{IRAS F17222-5953}}

{\includegraphics[width=\textwidth]{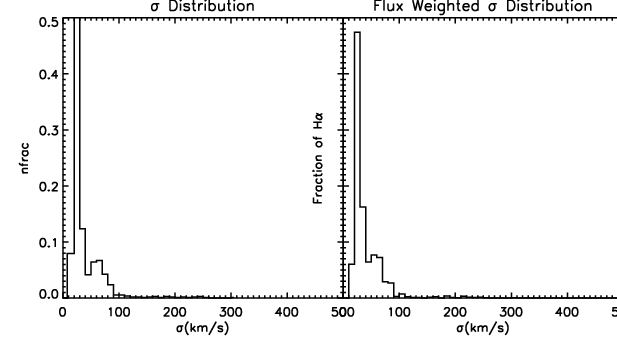}}
{\includegraphics[width=\textwidth]{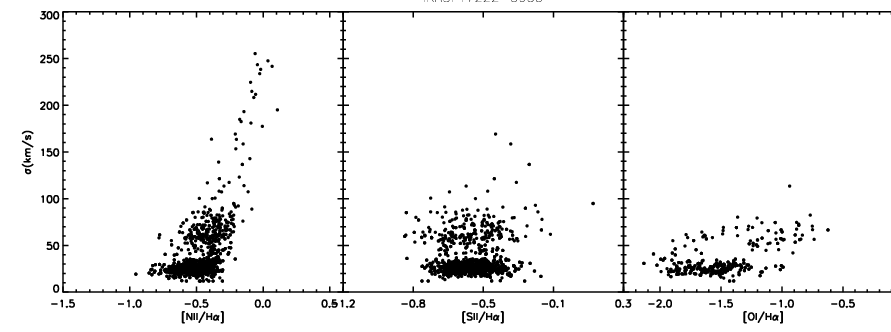}}
%{\includegraphics[width=\textwidth]{17222figBb_ind.png}}
\caption{Same as Figure B1, but for IRAS F17222-5953.}
\end{figure*}

\begin{figure*}%[htpb!]
\centering
{\Large \textbf{IRAS 17578-0400}}

{\includegraphics[width=\textwidth]{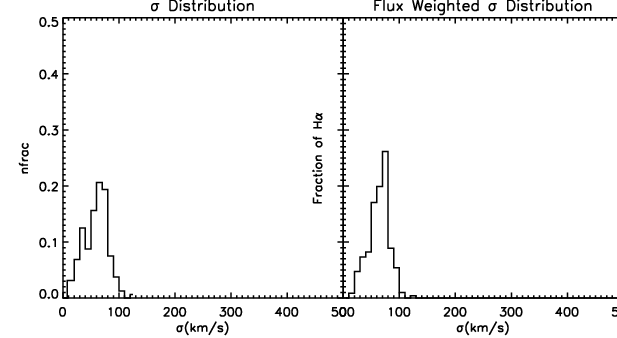}}
{\includegraphics[width=\textwidth]{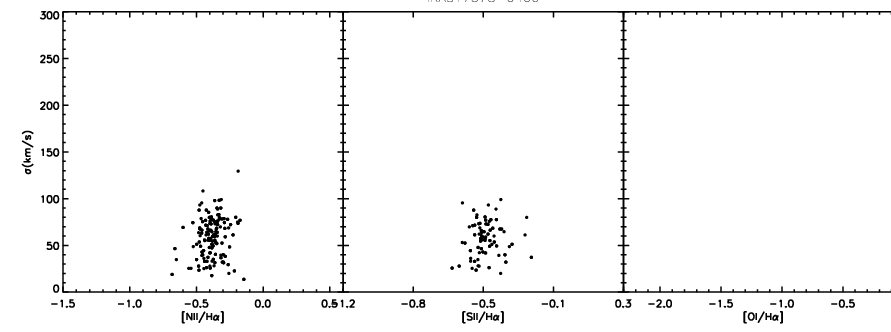}}
%{\includegraphics[width=\textwidth]{17578figBb_ind.png}}
\caption{Same as Figure B1, but for IRAS 17578-0400.}
\end{figure*}

\begin{figure*}%[htpb!]
\centering
{\Large \textbf{IRAS F18093-5744 N}}

{\includegraphics[width=\textwidth]{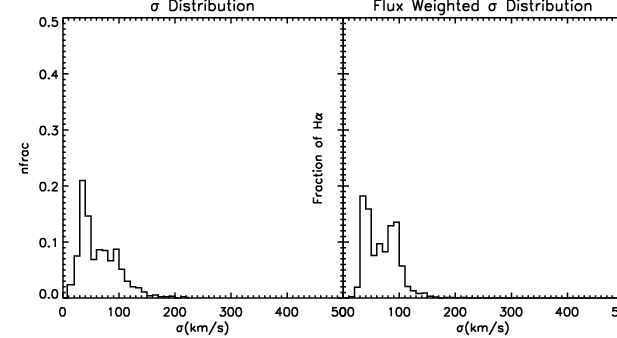}}
{\includegraphics[width=\textwidth]{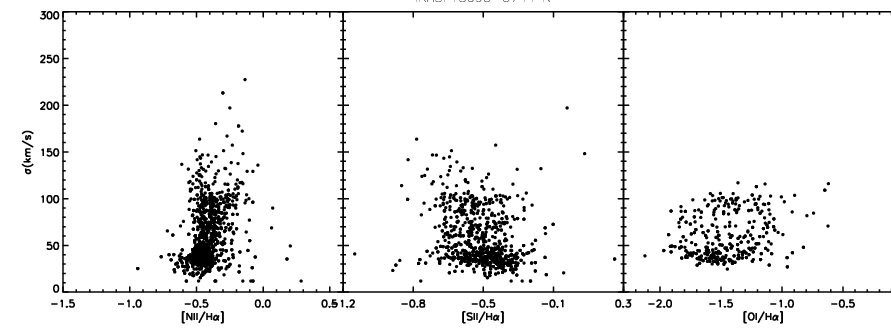}}
%{\includegraphics[width=\textwidth]{18093_nfigBb_ind.png}}
\caption{Same as Figure B1, but for IRAS F18093-5744 N.}
\end{figure*}

\begin{figure*}%[htpb!]
\centering
{\Large \textbf{IRAS F18093-5744 S}}

{\includegraphics[width=\textwidth]{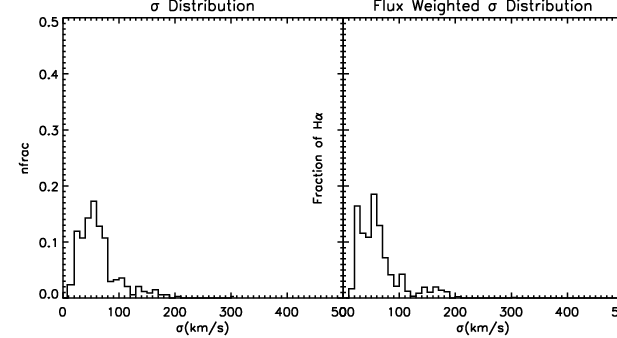}}
{\includegraphics[width=\textwidth]{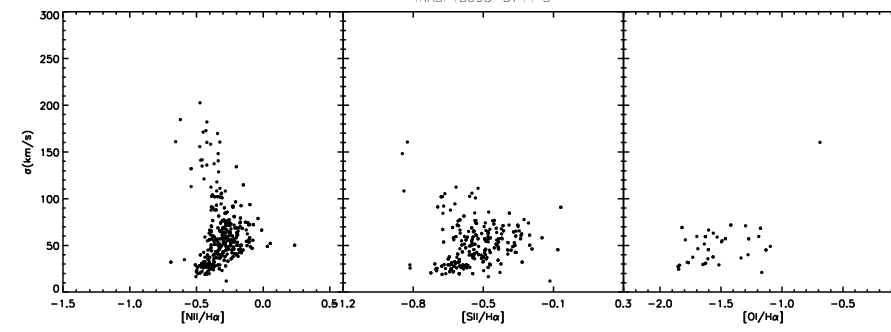}}
%{\includegraphics[width=\textwidth]{18093_sfigBb_ind.png}}
\caption{Same as Figure B1, but for IRAS F18093-5744 S.}
\end{figure*}

\begin{figure*}%[htpb!]
\centering
{\Large \textbf{IRAS F18093-5744 C}}

{\includegraphics[width=\textwidth]{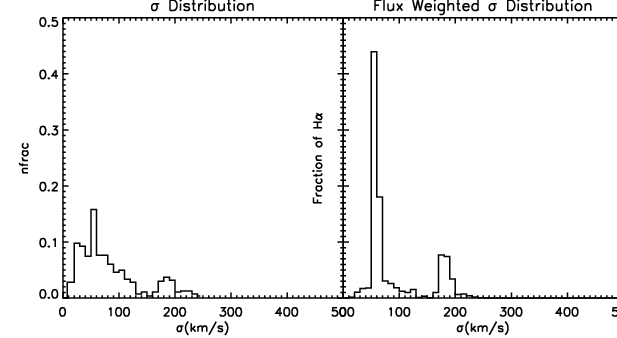}}
{\includegraphics[width=\textwidth]{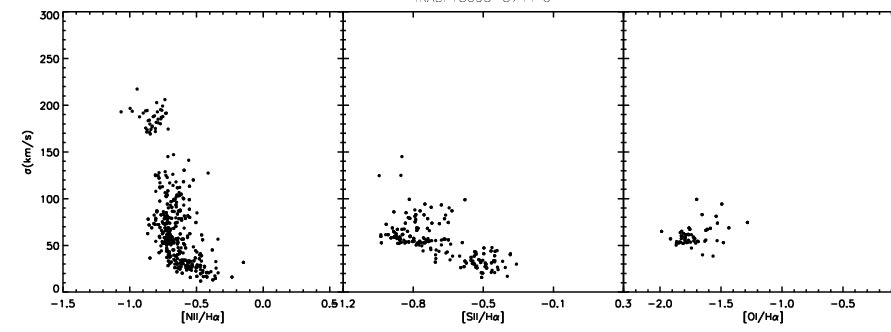}}
%{\includegraphics[width=\textwidth]{18093_cfigBb_ind.png}}
\caption{Same as Figure B1, but for IRAS F18093-5744 C.}
\end{figure*}

\begin{figure*}%[htpb!]
\centering
{\Large \textbf{IRAS F18293-3413}}

{\includegraphics[width=\textwidth]{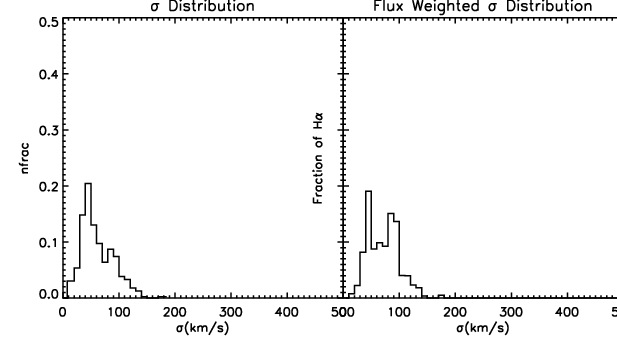}}
{\includegraphics[width=\textwidth]{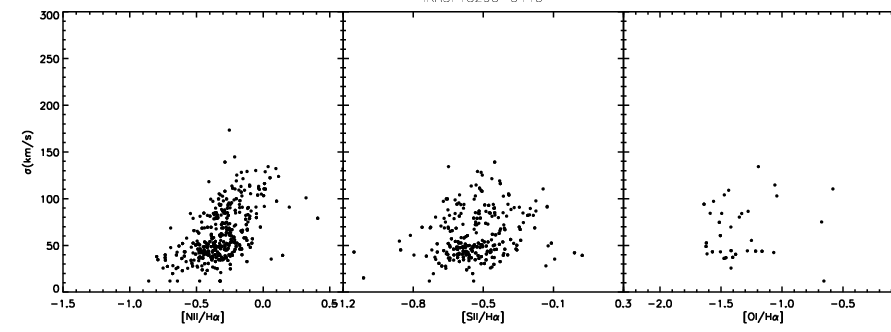}}
%{\includegraphics[width=\textwidth]{18293figBb_ind.png}}
\caption{Same as Figure B1, but for IRAS F18293-3413.}
\end{figure*}

\begin{figure*}%[htpb!]
\centering
{\Large \textbf{IRAS F18341-5732}}

{\includegraphics[width=\textwidth]{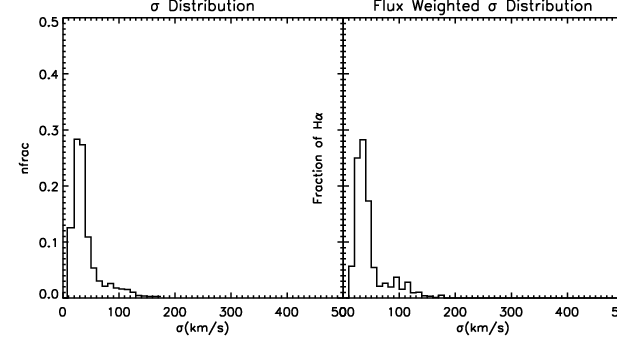}}
{\includegraphics[width=\textwidth]{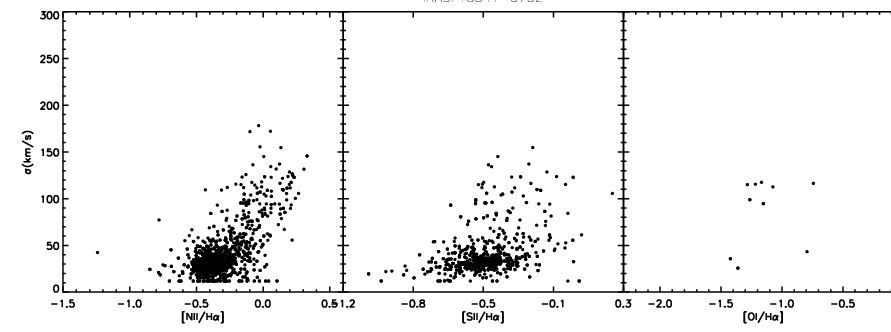}}
%{\includegraphics[width=\textwidth]{18341figBb_ind.png}}
\caption{Same as Figure B1, but for IRAS F18341-5732.}
\end{figure*}

\begin{figure*}%[htpb!]
\centering
{\Large \textbf{IRAS F19115-2124}}

{\includegraphics[width=\textwidth]{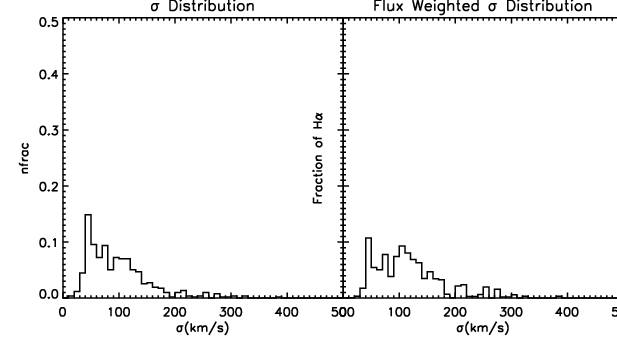}}
{\includegraphics[width=\textwidth]{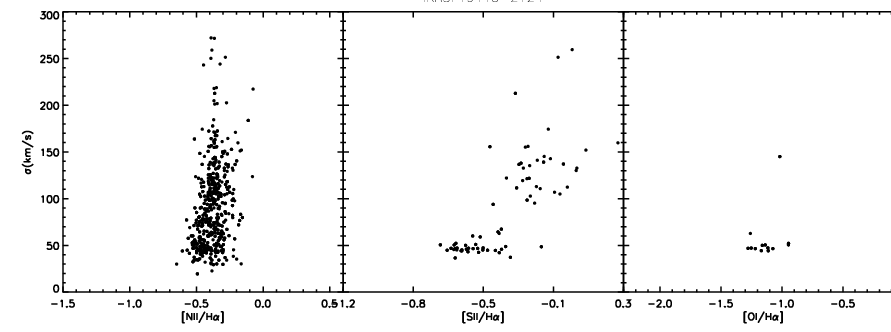}}
%{\includegraphics[width=\textwidth]{19115figBb_ind.png}}
\caption{Same as Figure B1, but for IRAS F19115-2124.}
\end{figure*}

\begin{figure*}%[htpb!]
\centering
{\Large \textbf{IRAS F20551-4250}}

{\includegraphics[width=\textwidth]{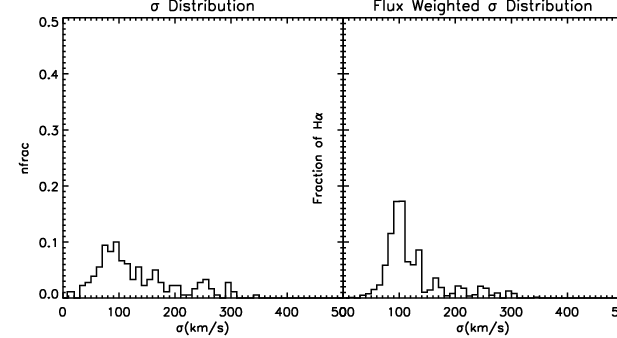}}
{\includegraphics[width=\textwidth]{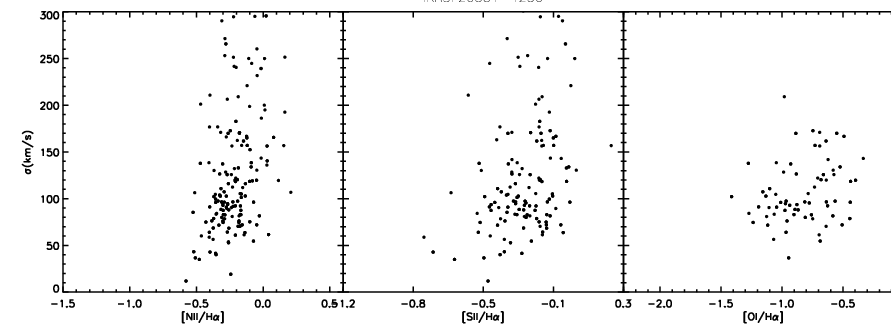}}
%{\includegraphics[width=\textwidth]{20551figBb_ind.png}}
\caption{Same as Figure B1, but for IRAS F20551-4250.}
\end{figure*}

\clearpage

\begin{figure*}%[htpb!]
\centering
{\Large \textbf{IRAS F21330-3846}}

{\includegraphics[width=\textwidth]{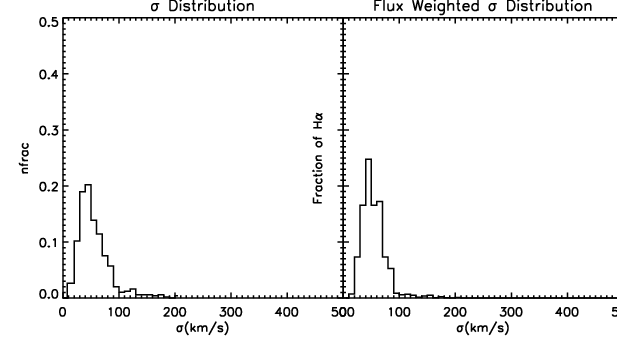}}
{\includegraphics[width=\textwidth]{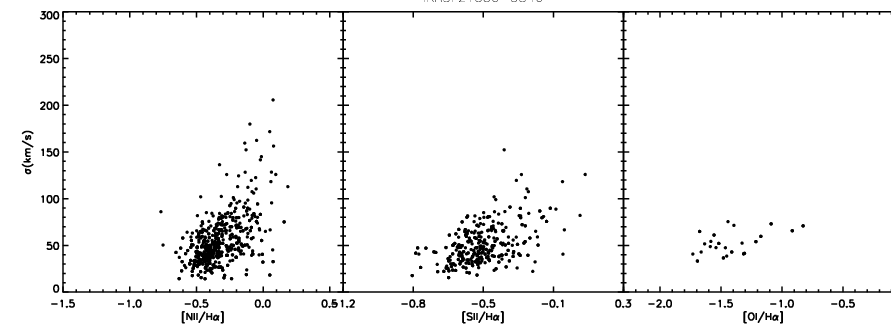}}
%{\includegraphics[width=\textwidth]{21330figBb_ind.png}}
\caption{Same as Figure B1, but for IRAS F21330-3846}
\end{figure*}

\begin{figure*}%[htpb!]
\centering
{\Large \textbf{IRAS F21453-3511}}

\clearpage

{\includegraphics[width=\textwidth]{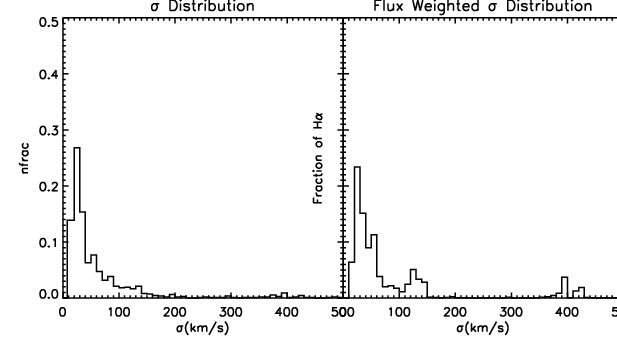}}
{\includegraphics[width=\textwidth]{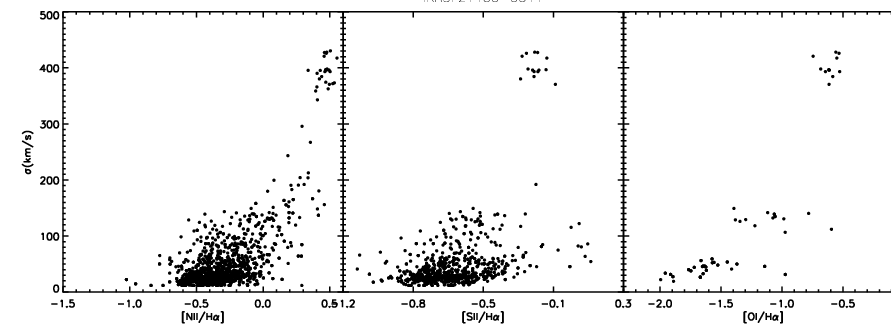}}
%{\includegraphics[width=\textwidth]{21453figBb_ind.png}}
\caption{Same as Figure B1, but for IRAS F21453-3511}
\end{figure*}

\begin{figure*}%[htpb!]
\centering
{\Large \textbf{IRAS F22467-4906}}

{\includegraphics[width=\textwidth]{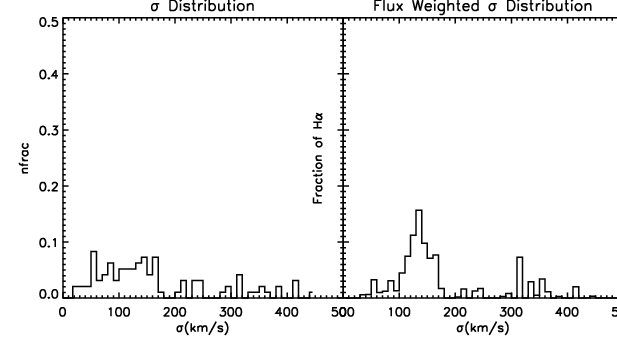}}
{\includegraphics[width=\textwidth]{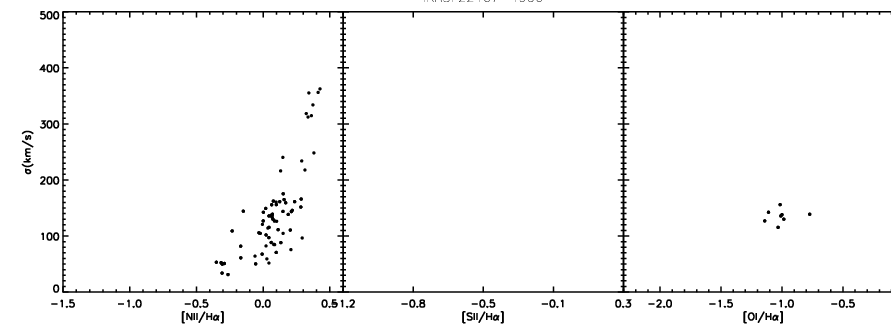}}
%{\includegraphics[width=\textwidth]{22467figBb_ind.png}}
\caption{Same as Figure B1, but for IRAS F22467-4906.}
\end{figure*}

\begin{figure*}%[htpb!]
\centering
{\Large \textbf{IRAS F23128-5919}}

{\includegraphics[width=\textwidth]{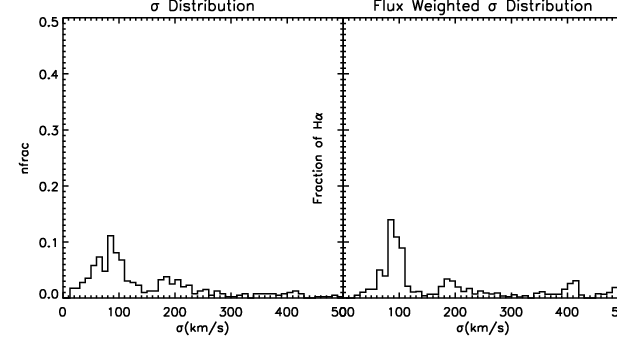}}
{\includegraphics[width=\textwidth]{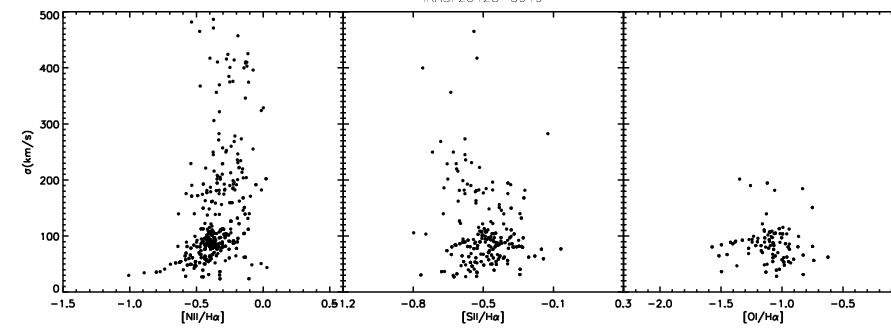}}
%{\includegraphics[width=\textwidth]{23128figBb_ind.png}}
\caption{Same as Figure B1, but for IRAS F23128-5919.}
\end{figure*}

\clearpage

%\onecolumn

\section{Velocity Dispersion Classification Maps}
This section contains maps of the distribution of low, moderate and high $\sigma$ for each velocity component fit. Isolated galaxies like IRAS F17222-5953 and widely separated pairs like IRAS F13373+0105 show primarily low and moderate (blue and green) velocity dispersion distributions, consistent with their emission being dominated primarily by star formation while the middle and late stage mergers show widespread high-$\sigma$ emission in at least one velocity component, consistent with a strong contribution from shocks. These maps provide a spatially resolved complement to the interpretation of Figs 5, 6, 8, 9 and 10 as discussed in Section 7. 

\begin{figure*}[htpb!]
\centering
{\Large \textbf{IRAS F01053-1746}}

{\includegraphics[height=0.55\textheight]{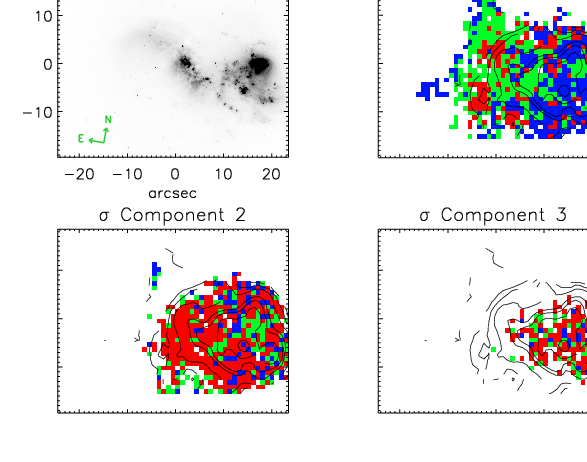}}
\caption{Maps of each velocity component color coded by velocity dispersion classification with cutoffs as shown in Figs. 5 \& 6 and discussed in Section 7: blue, low-$\sigma$ ($\sigma<50km~s^{-1}$), green mid-$\sigma (50km~s^{-1}<\sigma<90km~s^{-1})$ and high-$\sigma (\sigma>90km~s^{-1})$ velocity dispersion classification. IRAS F01053-1746 shows a strong combination of HII-region and turbulent star formation, seen as blue and green components as well as a widespread high-$\sigma$ component from shocks. Many of the mid to late stage mergers show a similar pattern. The upper left panel shows the galaxy image for comparison and the maps are shown with H$\alpha$ contours overlaid, as with the figures in Appendix A.}
\end{figure*}

\begin{figure}%[htpb!]
\centering
{\Large \textbf{IRAS F02072-1025}}

{\includegraphics[height=0.85\textwidth,angle=90]{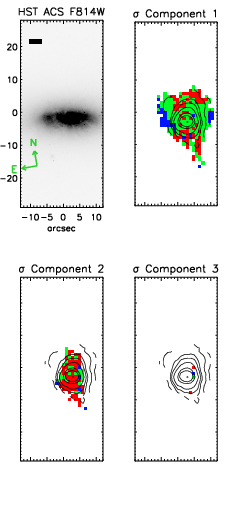}}
\caption{Same as Figure C1, but for IRAS F02072-1025.}
\end{figure}

\begin{figure}%[htpb!]
\centering
{\Large \textbf{IRAS F06076-2139}}

{\includegraphics[height=0.85\textwidth,angle=90]{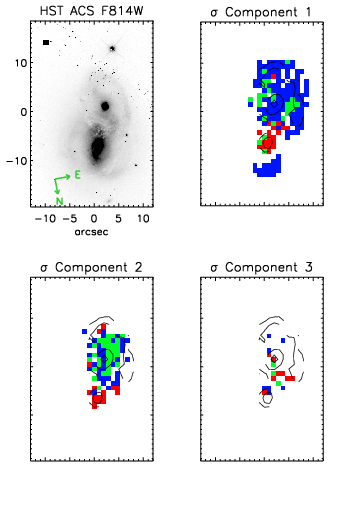}}
\caption{Same as Figure C1, but for IRAS F06076-2139.}
\end{figure}

\begin{figure*}%[htpb!]
\centering
{\Large \textbf{IRAS F08355-4944}}

{\includegraphics[height=0.85\textwidth,angle=90]{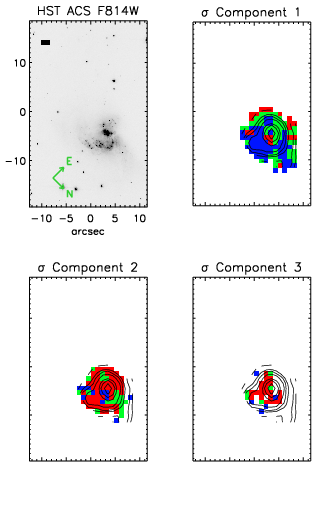}}
\caption{Same as Figure C1, but for IRAS 08355-4944.}
\end{figure*}

\begin{figure*}%[htpb!]
\centering
{\Large \textbf{IRAS F10038-3338}}

{\includegraphics[height=0.85\textwidth,angle=90]{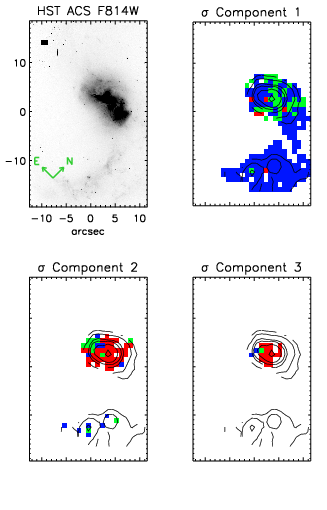}}
\caption{Same as Figure C1, but for IRAS F10038-3338.}
\end{figure*}

\begin{figure*}%[htpb!]
\centering
{\Large \textbf{IRAS F10257-4339}}

{\includegraphics[width=0.65\textwidth]{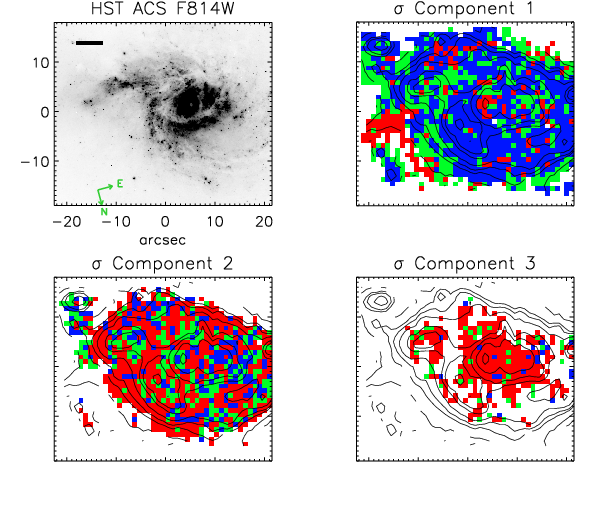}}
\caption{Same as Figure C1, but for IRAS F10257-4339.}
\end{figure*}

\begin{figure*}%[htpb!]
\centering
{\Large \textbf{IRAS F12043-3140}}

{\includegraphics[height=0.85\textwidth,angle=90]{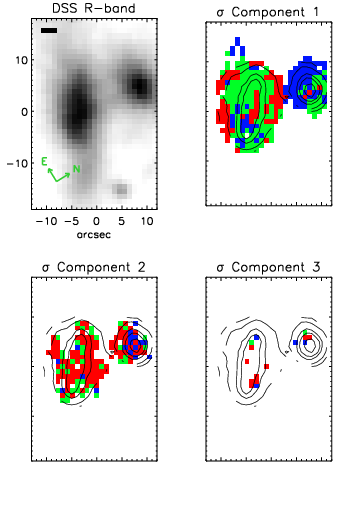}}
\caption{Same as Figure C1, but for IRAS F12043-3140.}
\end{figure*}

\begin{figure*}%[htpb!]
\centering
{\Large \textbf{IRAS F12592+0436}}

{\includegraphics[height=0.85\textwidth,angle=90]{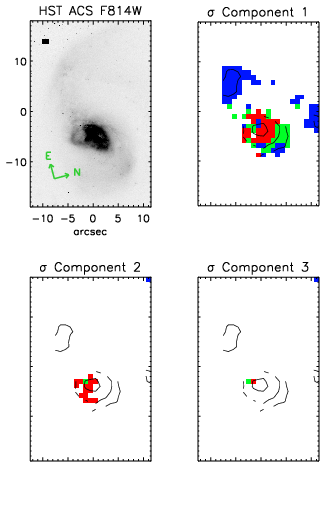}}
\caption{Same as Figure C1, but for IRAS F12592+0436.}
\end{figure*}

\begin{figure*}%[htpb!]
\centering
{\Large \textbf{IRAS 13120-5453}}

{\includegraphics[height=0.85\textwidth,angle=90]{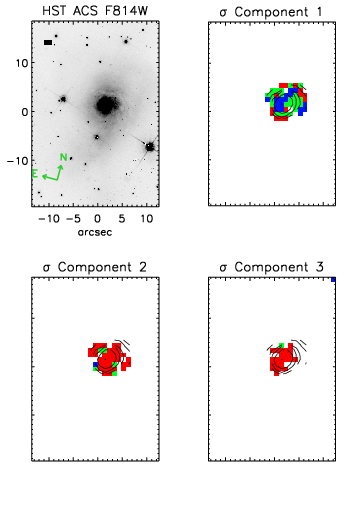}}
\caption{Same as Figure C1, but for IRAS 13120-5453.}
\end{figure*}

\begin{figure*}%[htpb!]
\centering
{\Large \textbf{IRAS F13373+0105 W}}

{\includegraphics[height=0.85\textwidth,angle=90]{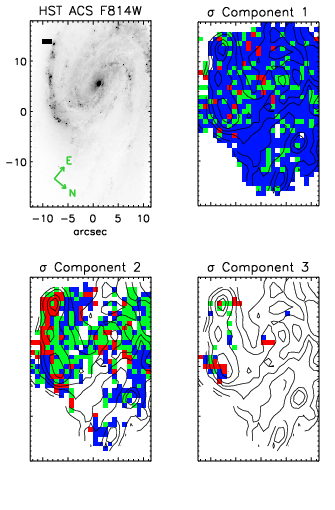}}
\caption{Same as Figure C1, but for IRAS F13373+0105 W.}
\end{figure*}

\begin{figure*}%[htpb!]
\centering
{\Large \textbf{IRAS F13373+0105 E}}

{\includegraphics[height=0.85\textwidth,angle=90]{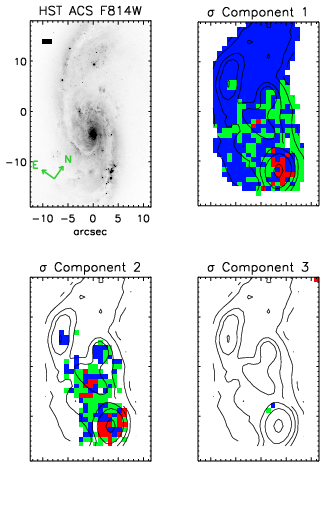}}
\caption{Same as Figure C1, but for IRAS F13373+0105 E.}
\end{figure*}

\begin{figure*}%[htpb!]
\centering
{\Large \textbf{IRAS F15107+0724}}

{\includegraphics[height=0.85\textwidth,angle=90]{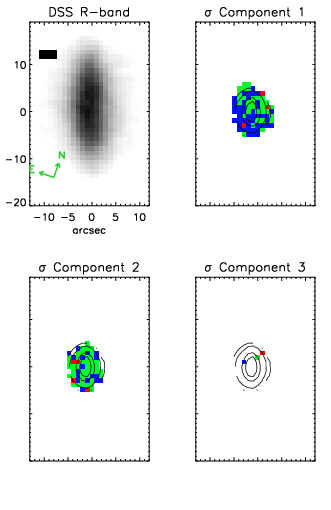}}
\caption{Same as Figure C1, but for IRAS F15107+0724.}
\end{figure*}

\begin{figure*}%[htpb!]
\centering
{\Large \textbf{IRAS F16164-0746}}

{\includegraphics[height=0.85\textwidth,angle=90]{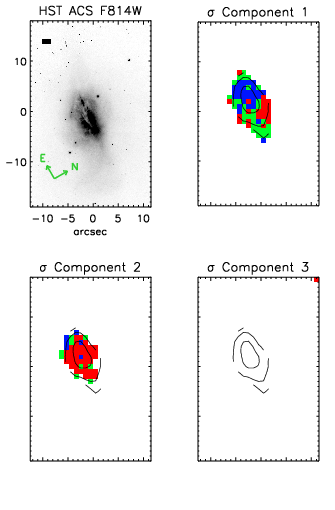}}
\caption{Same as Figure C1, but for IRAS F16164-0746.}
\end{figure*}

\begin{figure*}%[htpb!]
\centering
{\Large \textbf{IRAS F16399-0937}}

{\includegraphics[height=0.85\textwidth,angle=90]{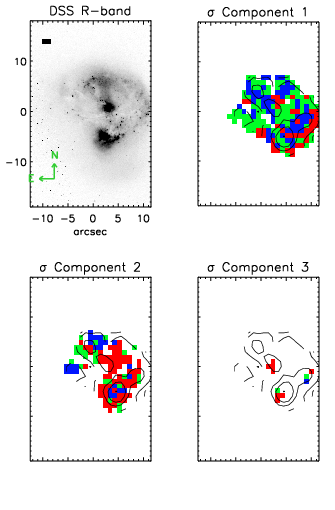}}
\caption{Same as Figure C1, but for IRAS F16399-0937.}
\end{figure*}

\begin{figure*}%[htpb!]
\centering
{\Large \textbf{IRAS F16443-2915 N}}

{\includegraphics[height=0.85\textwidth,angle=90]{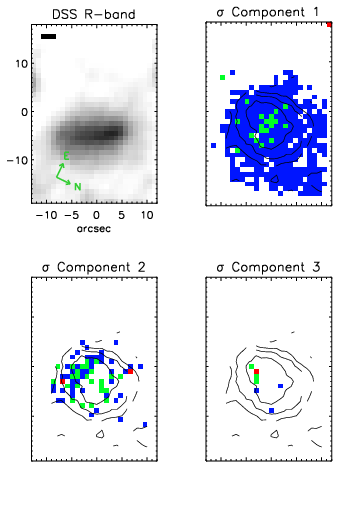}}
\caption{Same as Figure C1, but for IRAS F16443-2915 N.}
\end{figure*}

\begin{figure*}%[htpb!]
\centering
{\Large \textbf{IRAS F16443-2915 S}}

{\includegraphics[height=0.85\textwidth,angle=90]{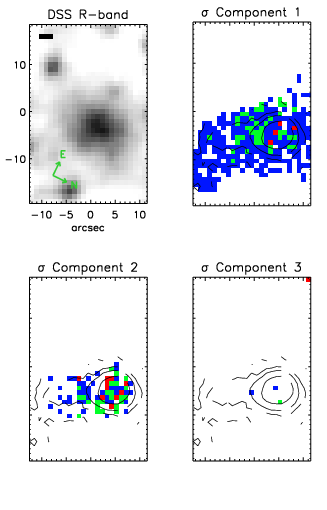}}
\caption{Same as Figure C1, but for IRAS F16443-2915 S.}
\end{figure*}

\begin{figure*}%[htpb!]
\centering
{\Large \textbf{IRAS F17138-1017}}

{\includegraphics[height=0.85\textwidth,angle=90]{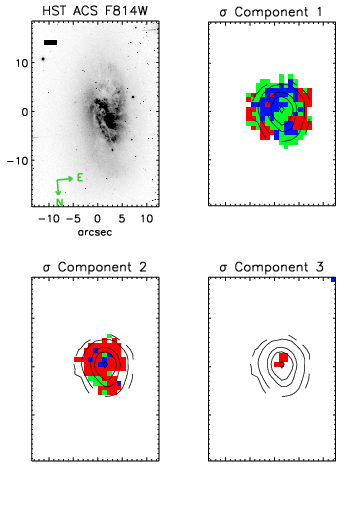}}
\caption{Same as Figure C1, but for IRAS F17138-1017.}
\end{figure*}

\clearpage

\begin{figure*}%[htpb!]
\centering
{\Large \textbf{IRAS F17207-0014}}

{\includegraphics[height=0.85\textwidth,angle=90]{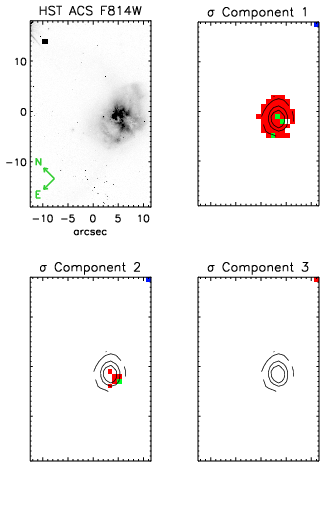}}
\caption{Same as Figure C1, but for IRAS F17207-0014.}
\end{figure*}

\begin{figure*}%[htpb!]
\centering
{\Large \textbf{IRAS F17222-5953}}

{\includegraphics[height=0.85\textwidth,angle=90]{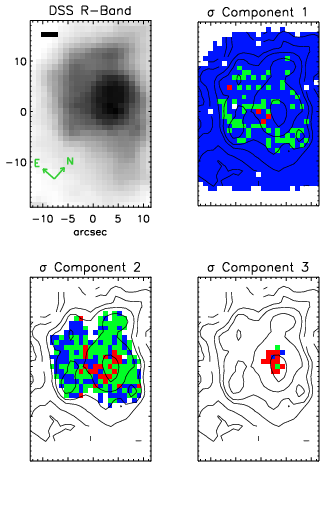}}
\caption{Same as Figure C1, but for IRAS F17222-5953.}
\end{figure*}

\begin{figure*}%[htpb!]
\centering
{\Large \textbf{IRAS 17578-0400}}

{\includegraphics[height=0.80\textwidth,angle=90]{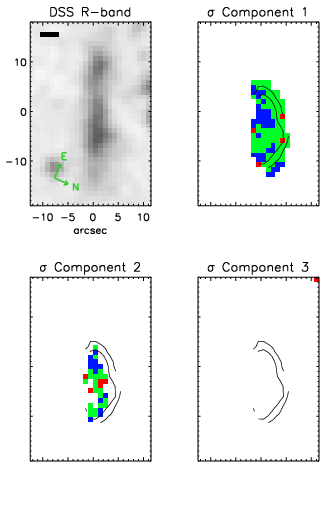}}
\caption{Same as Figure C1, but for IRAS 17578-0400.}
\end{figure*}

\begin{figure*}%[htpb!]
\centering
{\Large \textbf{IRAS F18093-5744 N}}

{\includegraphics[height=0.80\textwidth,angle=90]{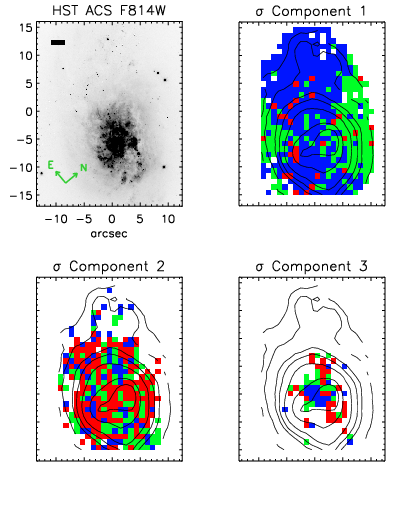}}
\caption{Same as Figure C1, but for IRAS F18093-5744 N.}
\end{figure*}

\begin{figure*}%[htpb!]
\centering
{\Large \textbf{IRAS F18093-5744 S}}

{\includegraphics[height=0.85\textwidth,angle=90]{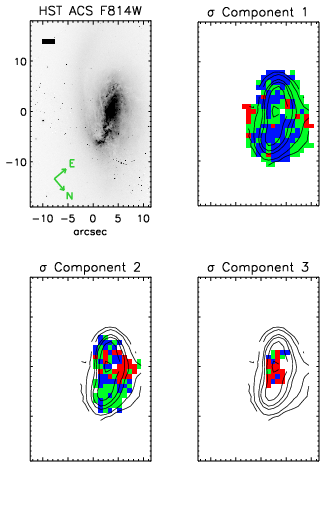}}
\caption{Same as Figure C1, but for IRAS F18093-5744 S.}
\end{figure*}

\begin{figure*}%[htpb!]
\centering
{\Large \textbf{IRAS F18093-5744 C}}

{\includegraphics[height=0.90\textwidth]{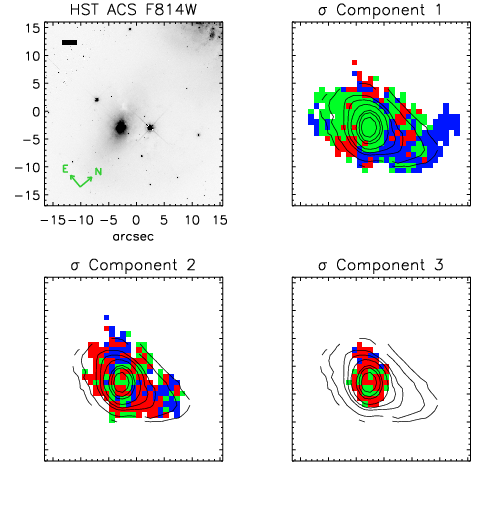}}
\caption{Same as Figure C1, but for IRAS F18093-5744 C.}
\end{figure*}

\begin{figure*}%[htpb!]
\centering
{\Large \textbf{IRAS F18293-3413}}

{\includegraphics[height=0.85\textwidth,angle=90]{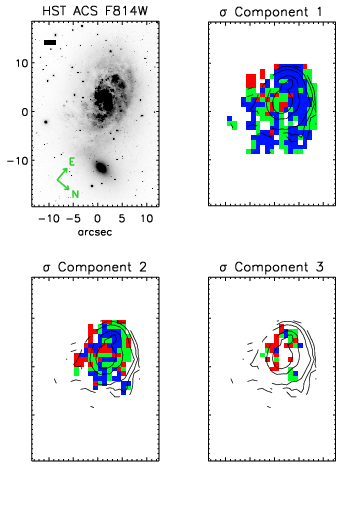}}
\caption{Same as Figure C1, but for IRAS F18293-3413.}
\end{figure*}

\begin{figure*}%[htpb!]
\centering
{\Large \textbf{IRAS F18341-5732}}

{\includegraphics[height=0.90\textwidth]{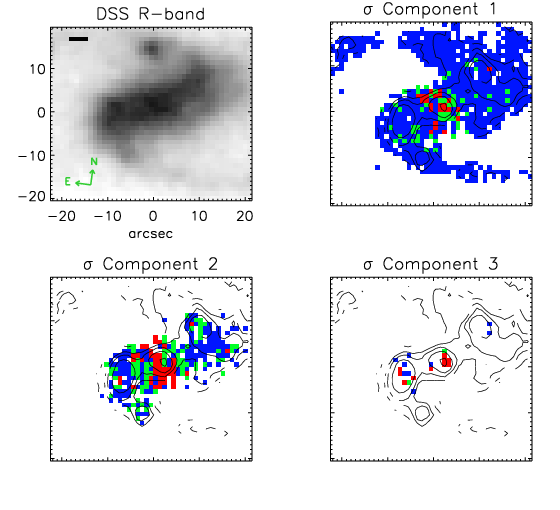}}
\caption{Same as Figure C1, but for IRAS F18341-5732.}
\end{figure*}

\begin{figure*}%[htpb!]
\centering
{\Large \textbf{IRAS F19115-2124}}

{\includegraphics[height=0.85\textwidth,angle=90]{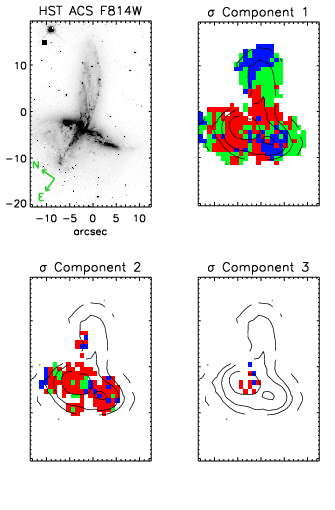}}
\caption{Same as Figure C1, but for IRAS F19115-2124.}
\end{figure*}

\begin{figure*}%[htpb!]
\centering
{\Large \textbf{IRAS F20551-4250}}

{\includegraphics[height=0.85\textwidth,angle=90]{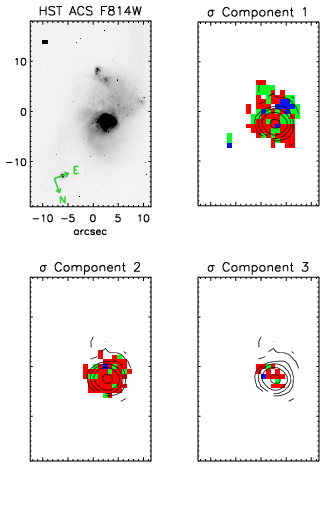}}
\caption{Same as Figure C1, but for IRAS F20551-4250.}
\end{figure*}

\begin{figure*}%[htpb!]
\centering
{\Large \textbf{IRAS F21330-3846}}

{\includegraphics[height=0.85\textwidth,angle=90]{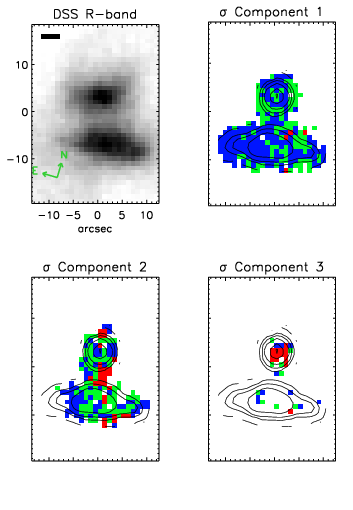}}
\caption{Same as Figure C1, but for IRAS F21330-3846}
\end{figure*}

\begin{figure*}%[htpb!]
\centering
{\Large \textbf{IRAS F21453-3511}}

{\includegraphics[height=0.85\textwidth,angle=90]{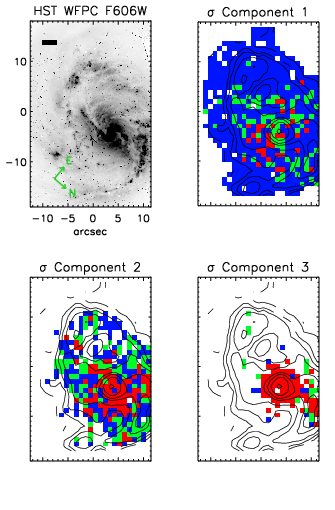}}
\caption{Same as Figure C1, but for IRAS F21453-3511}
\end{figure*}

\begin{figure*}%[htpb!]
\centering
{\Large \textbf{IRAS F22467-4906}}

{\includegraphics[width=0.80\textwidth]{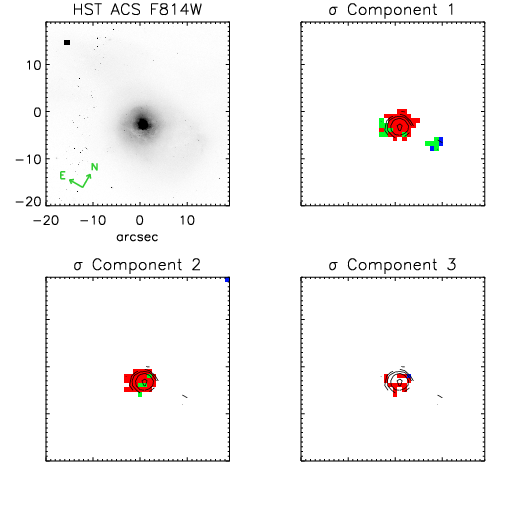}}
\caption{Same as Figure C1, but for IRAS F22467-4906.}
\end{figure*}

\begin{figure*}%[htpb!]
\centering
{\Large \textbf{IRAS F23128-5919}}

{\includegraphics[height=0.85\textwidth,angle=90]{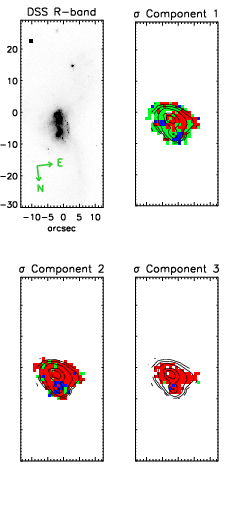}}
\caption{Same as Figure C1, but for IRAS F23128-5919.}
\end{figure*}

\clearpage

\end{document}